\documentclass[reprint, secnumarabic, amssymb, nobibnotes, aps, pra,superscriptaddress]{revtex4-1}
\usepackage[utf8]{inputenc}
\usepackage[left=1in,right=1in,top=2cm,bottom=2cm]{geometry}
\usepackage{graphicx,lipsum,amsmath}

\begin{document}

\title{Topological Chiral Edge States in a Synthetic Dimension of Atomic Trap States}
\author{David G. Reid}
\affiliation{School of Physics and Astronomy, University of Birmingham, Edgbaston, Birmingham B15 2TT, United Kingdom}
\author{Christopher Oliver}
\affiliation{National Quantum Computing Centre, Rutherford Appleton Laboratory, Harwell Campus, Didcot, Oxfordshire, OX11 0QX.}
\author{Patrick Regan}
\affiliation{School of Physics and Astronomy, University of Birmingham, Edgbaston, Birmingham B15 2TT, United Kingdom}
\author{Aaron Smith}
\affiliation{School of Physics and Astronomy, University of Birmingham, Edgbaston, Birmingham B15 2TT, United Kingdom}
\author{Thomas Easton}
\affiliation{School of Physics and Astronomy, University of Birmingham, Edgbaston, Birmingham B15 2TT, United Kingdom}
\author{Grazia Salerno}
\affiliation{Department of Applied Physics, Aalto University School of Science, FI-00076 Aalto, Finland}
\author{Giovanni Barontini}
\affiliation{School of Physics and Astronomy, University of Birmingham, Edgbaston, Birmingham B15 2TT, United Kingdom}
\author{Nathan Goldman}
\affiliation{Laboratoire Kastler Brossel, Collège de France, CNRS, ENS-Université PSL, Sorbonne Université, 11 Place Marcelin Berthelot, 75005 Paris, France}
\affiliation{International Solvay Institutes, 1050 Brussels, Belgium}
\affiliation{Center for Nonlinear Phenomena and Complex Systems, Université Libre de Bruxelles, CP 231, Campus Plaine, 1050 Brussels, Belgium}
\author{Hannah M. Price}
\affiliation{School of Physics and Astronomy, University of Birmingham, Edgbaston, Birmingham B15 2TT, United Kingdom}

\begin{abstract}
A key hallmark of quantum Hall physics is the existence of topological chiral states at the system boundary. Signatures of these edge states have been experimentally observed in cold atoms by using different approaches, including notably that of ``synthetic dimensions" in which internal states are coupled together and reinterpreted as sites along an artificial spatial dimension. However, previous atomic synthetic dimension implementations have been limited to relatively small system sizes with inflexible boundaries. In this paper, we propose instead how to use a synthetic dimension of atomic trap states to observe chiral edge states in a large quantum Hall system with a tunable edge. We present numerical simulations for relevant experimental parameters, showing how this scheme may be used to probe the properties and robustness of the edge states to defects. Our work opens the way for future experiments in topological physics with synthetic dimensions, while also providing new ways to manipulate and control highly-excited trap states.    
\end{abstract}
\maketitle

  \section{Introduction \label{sec:intro}}
  
Topological phases of matter are interesting materials which host robust states localised on the system surface~\cite{RMP_TI,RMP_TI2}. In the 2D quantum Hall effect, for example, these states are chiral and propagate one-way around the 1D edge of the system~\cite{klitzing1980new, VonKlitzing1986}. Such topological edge states are protected against disorder and fabrication imperfections as they cannot be removed without changing the topological invariants of the bulk material, either through closing the energy gap between bands or breaking symmetries~\cite{Chiu:2016RMP}.     

While first discovered in electronic materials~\cite{RMP_TI,RMP_TI2}, topological phases of matter have since become an important topic of research also in many other fields, including ultracold atoms~\cite{Goldman:2014ROPP,Goldman:2016NatPhys,cooper2018} and photonics~\cite{lu2016topological,khanikaev2017two,Ozawa_2019,price2022roadmap}. Within this context, the approach of ``synthetic dimensions" has emerged as a powerful way to implement artificial gauge fields for neutral atoms and photons, and hence to engineer topological phases of matter~\cite{Boada2012, Celi2014, Mancini2015, Stuhl2015, Livi2016, Kolkowitz2017, Roell2023, Chalopin2020, chen2024interaction,chen2024strongly,lu2024probing, lienhard2020realization,Kanungo2021, Ozawa:2017PRL, Price2017, Gadway2016, Viebahn2019, An2021, PRXQuantum.5.010310, bouhiron2024realization, cai2019experimental,Sundar2018, kang2020creutz,Ozawa_2019, lustig2019photonic,dutt2020single,balvcytis2021synthetic,chen2021real,lustig2021topological,oliver2023bloch,oliver2023artificial,dinh2024reconfigurable,leefmans2022topological,balvcytis2021synthetic,ehrhardt2023perspective,PhysRevLett.132.130601}. The central idea of this approach is to re-interpret some set of modes or degrees of freedom, such as internal atomic states~\cite{Boada2012, Celi2014, Mancini2015, Stuhl2015, Chalopin2020, lienhard2020realization,Kanungo2021, Livi2016, Kolkowitz2017,Roell2023, chen2024interaction,chen2024strongly,lu2024probing} or motional states~\cite{Gadway2016, Viebahn2019, An2021, PRXQuantum.5.010310,Price2017}, as sites along a synthetic spatial dimension by applying a suitable external coupling. This has led to investigations of many interesting phenomena, including quantum Hall ladders~\cite{Mancini2015, Stuhl2015, Gadway2016,PhysRevLett.132.130601}, non-Hermitian topological bands~\cite{Wang2021} and even 4D quantum Hall physics~\cite{Viebahn2019, Boada2012,price2015, bouhiron2024realization}. 

In ultracold atoms, synthetic dimensions notably  allowed for the first experiments on the ``skipping orbits" stemming from quantum-Hall chiral edge states~\cite{Mancini2015, Stuhl2015}; these were observed thanks to the presence of sharp boundaries along the synthetic dimension, resulting from only coupling together a finite number of hyperfine states. However, these early experiments were severely limited to only a few synthetic dimension sites making it impossible to distinguish edge and bulk physics. Since then, experiments in synthetic dimensions with cold atoms have expanded, but still remain limited to a couple of tens of sites along the synthetic dimension~\cite{Chalopin2020, An2021, Roell2023}. They have also recently been overtaken by 2D real-space cold-atom experiments (i.e. without synthetic dimensions), which have been able to explore larger systems with more tunable edge potentials~\cite{braun2024real, yao2023observation,Wang2024}.

In this paper, we propose how to overcome previous limitations to explore chiral topological states in a system with a very long synthetic dimension made up of atomic trap states~\cite{Price2017}. This builds on our recent experiment using this scheme, where we demonstrated 1D Bloch oscillations across many tens of states~\cite{oliver2023bloch}. We now theoretically show how to extend this set-up, by including another real spatial dimension and an artificial magnetic field, in order to realise a type of 2D quantum Hall system known as the coupled wire model~\cite{Kane2002, Budich_2017, Chalopin2020}. Our proposed scheme has the key benefit that we are able to engineer a tunable upper edge in the synthetic dimension, allowing us to easily change both the ``softness" of the boundary as well as the effective width of the resulting system, and hence the location and properties of the associated edge states. We also demonstrate how the controllablity of the long synthetic dimension can be exploited to explore both bulk and edge physics, and to imprint defects for testing the robustness of the chiral edge modes. 

Experimentally, our proposal is based on creating spatially- and temporally-varying optical potentials, e.g. using a digital micromirror device, and so does not require any additional apparatus as compared to the previous experiment~\cite{oliver2023bloch}. This work therefore paves the way for future experiments to explore chiral edge states in this type of set-up. In the future, it will also be especially interesting to explore the role of inter-particle interactions in this model, as the usual mean-field contact interactions in real space will lead to exotic long-range interactions along the synthetic dimension~\cite{Price2017}, and so may lead to new types of interacting ground states. Moreover, it is important to note that the topological edge states described in this work are physically very different from those found in other realizations, such as in real-space systems~\cite{braun2024real, yao2023observation} or with internal atomic states~\cite{Mancini2015, Stuhl2015, Chalopin2020}, as here they correspond to atoms moving in opposite real-space directions depending on whether they are in high-energy or low-energy states of a (perpendicularly-oriented) harmonic trap \cite{Salerno2019}. This therefore opens the way for new approaches to experimentally control different energy states, with potential applications in areas such as trapped and guided atom interferometry~\cite{Hu2018,Frank2014,Guarrera2015} and quantum thermodynamics~\cite{Vin2016,Quan2007,Uzdin2015}, when the control of highly-excited trapped states is required.     
\subsection*{Outline}
The outline of the paper is as follows: in Section~\ref{sec:scheme}, we introduce the proposed scheme and review how a synthetic dimension of atomic trap states can be engineered. We then show how to introduce a tunable upper edge along the synthetic dimension using, e.g. a digital micro-mirror device, in order to realise chiral edge state motion in the hybrid (real-synthetic) 2D plane. We discuss how to control the location and properties of the chiral edge states, firstly, via tuning the length of the synthetic dimension in Section~\ref{section:1a)}, and secondly, by changing the ``softness" of the effective edge potential in the synthetic dimension in Section~\ref{section:1b}. In Section~\ref{section:3}, we then show how to introduce static impurities along the upper edge of the synthetic dimension, and hence visualise the robustness of the chiral edge motion. Finally, in Section~\ref{section:4}, we propose how to experimentally observe bulk cyclotron orbits within the same set-up, before drawing conclusions in Section~\ref{sec:conclusions}.    

  \begin{figure}
              \label{fig:scheme}

\includegraphics[width=1\columnwidth]      
      {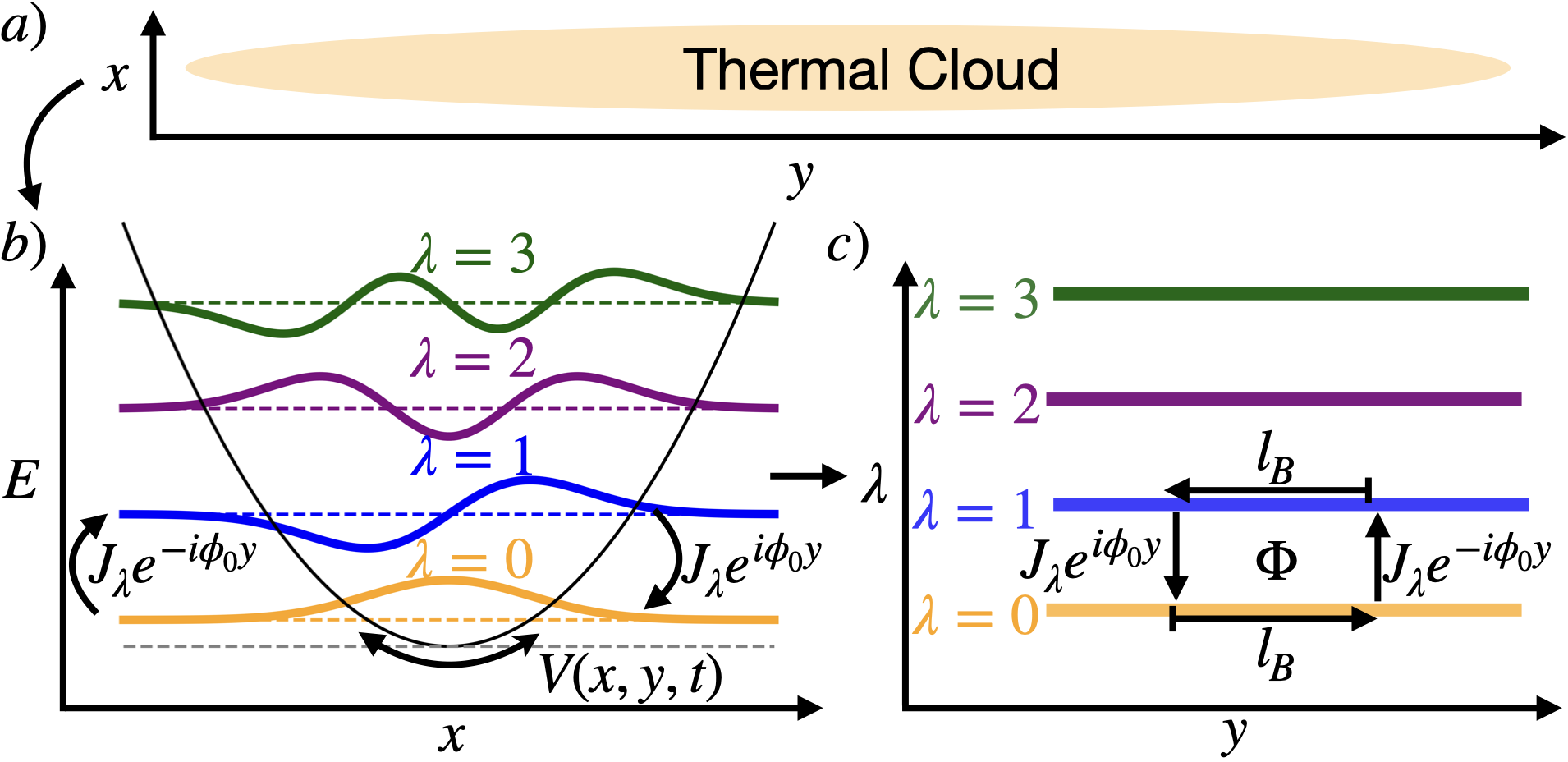}%
      \caption{A 2D schematic of the proposed experiment. a)~The elongated cigar-shaped harmonic trap, with $ \omega_x \!=\! \omega_z\!\equiv\! \omega \gg \omega_y$, projected into the relevant $x- y$ plane. 
      b)~The trap states along the $x$ direction are re-interpreted as sites in a synthetic dimension when the periodic driving potential $V(x, y, t)$ is applied. This driving potential is designed to introduce  nearest-neighbor hopping amplitudes with magnitude, $J_\lambda$ [Eq.~\ref{eq:J}], along the synthetic dimension. The initial driving phase, $\phi_0 y$, of $V(x, y, t)$ then appears as a $y$-dependent phase associated with the hopping between states. c) Combining the synthetic dimension indexed by $\lambda$ with the $y$ direction then leads to a 2D coupled wire model with an artificial magnetic flux quantum $\Phi$ through a plaquette of length $l_B\!\equiv\!2\pi  / \phi_0$ along the $y$-axis.}
      \label{fig:schematic}
     \end{figure}
      
              \section{Scheme} \label{sec:scheme}

In this Section, we will first review how to create a synthetic dimension of atomic trap states and then discuss how this approach can be used to realise the desired 2D quantum Hall coupled wire model~\cite{Kane2002, Budich_2017}. We will then briefly introduce the numerical simulations that will be used in the remainder of the paper to explore the behavior of this model.

\subsection{Realising the 2D Coupled Wire Model}
              
      Motivated by the recent experiment realising a synthetic dimension of atomic trap states~\cite{oliver2023bloch}, we start from an atomic gas in
        an elongated `cigar-shaped' harmonic trap with trapping frequencies along the $x$, $z$ and $y$ direction chosen to be, respectively,  $ \omega_x \!=\! \omega_z\!\equiv\! \omega \gg \omega_y$ [see Fig.~\ref{fig:schematic}(a)]. Hereafter, we assume that we can neglect the particle motion in the $z$ direction due to the strong trapping confinement, and hence we restrict our discussion to dynamics in the $x\!-\!y$ plane~\cite{oliver2023bloch}. In this setting, the motion along $x$ will be activated by external drive. The full 2D Hamiltonian is then given by
        \begin{eqnarray} \label{eq:full}
\hat{\mathcal{H}} (x, y, t) =\frac{\hat{p}^2_x + \hat{p}^2_y}{2m} + \frac{m}{2}(\omega^2 x^2 + \omega_y^2 y^2) + V(x, y, t) , 
\end{eqnarray}
where $m$ is the mass of a single atom and we choose the periodic shaking potential as~\cite{Price2017}
        \begin{equation}
        V(x, y, t) = \kappa x \cos(\omega_D t + \phi_0 y) ,
        \label{Eq.couplingpotential}
        \end{equation}
      where $ \phi_0 y$ is the $y$-dependent initial driving phase,  $\omega_D$ is the driving frequency and $\kappa$ is the driving strength. As articulated in section \ref{sec:intro}, we believe that a digital micro-mirror device is ideal for this style of potential due to its ability to create complex $x$ and $y$ spatially dependent potentials \cite{Gauthier_2016}. Additionally, the device can create linear potentials by utilizing its rapid update rates, over 20 kHz, to time average, which can additionally be used in conjugation with half toning to increase linearity further \cite{Amico_2021}. 

Throughout, we neglect the role of inter-particle interactions, focusing instead on the single-particle physics; this is a good approximation for sufficiently dilute thermal atomic clouds, as in the experiment of Ref.~\onlinecite{oliver2023bloch}, although it will be very interesting in the future to include interaction effects~\cite{Price2017}.

    Indexing the harmonic trap states along $x$ with $\lambda\! =\!0,1,2...$, the stroboscopic motion in the $x$-direction can be captured by an effective time-independent Floquet tight-binding model~\cite{Price2017}
      \begin{eqnarray}
      \hat{{H}}_{\text{synth}}(\lambda, y) &\approx &     \sum_{\lambda}J_{\lambda}| \lambda -1, y\rangle \langle \lambda, y|e^{i \phi_0 y} + \text{h.c.} \nonumber \\
      & & +\Delta \sum_{\lambda} \lambda |\lambda, y \rangle \langle \lambda, y| ,
      \label{eq:main}
      \end{eqnarray} where $| \lambda, y\rangle$ denotes the trap-state basis, $\hbar=1$ throughout, $\Delta \!\equiv \! \omega - \omega_D$ is the (small) drive detuning and $J_{\lambda}$ is the effective hopping amplitude,
      \begin{equation}
      J_{\lambda} = \kappa \sqrt{\frac{\lambda }{8m \omega}}. \label{eq:J}
      \end{equation}
     This effective Hamiltonian therefore describes nearest-neighbor couplings along a synthetic dimension of atomic trap states [see Fig.~\ref{fig:schematic}(b)] together with a force along the synthetic dimension, $F\!\equiv\! -  \Delta$, which can induce Bloch oscillations as observed experimentally in Ref.~\cite{oliver2023bloch} and as will be used in Section~\ref{section:4}.  
      Note that Eqs.~\ref{eq:main} and~\ref{eq:J} are only valid within a rotating-wave approximation, which holds for $ \omega_D \approx \omega\! \gg\! J_{\lambda}$ and $\Delta$~\cite{Price2017}. Other driving potentials can also be used to couple the atomic trap states in a similar manner, leading to different forms of $J_\lambda$, as in the experiment of Ref.~\onlinecite{oliver2023bloch} and as discussed further in Appendix A.  
      
      Setting $\Delta=0$ and including the kinetic energy along the $y$ direction, we then obtain the desired effective stroboscopic 2D model given by
          \begin{equation}
       \hat{\mathcal{H}}_{CWM}(\lambda, y) = \frac{\hat{p}_y^2}{2m} + \hat{{H}}_{\text{synth}}(\lambda, y),
       \label{eq:2dnotrap}
      \end{equation}
      which is represented schematically in Fig.~\ref{fig:schematic}(c). This is a coupled wire model~\cite{Kane2002, Budich_2017, Chalopin2020}, as the particles move within a discrete array of ``wires" that are oriented along $y$, with unit spacing along the $\lambda$ direction. The $y$-dependent hopping phases in Eq.~\ref{eq:main} can then be interpreted as the Peierls phases acquired by a charged particle (with charge $q=1$) hopping in the presence of a perpendicular magnetic field in the $\lambda\!-\!y$ plane, such that a magnetic flux quantum $\Phi_0$ is enclosed within an effective plaquette of magnetic length $l_B \equiv 2\pi  / \phi_0$ along the $y$ direction [see Fig \ref{fig:schematic}~(c)]. 

      If the hopping amplitudes are translationally-invariant [i.e. if $J_\lambda$ is independent of $\lambda$], it can be easily shown that the resulting 2D coupled wire model [Eq.~\ref{eq:2dnotrap}] has a topologically non-trivial band structure with the lowest energy bands being classified with non-trivial Chern numbers $|\mathcal{C}|=1$ with associated robust chiral edge states inside the band-gaps~\cite{Budich_2017}. Introducing non-uniformity in the hopping amplitudes [c.f.~Eq.~\ref{eq:J}] distorts the band-structure [see Appendix~\ref{sec:appdriving}], but does not destroy the desired topological chiral edge state physics provided the band-gap remains open~\cite{Price2017, Chalopin2020}. 

     Including the weak harmonic trapping potential along the $y$ direction leads finally to the full stroboscopic 2D effective Hamiltonian associated with Eq.~\ref{eq:full} as given by
        \begin{equation}
       \hat{\mathcal{H}}_{2D}(\lambda, y) =\frac{\hat{p}_y^2}{2m} + \frac{1}{2} m \omega_y^2 y^2 + \hat{{H}}_{\text{synth}}(\lambda, y). 
       \label{eq:2dtrap}
      \end{equation} 
    The presence of the weak trap along $y$ confines the atoms in this direction and serves as an effective ``soft" boundary on which to observe chiral edge state behavior [c.f.~Ref.~\onlinecite{Goldman2013} and discussion below]. We therefore control the effective length of the system along $y$ by changing the value of $\omega_y$; physically, if the trapping frequency is weakened then atoms can move further away from the center of the system, which should lead to larger experimental signatures of chiral edge state behavior. Note that the choice of a harmonic trap along $y$ is motivated by the existing experimental set-up of Ref.~\onlinecite{oliver2023bloch}, but in principle, other (e.g. sharper) confining potentials along $y$ could be used to probe the edge state behavior.
    
    Throughout the paper, we focus on observables that could feasibly be measured utilizing the techniques developed in Ref.~\onlinecite{oliver2023bloch}. In this previous work, it was demonstrated that the real space observables $\langle x \rangle$, the centre of mass in $x$, and $\sigma_x^2$, the (square of the) width of the cloud in $x$, can be related to the average occupation in the synthetic dimension $\langle \lambda \rangle$. The central result of this investigation is
    \begin{equation}
    \langle \lambda \rangle = \langle x \rangle^2 + \sigma_x^2 - \frac{1}{2}
    \label{Eq:realspacetosynth}
    \end{equation}, where the position and width are measured in units of the usual harmonic oscillator lengthscale. This result was successfully utilized to compare experimental data and numerical simulations with alterations to compensate for time-of-flight measurement. As such, our results focus on obtaining $ \langle \lambda (t)\rangle $ v.s. $\langle y(t) \rangle$ simulations which are quantities that can be experimentally measured directly or via the quantities in Eq. \ref{Eq:realspacetosynth}.

    Moreover, in a previous experimental investigation \cite{oliver2023bloch}, a long synthetic dimension of tens of sites was achieved despite possible constraints effects, including anharmonicity of the trapping potentials and heating from interactions. Although these effects will become important for even longer synthetic dimensions, they did not dominate the physics for a dimension of tens of sites. Thus, in building from this base, we are confident that we can study two dimensional physics with a synthetic dimension of several tens of sites. For now, we do not investigate a system with a synthetic dimension of hundreds of sites, as the aforementioned effects may become limiting for such a long dimension.
    
    The nature of the edges in the synthetic dimension are also highly relevant for studying this behavior and we discuss how these can be engineered to assist our investigation into edge-state physics in Section~\ref{sec:tunableedge}.

    \subsection{Numerical Simulations }

    In the following, we study the time dynamics of the 2D time-dependent Hamiltonian [Eq.~\ref{eq:full}] with additional potential terms, which are introduced below to engineer a tunable upper edge in the synthetic dimension [c.f.~Section~\ref{section:1a)}]. By considering the full time-dependent Hamiltonian, we are simulating the full dynamics, without a stroboscopic or rotating-wave approximation, allowing us to verify that the desired chiral behavior of the 2D coupled wire model may be observed using this experimental scheme. To carry out the simulations, we express all the $x$-dependent terms in the Hamiltonian in terms of $\lambda$ and perform the evolution in the synthetic $\lambda \!-\! y$ space. While we primarily present and analyze our results in synthetic space, we can also map the wavefunction back to the real $x\!-\!y$ space after every time step to obtain the full real-space dynamics. Note that, experimentally, measurements, e.g. of the atomic density, will be made in real space, but then can be translated into synthetic space as discussed in Ref.~\cite{oliver2023bloch}.    
    
    To illustrate the interesting physics, we focus principally on two different types of (non-interacting) simulations; firstly, we consider idealised zero-temperature simulations for an initial Gaussian-like wave-packet, and secondly, we simulate a thermal cloud that is initialised at an experimentally-realistic temperature~\cite{oliver2023bloch}. Firstly, for the zero-temperature simulations, we chose the initial state to have the Gaussian form
    \begin{equation}
\psi_G\left(y, \lambda \right) = Ae^{-\frac{\left(y- y_0\right)^2}{2\sigma_y^2}} e^{ik_y\left(y - y_0\right)}    e^{\frac{\left(\lambda - \lambda_0\right)^2}{2\sigma_{\lambda}^2}}e^{ik_{\lambda}\left(\lambda - \lambda_0\right)}  
\label{Eq:gaussian_initstate}
\end{equation} 
where  $A$ is a normalisation constant and $y_0$ and $\lambda_0$ are the initial center-of-mass positions along $y$ and $\lambda$ respectively. Similarly, $\sigma_y$ ($\sigma_\lambda$) and $k_y$ ($k_\lambda$) are respectively the initial widths and the initial momenta along the $y$ ($\lambda$) directions. Experimentally, it is natural for the wave-packet to start at the center of the weak trap along $y$, and so in the simulations we set $y_0 = 0$. Additionally, we select $\lambda_0=0$, as we physically expect that atoms will predominantly occupy the lowest energy states before the shaking potential is switched on. One advantage of our scheme is that $\lambda=0$ is also naturally the lower boundary of the synthetic dimension, meaning that we can expect good coupling into the desired chiral edge state. 

For all simulations, we chose experimentally-realistic parameters~\cite{oliver2023bloch} corresponding to, unless otherwise stated, $m=86.9\text{amu}$ (for $^{87}$Rb where amu is the atomic mass unit), $\phi_0 = 0.2 \mu \text{m}^{-1}$, $\omega =\omega_D= 2\pi \times 200 \text{Hz}$, $\omega_y = 2\pi \times 5\text{Hz}$ and $\kappa = -98\text{Hz}$. Note, to represent $\kappa$ in these units we have multiplied it by the harmonic oscillator length along $x$. We have also numerically tuned the initial wave-packet widths and the initial momenta to enhance the coupling to the edge-state, in order to maximise the amount of the wave-packet which we observe moving chirally around the system; for the above parameters, this leads us to set $\sigma_{\lambda} =0.5 $, $\sigma_y = 2.8 y_{\text{ho}}$ (where $y_{\text{ho}} \!\equiv\!\sqrt{ 1/ m \omega_y}$), and $k_y\! =\! -0.1 (y_{\text{ho}})^{-1}$. We set $k_{\lambda} = 0$ which is a natural choice for the edge-state investigation, see also \cite{Price2017}.

Secondly, we apply the methodology of Ref.~\onlinecite{oliver2023bloch} to simulate the dynamics of a non-interacting thermal gas, i.e. a non-interacting cloud of atoms which are initially distributed over the levels of the 2D harmonic trap according to a classical Boltzmann distribution. Numerically, an initial state has the form 
\begin{equation}
\psi_T (x, y) = B \sum_{i=0}^{N-1} \sqrt{\frac{\text{exp}\left(-\beta E_i\right)}{\mathcal{Z}}}\text{exp}\left(i\theta_i\right)|\varphi_i \rangle
\label{eq:therm}
\end{equation}
where $B$ is a normalization constant, $\beta$ is the inverse temperature, $E_i$ corresponds to the energy of the i$^{\text{th}}$ 2D harmonic oscillator eigenstate $|\varphi_i \rangle$ and $\mathcal{Z}$ is the associated partition function. Throughout, we set the initial temperature to $T= 20$nK, as in the experiment of Ref.~\onlinecite{oliver2023bloch}. For each initial state, the phases $\theta_i$ are randomly selected for every $|\varphi_i \rangle$ from a uniform distribution between $0$ and $2 \pi$. Note that only a finite number, $N$, of the 2D harmonic trap states are used in this construction; we verified that $N$ is always large enough that it does not affect the results. Numerically, we generate many such initial states, each with a different set of random phases $\{\theta_i\}$, and time-evolve each one under the full time-dependent Hamiltonian, with the same experimentally-relevant parameters as above. The resulting atomic densities are then averaged together, destroying the phase coherence of the atomic distribution to mimic the behavior of a thermal cloud~\cite{oliver2023bloch}.

    \section{Engineering a Tunable Edge in Synthetic Space\label{sec:tunableedge}} 
    To study topological edge states it is highly desirable to be able to control the properties of the system's edges, such as their location and ``softness". The latter typically refers to the power, $\gamma$ of the power law potential $V \!=\! |r|^{\gamma}$ that induces the edge along the given $r$ direction~\cite{Goldman2013}, with a $\gamma\!=\!1$ potential corresponding to the softest edge investigated in this work and a square-well potential ($\gamma \!=\! \infty$) to the hardest possible edge. From previous theoretical and experimental investigations~\cite{braun2024real, PRL108255303, Goldman2013, PhysRevA.94.043611}, the softness of the edge is known to affect, for example, the efficiency of coupling a wave-packet into the edge state as well as the wave-packet's resulting group velocity.
    
    In many synthetic dimension schemes, such as those based on coupling internal atomic states~\cite{Boada2012, Celi2014, Mancini2015, Stuhl2015, Chalopin2020, lienhard2020realization,Kanungo2021, Livi2016, Kolkowitz2017,Roell2023, chen2024interaction,chen2024strongly,lu2024probing}, there is a natural but fixed hard boundary at both ends of the synthetic dimension reflecting that the synthetic dimension is built out of a finite number of states~\cite{Ozawa_2019}. Conversely, a synthetic dimension of harmonic trap states is, in principle, a semi-infinite ladder of states, which starts from a ``hard wall" defined by the ground state ($\lambda=0$) and extends indefinitely as $\lambda \rightarrow \infty$. In practice, the anharmonicity of experimental traps adds, at lowest order, a harmonic anti-trapping term along the synthetic dimension~\cite{Price2017}. However, while this effectively limits the ``length" of the synthetic dimension over which Eq.~\ref{eq:main} is sufficient to describe the physics, it does not lead us to observe clean chiral edge state behavior. As we shall now propose, we can instead introduce an additional real-space potential to engineer a tuneable upper edge in synthetic space. We shall firstly demonstrate that we can use this to control the number of sites of the synthetic dimension, and then secondly discuss the extent to which we can tune how ``soft'' the upper edge is in the synthetic dimension. 

\begin{figure}
\centering
\includegraphics[width=1\columnwidth]
      {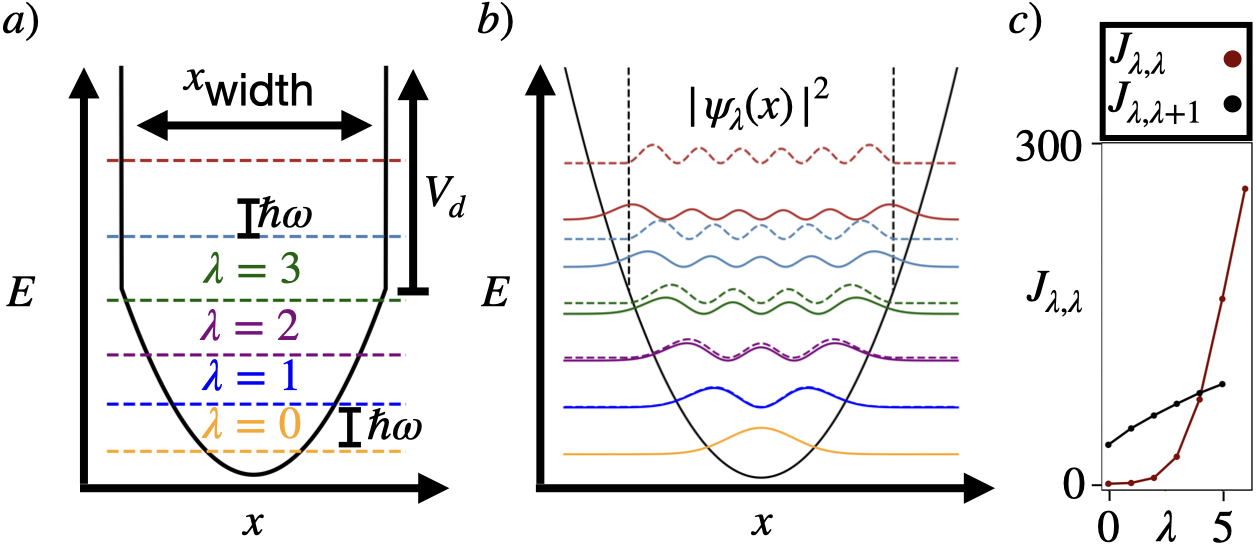}%
      \caption{(a)\label{fig:impl_squarewell}The lowest energy eigenvalues of Eq.~\ref{eq:E1} plotted as horizontal dashed lines for the illustrative example of $V_d=10  \hbar \omega$ and $x_{\text{width}}=\sqrt{32} x_{\text{ho}}$, in units where $m\!=\!\omega\!=\!1$, showing that the square-well potential shifts the higher $\lambda$ states off resonance as compared to the usual energy spacing $ \omega$. b) The corresponding eigenstate probability densities ({\it dashed lines}) as compared to those of the usual harmonic trap ({\it solid lines}), showing that higher energy states are increasingly modified by the square-well potential. Here, the colors match those used in panel (a), while the bare harmonic trapping potential is represented by a solid black line, and the addition of the square-well potential is indicated by the vertical dashed black lines. (c) The numerically-calculated Floquet Hamiltonian matrix elements [Eq.~\ref{eq:Heff}] for onsite ({\it orange dots}) and nearest-neighbor-hopping terms ({\it black dots}), for the illustrative example $V_d = 590 \text{Hz}$, and other parameters as described in Section~\ref{sec:scheme}.}
     \end{figure}

    \subsection{Tuning the length of the synthetic dimension \label{section:1a)}}

    To realise a tunable upper boundary in the synthetic dimension, we  propose to shift higher harmonic states off resonance towards higher energies. In our scheme,  the trap states along $x$ are coupled with a single frequency $\omega_D$ [c.f.~Eq.~\ref{Eq.couplingpotential}] exploiting the fact that the states are equally spaced in frequency with $\omega\simeq \omega_D$. If the spacing between states becomes unequal and much larger than $\omega_D$, then it becomes energetically unfavourable to hop between sites. Thus, by only modifying the energy spacing of the higher sites we can truncate the length of the synthetic dimension. To implement this, we introduce a square-well potential in the $x$ direction as
   \begin{equation}
   V_{e}\left( x \right) = V_d\left[1 - \Theta\left(x - x_{\text{e}}\right) +\Theta \left(x + x_{\text{e}}\right)\right] ,
   \label{Eq.potentialedge}
   \end{equation} where $V_d$ is the depth of the potential, $\Theta \left(x \right)$ is the Heaviside function and $x_{\text{e}}\equiv x_{\text{width}}/2$ where $x_{\text{width}}$ is the potential width [see Fig. 2(a)]. Such a potential could be straightforwardly implemented, for example, by using a DMD in the same way as the other driving and trapping potentials in Eq.~\ref{eq:full}, meaning that no additional equipment is required as compared to the experimental set-up of Ref.~\onlinecite{oliver2023bloch}. 

   We can see that the square well shifts the higher harmonic states off resonance as desired by considering the 1D time-independent Schr\"{o}dinger equation along $x$ in the absence of the driving potential,
   \begin{equation}
       E  \psi(x) = \left[ -\frac{1}{2m}\frac{d^2}{dx^2}  + \frac{1}{2}m \omega^2{x}^2  + V_e\left(x\right) \right] \psi(x) .
       \label{eq:E1}
   \end{equation}
The resulting energy eigenvalues, calculated in units where $m\!=\!\omega\!=\!1$, for the illustrative example of $V_d= 10 \hbar \omega$ and $x_e= \sqrt{8} x_{\text{ho}}$ (where $x_{\text{ho}} \!\equiv\!\sqrt{ 1/ m \omega}$) are indicated for the first few states in Fig.~2(a); as one can see, the lowest eigenvalues remain approximately equispaced, while above $\lambda=2$, the energy offset becomes significant. In Fig. 2(b), we also compare the probability density of the eigenstates ({\it dashed lines}) of Eq.~\ref{eq:E1} with those of the usual harmonic trap states ({\it solid lines}), showing decreasing agreement with increasing $\lambda$. 

More formally, we can calculate the effective 1D time-independent Floquet Hamiltonian $\hat{H}_{\text{eff}}$ that captures the stroboscopic physics of the periodically-driven $x$ direction with the square well trap 
\begin{equation}
\mathcal{H}_{\text{Floq}} = \frac{\hat{p}_x^2}{2m} + \frac{m}{2}\omega^2x^2 + V(x, y=0, t) + V_e(x)
\end{equation}
~\cite{rudner2020floquet, PhysRevX.4.031027, Price2017}. 
To evaluate this, we numerically construct the stroboscopic evolution operator for the 1D Hamiltonian $\hat{U}(T, 0)$, starting from an initial time $t=0$ up to a final time $t=T$, where $T$ is the period of the drive. This 1D Hamiltonian is as described in Eq. \ref{eq:full}, neglecting the kinetic energy along $y$ and setting $\omega_y\! =\!y\!=0$. We then calculate the effective Hamiltonian using
\begin{equation}
    \hat{U}(T, 0) = e^{-i T\hat{H}_{\text{eff}}}.
\end{equation}  
To a good approximation for the $\lambda$ values of interest, we find
\begin{equation}
\hat{H}_{\text{eff}} \approx \sum_{\lambda} \left[ J_{\lambda+1, \lambda}
| \lambda+1 \rangle  \langle  \lambda| + \text{h.c.} +  J_{\lambda, \lambda}| \lambda \rangle \langle \lambda | \right], \label{eq:Heff}
\end{equation}
where the first two terms are the nearest-neighbor hoppings and the third term is an on-site potential. The values for the corresponding matrix elements, $J_{\lambda, \lambda^{'}}$, are plotted in Fig.~\ref{fig:impl_squarewell}(c) as a function of $\lambda$, for the example of $V_d \!=\! 590 \text{Hz}$, $x_{\text{width}}\!=\!\sqrt{32} x_{\text{ho}}$, with other parameters as stated in Section~\ref{sec:scheme}. As can be seen, the nearest-neighbor terms vary as $\sqrt{\lambda}$ as predicted by Eq.~\ref{eq:J}, while the square-well trap induces a large on-site energy for higher $\lambda$ sites. This energetic penalty grows from being small to very large over just 3 sites implying a relatively ``hard'' edge; ways to make this a ``softer'' edge are discussed in the next section. Note also that if $V_d$ is significantly larger than the value above, then other longer-range (e.g. next-nearest-neighbor) hopping terms, which have so far been neglected, will also be significantly affected by the addition of the square well. However, these terms only become sizable for values of $\lambda$ where $J_{\lambda, \lambda}$ is large compared to the nearest-neighbor hopping terms.

\begin{figure*}
\includegraphics[width=\textwidth]{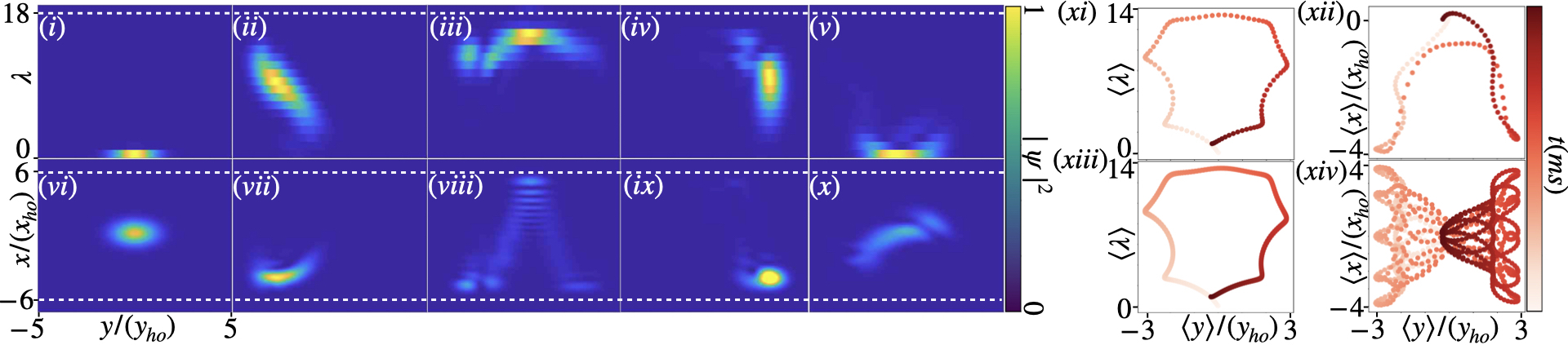}
\caption{\label{Fig:gauss_synth_real} (i)-(v) 
Snapshots for the numerical evolution of the atomic density, $|\psi|^2$, for an initial Gaussian wave-packet in the synthetic $\lambda\!-\!y$ plane with parameters as discussed in the main text and as stated in section~\ref{sec:scheme}. For these parameters, the predicted position of the upper edge along the synthetic dimension is $\lambda_e = 18$ [c.f.~Eq.~\ref{eq:le}] (\textit{white dotted line}). The snapshots correspond to times $t\!=\! 0\text{ms}$, $150 \text{ms}$, $265 \text{ms}$, $365 \text{ms}$ and $560 \text{ms}$ respectively, showing clear chiral motion of the wave-packet around the edge of the synthetic plane. (vi)-(x) Corresponding snapshots of the atomic density in real $x \!-\! y$ space for the same times as  the upper panels. These times correspond to stroboscopic measurements of the wave-packet; meaning any micromotion has been removed. As the wave-packet climbs the synthetic dimension, the wave-packet shifts to negative $x$ and $y$ until it approaches the square well at $x_e = 6 x_{\text{ho}}$ (\textit{white dotted lines}). The atoms are then transported back along the positive $y$ direction, while occupying the highest available $\lambda$-states; note during this process the wave-packet $\textit{zig-zags}$ between $x=\pm 6$, as visible in panel (viii). Upon reaching the ``soft" edge in $y$ (due to the weak harmonic trap), the atoms move from $x=-6$ to $x=0$, which is associated with a descent down the synthetic dimension.(xi), (xii), (xiii) and (xiv) are center-of-mass trajectories with the upper and lower panel illustrating the stroboscopic and full (including micromotion) dynamics. (xi) and (xiii) are the synthetic space case with (xii) and (xiv) corresponding to the real space equivalents respectively. Finally, in (xi), (xii), (xiii) and (xiv) the red becomes darker further into the trajectory from $t=0 ms$ to $t =600 ms$ }
\end{figure*}

From the above arguments, it is natural to anticipate that changing the width, $x_{\text{width}}$, of the square well potential in real space can be used to directly tune the effective location of the edge along $\lambda$, by shifting different numbers of states off resonance. Analytically, we can make a prediction for the location of the edge along $\lambda$, by considering a semiclassical wave-packet moving in a 1D harmonic trap and a large square-well-like potential within the WKB approximation. Inside the well, the semiclassical (i.e. center-of-mass) position, $\lambda_\text{com}$, can be expressed as~\cite{Price2017}
\begin{equation}
\lambda_\text{com} =  \frac{m \omega}{2 } {x}_{\text{com}}^2 + \frac{1}{2 m \omega  } {p}_{\text{com}}^2
\label{Eq:synth_to_real}
\end{equation} 
where ${x}_\text{com}$ and ${p}_\text{com}$ are the semiclassical (center-of-mass) real-space position and momentum respectively. This originates from introducing a classical complex variable
\begin{equation}
\alpha = \sqrt{\frac{m\omega}{2}}x_{com} + i \frac{p_{com}}{\sqrt{2m\omega}},
\end{equation}
 where $|\alpha|^2 = \lambda_{\text{com}}$. Then, at the turning points at $\pm x_\text{width}/2$, the momentum ${p}_\text{com}=0$, and hence the position along the $\lambda$ direction is given by
\begin{equation}
 \lambda_e = \frac{x_{\text{width}}^2}{8 x_{\text{ho}}^2}. \label{eq:le}
\end{equation}
 We can cross-reference this equation with the calculated Floquet-Hamiltonian hopping elements and we see that that $\lambda_e$ is within a couple sites of where the onsite potential becomes comparable to the nearest-neighbor hopping amplitudes.
This suggests that the effective length of the synthetic dimension can be tuned from a few up to e.g. 50 sites by choosing suitable values of $x_{\text{width}}$ up to 20$x_{\text{ho}}$ (i.e. $15.3 \mu$m for these parameters). 

We now verify that the above prediction [Eq.~\ref{eq:le}] for tuning the length of the synthetic dimension holds for the full 2D Schr\"{o}dinger dynamics, and show that we can indeed use this scheme to observe chiral edge state behavior in our system. To this end, we numerically simulate the full real-space time-evolution under the total Hamiltonian consisting of $\mathcal{H}$ [Eq.~\ref{eq:full}] together with $V_e$ [Eq.~\ref{Eq.potentialedge}]. We perform this process firstly in idealised Gaussian wave-packet simulations and then for a thermal cloud. 

In Fig.~\ref{Fig:gauss_synth_real}, we show snapshots  of the numerical atomic density, $|\psi|^2$, for an initial Gaussian wave-packet  at successive times in (i)-(v) synthetic $\lambda\!-\!y$ space and (vi)-(x) real $x\!-\!y$ space. Here, we set $V_{\text{d}} =0.13\text{MHz}$ and $x_{\text{width}}=12 x_{\text{ho}}$, with other parameters as in Section~\ref{sec:scheme}. We have taken the above snapshots stroboscopically, at full periods of $T_D\!\equiv\! 2 \pi /\omega_D$,  such that the effects of micromotion  are removed. In Fig.~\ref{Fig:gauss_synth_real}, we include center of mass trajectories for the stroboscopic ((xi) and (xii)) and full dynamics - including micromotion ((xiii) and (xiv)). (xi) and (xiii) are synthetic space trajectories, whilst  (xii) and (xiv) are the equivalent real space trajectories.

When measured stroboscopically the synthetic space picture, Fig.~\ref{Fig:gauss_synth_real} (i-v) and (xi), exhibits a clear clockwise trajectory around the $\lambda\!-\! y$ plane, as expected for the chiral edge state behavior in a 2D quantum Hall system. We have checked that reversing the sign of the artificial magnetic flux, $\phi_0$, also reverses the direction of the trajectory as expected. As predicted, we also see that the wave-packet does not travel above $\lambda_e$ ({\it white dotted line}), showing that we have indeed created an upper edge along the synthetic dimension. Note that, due to the wave-packet width, the wave-packet peak is observed a few sites below $\lambda_e$, as we have also used in our quantitative analysis later in this section.  When micromotion is included the synthetic space trajectories remain essentially unaltered Fig.~\ref{Fig:gauss_synth_real} (xiii).

During the corresponding stroboscopic motion in real space, we observe that the atoms shift to the left along $y$ whilst performing downward hops in the $x$ direction. This hopping corresponds to the spatial extent of the atomic trap state; as the wave-packet travels up synthetic space it descends in $x$ direction. The presence of the additional square-well ({\it dotted white lines}) prevents the wave-packet from reaching lower values of $x$ hence enforcing a maximum $\lambda$ state. When measured stroboscopically the ideal chiral edge state behavior would therefore translate into the atomic density sequentially moving in the negative $y$ direction, stroboscopically travels down $x$, moving in the positive $y$ direction whilst performing a ``$\textit{zig-zag}$" between $x \pm 6$, moving back to $x = 0$, and moving back towards $y\!=\!0$. If the sign of $\kappa$ is switched the wave-packet undergoes the same trajectory but travels upwards in the $x$ direction instead. Numerically, we observe that the motion along $y$ and the along $x$ are not entirely decoupled, reflecting the presence of the weak harmonic trap along $y$, as discussed further below. Nevertheless, we can still see the clear signatures of this physics [c.f.~Fig.~\ref{Fig:gauss_synth_real}], including the nodal structure associated with the high-lying trap states in panel (viii). 

Unlike the synthetic space picture the inclusion of micromotion has a significant effect on the trajectory of the wave-packet in real space. To illustrate the effect of the micromotion we plot the center mass, Fig~\ref{Fig:gauss_synth_real} (xiv), ten times per period $T_D$. We observe that the stroboscopic movement downwards is replaced with oscillations centered on the origin in $x$ with varying amplitude. The stroboscopic motion of the wave-packet in $x$ acts as an envelope for the micromotion oscillations. Thus the amplitude increases as the spatial extent of the harmonic states grows.
 \begin{figure}
\includegraphics[width=1\columnwidth]
      {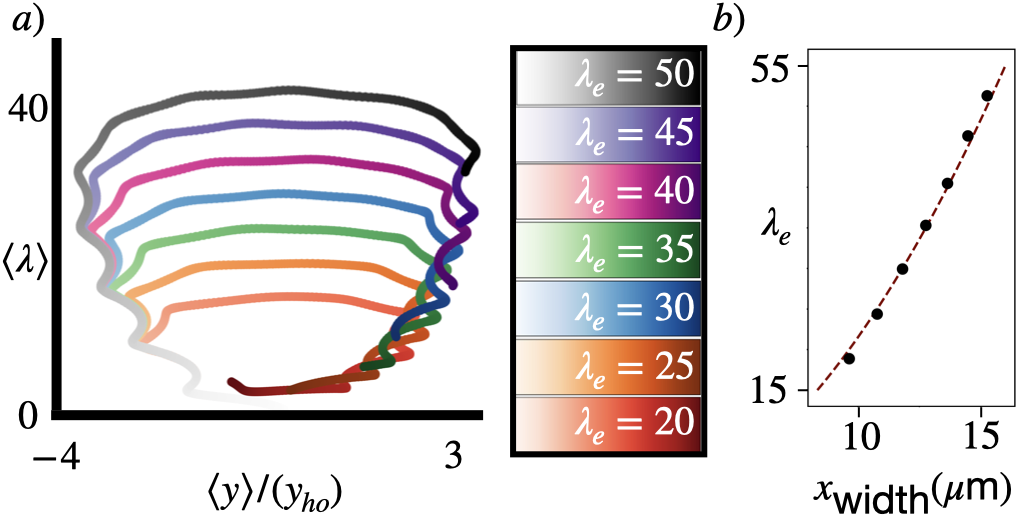}%
      \caption{(a) The numerical evolution of the center of mass for a Gaussian wave-packet in the $\lambda \! -\!y$ plane, for different values of $\lambda_e$ (as determined from Eq.~\ref{eq:le}). The colorbar for each trajectory runs over time from $t=0 \text{ms}$ ({\it pale color}) to $t=701 \text{ms}$ ({\it dark color}). Each trajectory shows the characteristic chiral motion associated with the edge of a quantum Hall system, with orbits for higher values of $\lambda_e$ moving further along the synthetic $\lambda$ direction as expected. (b) Numerical estimates for $\lambda_e$ from Eq.~\ref{eq:gmax} ({\it black dots}) as compared to the analytical prediction from Eq.~\ref{eq:le} ({\it red dashed line}), showing excellent agreement. All parameters are as stated in Section~\ref{sec:scheme}. 
      }
      \label{fig:comgedge}
     \end{figure}

 \begin{figure}
   \centering
      \includegraphics[width=1\columnwidth]
      {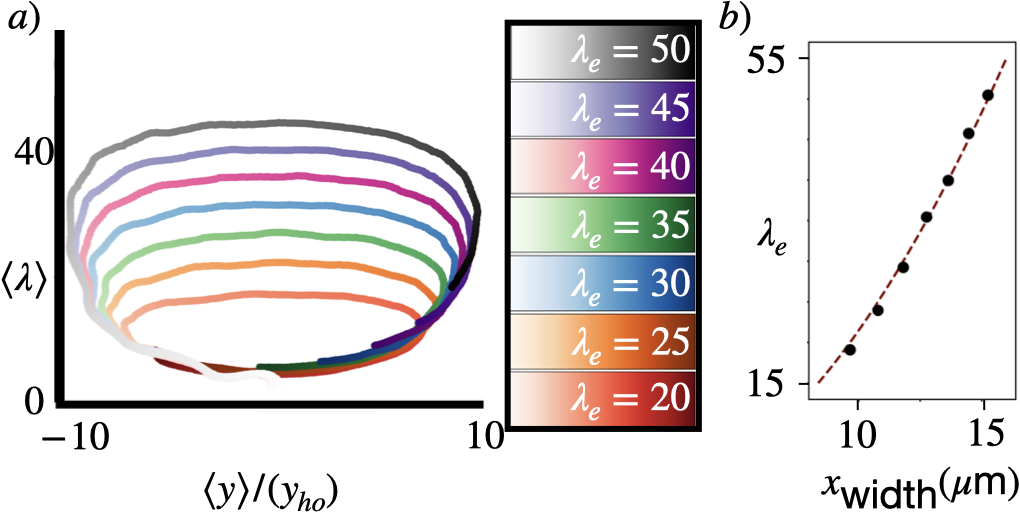}%
      \caption{(a) The numerical evolution of the center of mass for a thermal cloud in the $\lambda \! -\!y$ plane, for different values of $\lambda_e$ (as determined from Eq.~\ref{eq:le}). The colorbar for each trajectory runs over time from $t=0 \text{ms}$ ({\it pale color}) to $t=970 \text{ms}$ ({\it dark color}). Each trajectory again shows the characteristic chiral motion associated with the edge of a quantum Hall system, with orbits for higher values of $\lambda_e$ moving further along the synthetic $\lambda$ direction as expected. (b) Numerical estimates for $\lambda_e$ ({\it black dots}), as calculated according to Appendix~\ref{sec:appupper},  as compared to the analytical prediction from Eq.~\ref{eq:le} ({\it red dashed line}), again showing excellent agreement. All parameters are as stated in Section~\ref{sec:scheme} and the main text.
      }
      \label{fig:thermal}
      \end{figure}

To confirm the tunability of the upper edge along the synthetic dimension, we now vary $x_{\text{width}}$, with all other parameters as specified above unless otherwise stated. In Fig.~\ref{fig:comgedge}(a) we plot the resulting numerical center of mass with respect to $\lambda$ and $y$ over time for seven values of $\lambda_e$, as implemented by selecting the appropriate values of $x_{\text{width}}$ in Eq.~\ref{eq:le}. In each case, we see a clockwise trajectory around the $\lambda\!-\! y$ plane, as expected. Initially, all seven trajectories coincide while the wave-packet is exploring small values of $\lambda$, as the influence of the real-space square-well potential is weak. However, the trajectories then deviate from each other sequentially as those with a larger $x_{\text{width}}$ and hence larger $\lambda_e$ continue further up the synthetic dimension before reaching the effective edge. Note that the edge along $\lambda$ is relatively ``hard" [see Fig.~\ref{fig:impl_squarewell} and Section~\ref{section:1b} below], while that along $y$ is ``soft", due the weak harmonic trap along $y$. This is reflected in the shape of the chiral trajectory, where we see that atoms continue to move in $y$ when they are skipping along $\lambda$ on the right or left side of the system, but they do not significantly move in $\lambda$ while they are traveling at the top or bottom of the system along $y$ [c.f.~Fig.~\ref{Fig:gauss_synth_real}]. For these parameters, the effects of micromotion along the synthetic dimension are also small and not clearly visible on this scale. 

We can also use the numerical center-of-mass results from Fig.~\ref{fig:comgedge}(a) to quantitatively check the tunability of the length of the synthetic dimension that is predicted by $\lambda_e$ [Eq.~\ref{eq:le}]. To do so, we first find the maximum center-of-mass position, $\langle \lambda \rangle_{\max}$, for each trajectory. We then calculate the width of the wave-packet density distribution at this maximum position to extract the corresponding standard deviation $\sigma_{\lambda \text{max}}$. The location of the upper synthetic edge can then be estimated as
\begin{equation}
\lambda_e \approx \langle \lambda \rangle_{\max} + \sigma_{\lambda \text{max}}. \label{eq:gmax}
\end{equation}

The corresponding numerical estimates of $\lambda_e$ are plotted in Fig.~\ref{fig:comgedge}(b)({\it black dots}), where they are compared to the analytical prediction [Eq.~\ref{eq:le}] ({\it dashed red line}). As can be seen, there is very good agreement, verifying that we can reliably tune the length of the synthetic dimension by changing $x_{\text{width}}$. 
    
We now repeat this analysis for the more experimentally relevant case of an initial thermal cloud [c.f.~Section~\ref{sec:scheme}], as plotted in Fig.~\ref{fig:thermal} and shown in more detail in Appendix~\ref{sec:appreal}. All parameters are the same as in the Gaussian wave-packet simulations above, except that the driving strength is increased to $\kappa = -500 \text{Hz}$; this is because we observe empirically that a higher driving strength is more effective at coupling the atoms into the synthetic dimension in the thermal simulations. As can be seen in Fig.~\ref{fig:thermal} (a), the resulting center-of-mass trajectories are smoothed out in the thermal simulations as compared to those in Fig.~\ref{fig:comgedge}. However, importantly, the critical features remain and the chiral edge state behavior is clearly visible. Note that as compared to the Gaussian wave-packet simulations, a reduced percentage of atoms participate in the chiral motion [c.f.~Appendix~\ref{sec:appreal}]; this is because the width of the thermal cloud is set by the temperature and strength of the weak trap, and so cannot be easily optimised as for the idealised simulations. We could, in principle, increase the number of atoms participating in the chiral dynamics by lowering the temperature, which would be more experimentally restrictive, or increasing $\omega_y$, which would reduce the desired motion along the $y$ direction. 
Finally, we can again use the numerical trajectories to quantitatively confirm the tunability of the synthetic dimension length. The data analysis is detailed in Appendix~\ref{sec:appupper}, and the corresponding numerical results are plotted in Fig.~\ref{fig:thermal}(b). Here, we again see clear quantitative agreement between the numerics and analytics, demonstrating that this approach can be successfully used to tune the length of the synthetic dimension over many sites. 

\subsection{Tuning the softness of the upper edge in the synthetic dimension \label{section:1b}}

As introduced above, another key feature of a boundary is its ``softness", corresponding to how quickly the boundary potential changes spatially. In a recent real-space experiment, the softness of the edge was shown to affect the properties of chiral quantum Hall trajectories, such as the group velocity of the particles~\cite{braun2024real}. As we shall now investigate, we can also effectively change the softness of the upper edge in our synthetic dimension by modifying the square-well potential introduced above. We shall firstly consider the effect of simply softening the real-space profile of the square-well potential across certain regions (as specified below), before exploring a wider range of parameters to show how to create sharp and soft edges in the synthetic dimension.    

\begin{figure}
  \centering
      \includegraphics[width=1\columnwidth]
{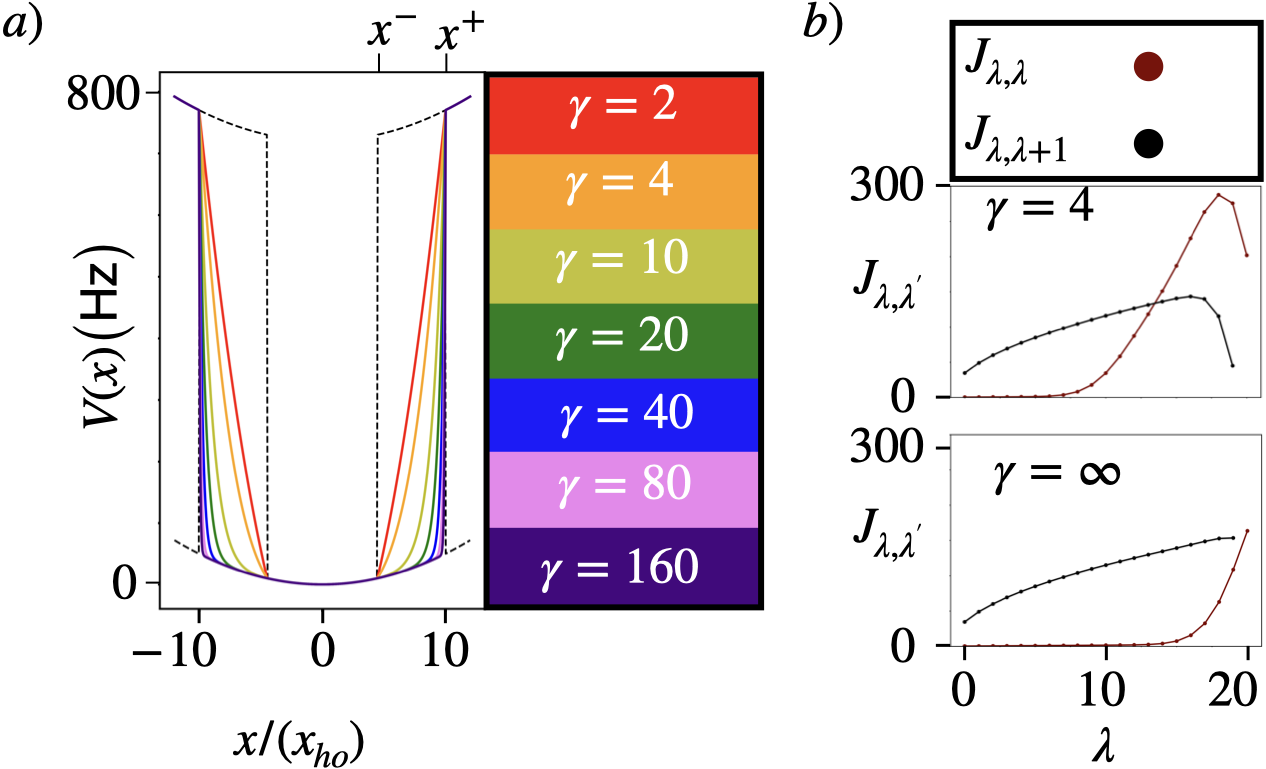}%
\caption{(a) The real-space potential as a function of $x$ as introduced to tune the softness of the upper edge in the synthetic dimension. Around $x=0$, this potential is simply given by the bare harmonic trap, while between positions $|x|=x^-$ and $|x|=x^+$ ({\it vertical dashed lines}), it increases according to a power law with exponent $\gamma$ up to a fixed height, $V_d$, above the trap. Here, we have taken $x^{-}= 4 x_{\text{ho}}$,  $x^{+} = 10 x_{\text{ho}}$ and $V_d =980 \text{Hz}$ , with other parameters as given in Section~\ref{sec:scheme}. (b) Numerically-calculated matrix elements of the corresponding effective 1D Floquet Hamiltonian for $\gamma = 4$ and $\gamma=\infty$, with  $x^{+} = 6 x_{\text{ho}}$ and other parameters as in panel (a). As can be seen, the onsite terms rise more gradually for $\gamma=4$ than $\gamma=\infty$, indicating a softer boundary in the synthetic dimension. In both cases, the nearest-neighbor terms ({\it black dots}) vary as $\sqrt{\lambda}$, as expected from Eq.~\ref{eq:J}. Note that for $\gamma=4$, we observe a marked deviation from this behavior at large $\lambda$, where we observe that other terms in the Floquet Hamiltonian become important. This is indeed true more generally but can-not be observed in the $\gamma=\infty$ case due to number of $\lambda$ sites included. Lines are a guide to the eye. Note that a similar treatment has been undertaken in \cite{PhysRevResearch.6.L012054} for a fractional Chern insulator.}
\label{fig:soften}
\end{figure}

As shown in Fig.~\ref{fig:soften}(a), instead of the square-well potential discussed above, we can apply a softer real-space potential to tune higher $\lambda$ states away from resonance. Experimentally, this could be applied using a DMD, as in the previous section. Physically, we expect that softening the real-space potential will mean that the higher trap states are less far-detuned, and hence that the onsite potential in synthetic space rises more gradually. To explore this systematically, we define symmetric spatial regions on either side of the origin that go from $|x|=x^-$ to $|x|=x^+$, where $x^{+} > x^{-}$. Within these regions, the optical potential then rises to a fixed height, $V_d$, above the bare harmonic trap, according to a power-law spatial dependence $\propto |x|^\gamma$ [see Fig.~\ref{fig:soften}(a)]. When $\gamma \rightarrow \infty$, this construction reduces to the square-well potential considered in Section~\ref{section:1a)} with $x_e = x^+$.  

\begin{figure}
  \centering
      \includegraphics[width=1\columnwidth]
{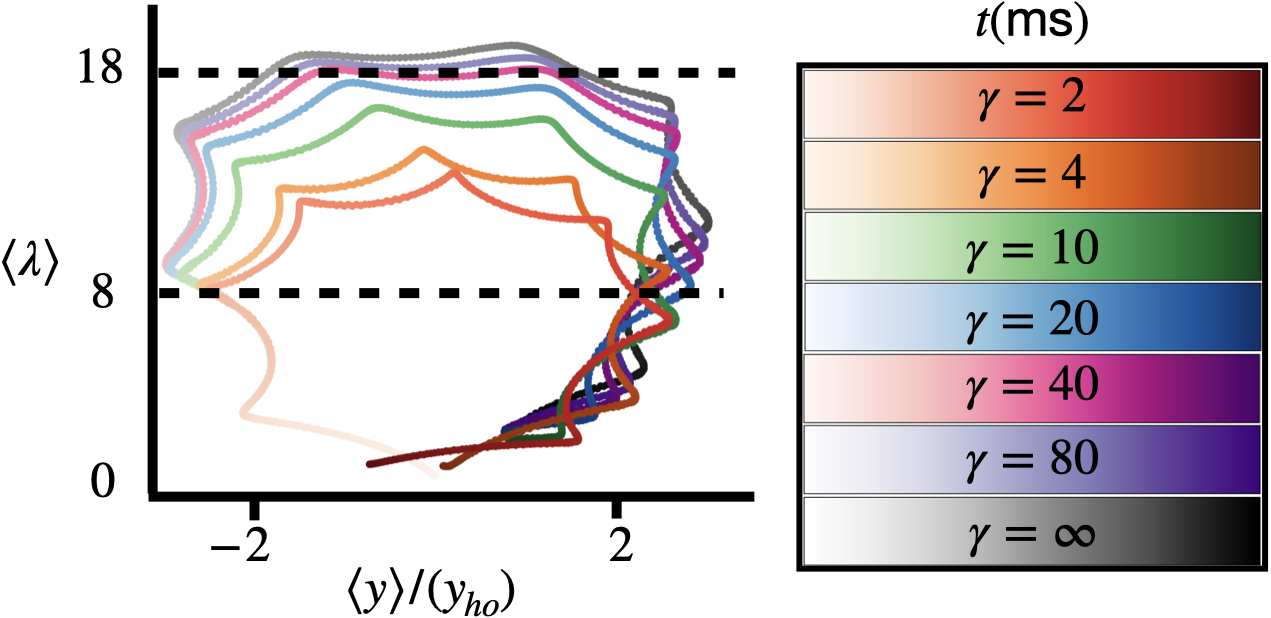}%
\caption{The numerical evolution of the center-of-mass for a Gaussian wave-packet in the $\lambda \!-\!y$ plane, for different values of $\gamma$. Parameters are as given in the main text and in Section~\ref{sec:scheme}. The colorbar for each trajectory runs over time from $t=0 \text{ms}$ ({\it pale color}) to $t=620 \text{ms}$ ({\it dark color}). The lower and upper dashed vertical lines indicate the values of $\lambda_{e-}$ and $\lambda_{e+}$ respectively. As can be seen, the chiral trajectories coincide until they reach $\lambda_{e-}$, after which the effect of the power-law real-space potential becomes important. For larger values of $\gamma$, the upper part of the trajectories is further up the synthetic dimension and becomes noticeably flatter, reflecting a sharper edge in the synthetic dimension.}
\label{fig:softcom}
\end{figure}

To explore how softening the real-space potential affects the edge in synthetic space, we can numerically calculate the 1D effective Floquet Hamiltonian as in Section~\ref{section:1a)}. The resulting matrix elements are shown in Fig.~\ref{fig:soften}(b) for $\gamma=4$ and for the square-well with $\gamma=\infty$, with $x^{-} = 4 x_{\text{ho}}$ and $x^{+} = 6 x_{\text{ho}}$ and other parameters as in panel (a). As can be seen, the onsite terms ({\it red dots}) rise more gradually for $\gamma=4$ than for $\gamma=\infty$, indicating a softer upper edge in the synthetic dimension. Motivated by Eq.~\ref{eq:le}, we can also introduce two characteristic positions along the synthetic dimension, namely
\begin{equation}
\lambda_{e-} \equiv \frac{(2x^-)^2}{8 x_{\text{ho}}^2}, \qquad \lambda_{e+} \equiv \frac{(2x^+)^2}{8 x_{\text{ho}}^2}. 
\label{eq:lepm}
\end{equation}
We observe that for $\gamma=4$, the onsite terms begin to increase near $\lambda_{e-}$, where the real-space potential starts to significantly affect the trap states. These terms then peak close to $\lambda_{e+}$, after which both the onsite potential and the nearest-neighbor couplings ({\it black dots}) decrease and other longer-range terms become important. In general these terms may peak before $\lambda_e^{+}$ if the $V_d$ is sufficiently large for the real space potential to have increased significantly before $x^{+}$. For $\gamma=\infty$, instead the onsite terms do not increase until near $\lambda_{e+}$, where the square-well potential first becomes important.

 \begin{figure}
    \centering
      \includegraphics[width=1\columnwidth]
      {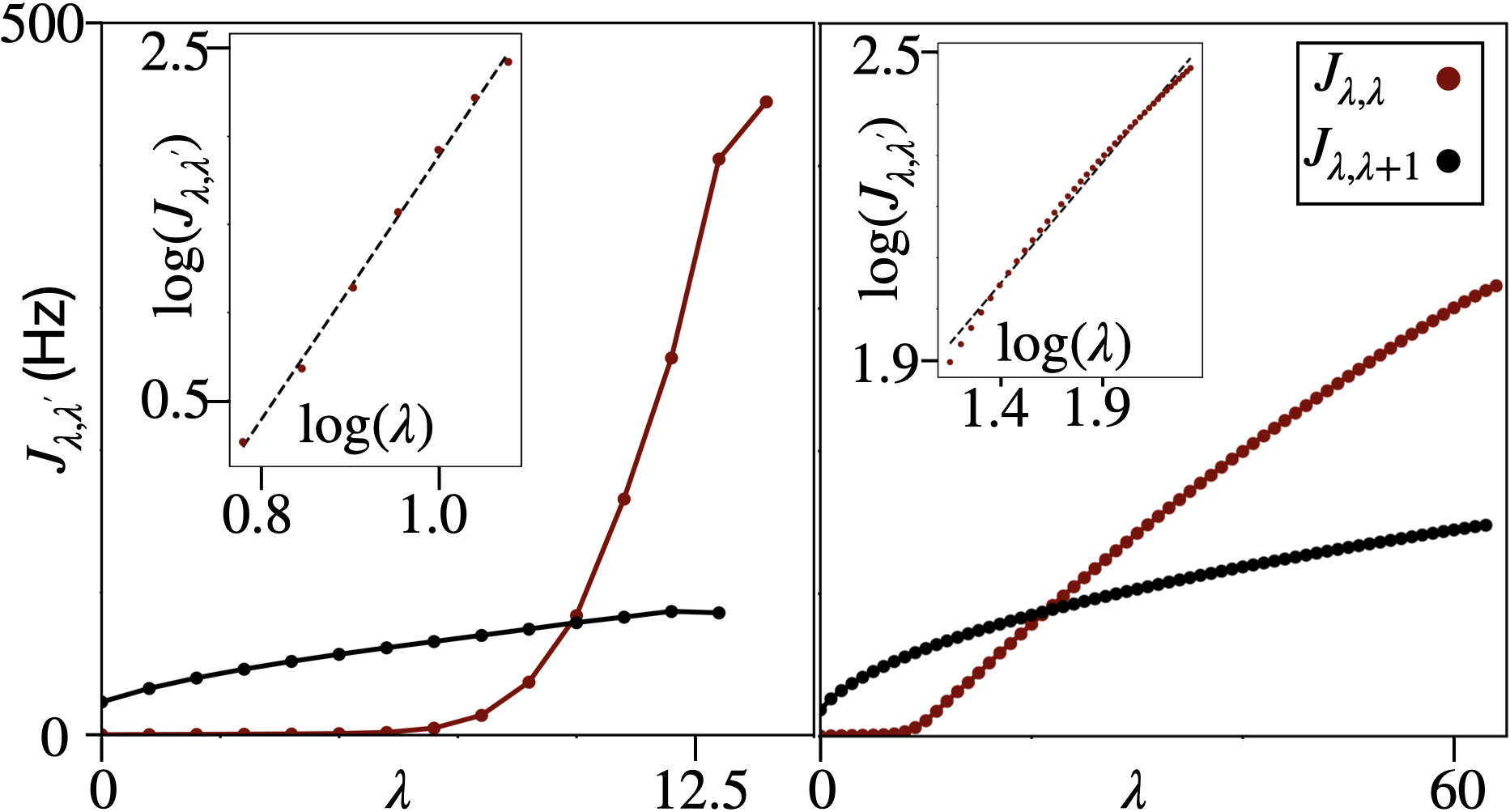}%
      \caption{Numerically-calculated matrix elements of the effective 1D Floquet Hamiltonian for the examples of a sharp edge ({\it left panel}) and a soft edge ({\it right panel}) along the synthetic dimension, as found by scanning over various parameters (see main text for details). For the sharp edge, the onsite terms ({\it red dots}) become comparable to the nearest-neighbor hopping energy ({\it black dots}) over $\sim 4$ sites, while for the soft edge, this requires more than 10 sites. Lines plotted are a guide to the eye. To quantify the edge softness, we plot the onsite terms on a logarithmic scale for $\lambda \in \left[6, 13 \right]$ ({\it left inset}) and $\lambda \in \left[20, 60 \right]$ ({\it right inset}). A linear fit ({\it dashed line}) is used to extract the power-law exponent, which is found to be $\gamma_\lambda=7.4$ for the sharp edge and $\gamma_\lambda=1.2$ for the soft edge, demonstrating the tunability of this upper boundary.}
      \label{fig:flsharpvssoft} 
\end{figure}

To explore the effects of changing $\gamma$ on the chiral edge state dynamics, we numerically simulate the full time-evolution of an initial Gaussian wave-packet [c.f.~Section~\ref{sec:scheme}]. The resulting center-of-mass evolution in the $\lambda\!-\!y$ plane is shown in Fig.~\ref{fig:softcom} for different values of $\gamma$. Here, we have chosen $V_d \!=\! 981\text{Hz}$, $x^-\!=\!4 x_{\text{ho}}$ and $x^+\!=\!6 x_{\text{ho}}$ (corresponding to $\lambda_{e-}\!=\!8$ and $\lambda_{e+}\!=\!18$ respectively), with all other parameters as stated in Section~\ref{sec:scheme}. 
As can be seen, all the orbits coincide initially, exhibiting clockwise skipping trajectories as they travel left along $y$ and up the first few $\lambda$ states [c.f.~Fig \ref{fig:softcom}]. Upon reaching $\lambda_{e-}$ ({\it lower dashed line}), the trajectories begin to diverge, with those for lower $\gamma$ being more significantly affected by the power-law potential and thus extending less far up the synthetic dimension. Notably, we observe that the shape of the upper part of the orbit becomes much flatter as $\gamma$ is increased, confirming our expectation that the effective upper edge in the synthetic dimension becomes sharper.  

We have also explored numerically how varying more parameters in our model can be used to maximize the sharpness/softness of the edge along the synthetic dimension. In Fig.~\ref{fig:flsharpvssoft}, we show the calculated Floquet matrix elements for examples found of a very sharp edge ({\it left panel}) and a very soft edge ({\it right panel}). The former corresponds to a square-well potential with $V_d = 1040$Hz and $\lambda_e=10$, while the latter is for a real-space potential with $\gamma=1$, $V_d = 590$Hz, $\lambda_{e-}=8$ and $\lambda_{e+}=60$. We have also chosen $\kappa = -65.5\text{Hz}$ for both, with other parameters as in Section~\ref{sec:scheme}. As can be seen, for the sharp edge, the onsite terms rise rapidly from zero, while for the soft edge, a similar rise requires tens of sites. To quantify the softness of the edge, we plot the onsite terms on a logarithmic scale [see insets in Fig.~\ref{fig:flsharpvssoft}] in order to extract the power-law dependence as a function of $\lambda$. We find that these onsite terms vary $\propto \lambda^{\gamma_\lambda}$, with an exponent of $\gamma_\lambda=7.4$ for the sharp edge and $\gamma_\lambda=1.2$ for the soft edge. This provides the upper and lower bound of the edge sharpness in the regime of parameters investigated throughout this paper. Additionally, we observe that the curvature of this edge potential can be controlled via this method too with positive and negative curvatures being seen respectively in the two cases shown in Fig~\ref{fig:flsharpvssoft}. It is possible to interpolate between these two cases by further varying the parameters, showing that the softness of the upper edge can be tuned over a wide range.  

\begin{figure}
    \centering
      \includegraphics[width=1\columnwidth]
      {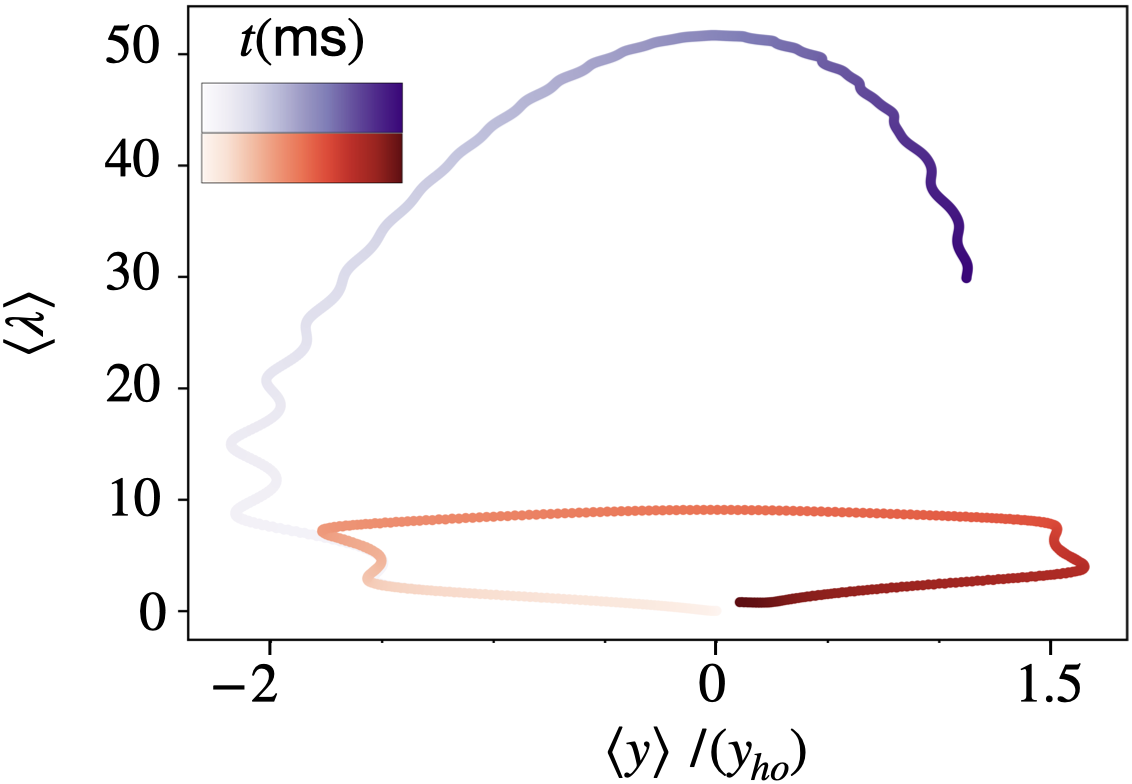}%
      \caption{The numerical evolution of the center-of-mass for a Gaussian wave-packet in the $\lambda \!-\!y$ plane, for the two cases shown in Fig.~\ \ref{fig:flsharpvssoft}. As can be seen, the upper part of the chiral trajectory for the sharp edge ({\it red}) is much flatter with respect to $\lambda$, as the atoms travel from left to right along $y$, than for the soft edge ({\it blue}), reflecting the different softness of the boundary. The colorbar for the red and blue trajectories
runs over time from t = 0ms (pale color) to t =638ms and t =2380ms
(dark color) respectively. }
      \label{fig:comsoftsharp} 
\end{figure}

To probe the effects on dynamics, we have numerically simulated the full time-evolution of an initial Gaussian wave-packet for each of the above two cases, with the resulting center-of-mass trajectories plotted in Fig. \ref{fig:comsoftsharp}. Here, we observe that the two trajectories, while both being chiral, have very different shapes, reflecting the differing softness of the upper edge. In particular, the upper part of the trajectory for the sharp edge ({\it red}) is much flatter with respect to $\lambda$ than that for the soft edge ({\it blue}), as expected. Note that for the soft edge, the chosen $V_d$ is weak enough that some of the wave-packet continues moving even further up the synthetic dimension; this is because there is a natural trade-off between how soft the upper edge is and its ability to prohibit all of the wave-packet from coupling into higher $\lambda$ states. This effect is reflected in Fig \ref{fig:comsoftsharp} by the asymmetry between the ascent and descent of the center-of-mass trajectory in synthetic space. 

\section{Constructing Impurities in the Synthetic Dimension \label{section:3}}

  \begin{figure}
\includegraphics[width = \columnwidth]{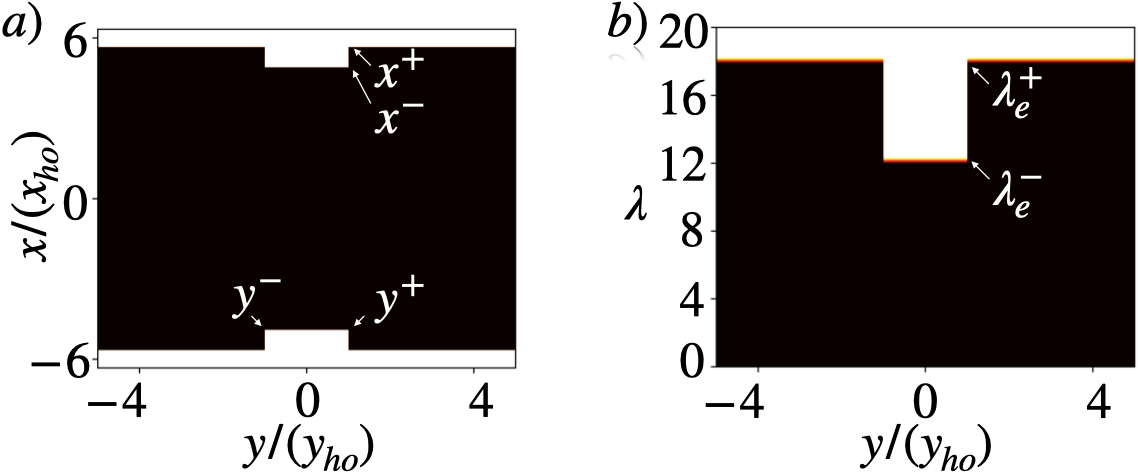}
\caption{Constructing an impurity along the upper edge of the synthetic dimension. (a) The real-space  potential as a function of $x$ and $y$ which is used to engineer both the upper edge and the impurity in synthetic space. The details and parameters of this construction are described in the main text and further specified in Appendix~\ref{sec:appedge}. Black indicates regions with vanishing potential, while white indicates a total potential strength of $V_d$. (b)  The potential from panel (a) represented schematically. Here black indicates  $\lambda \!<\! \lambda_{e-}$ for $y^- \!\leq\! y \!\leq\! y^+$, and $\lambda \!<\! \lambda_{e+}$ otherwise [c.f. Eq.~\ref{eq:lepm}]. As can be seen, the narrowing of the potential well in panel (a) leads to a narrowing of the system length along the synthetic dimension in panel (b); this can be interpreted as having a defect on the upper edge of the synthetic dimension. }
\label{fig:constructionimpurity}
\end{figure}

One of the reasons for the substantial research interest in topological physics is the robustness of topological edge states to back-scattering from defects and imperfections~\cite{RMP_TI, RMP_TI2,Ozawa_2019,price2022roadmap}. Having demonstrated above how to induce a tuneable upper edge in the synthetic dimension of harmonic trap states, we now show how the same toolkit can be extended to implement edge impurities, and hence to probe this intrinsic robustness in future experiments.  

To construct a boundary defect in the $\lambda\!-\!y$ plane, we consider a real-space potential like that shown in Fig.~\ref{fig:constructionimpurity} (a). As further detailed in Appendix~\ref{sec:appedge}, this potential is simply composed of a square-well potential, $S_1 (x,y)$, with height $V_d$ and $x_\text{width}\! =\! 2 x^+$, together with a second square-well potential, $S_2 (x, y)$, also with height $V_d$ but narrower width $x_\text{width} \!=\! 2 x^-$ that only exists within a region $y^- \!\leq\! y \!\leq\! y^+$ and $|x| \!\leq\! x^+$. The first square well is used to create the upper edge in the synthetic dimension [c.f.~Section~\ref{section:1a)}], while the second represents the defect. Note that for the Fig.~\ref{fig:constructionimpurity} (a), we have chosen $x^+ \!=\!6 x_\text{ho}$, $x^- \!=\!\sqrt{24} x_\text{ho}$, and $y^+\!=\! y^-\!=\!1 y_\text{ho}$. Experimentally, this type of potential could again be engineered using a DMD as in the previous sections, meaning this scheme requires no additional equipment. 

To visualise what the above means in synthetic space, we can use the logic that we developed when discussing the edge implementation in Section~\ref{section:1a)}. As before, we expect there to be a relationship between $x_{\text{width}}$ and where the effective on-site potential increases in the synthetic dimension. However, now $x_{\text{width}}$ has a simple $y$-dependence, which will in turn naturally imprint a $y$-dependence in the on-site potential for the synthetic dimension. Specifically, the narrower square-well potential [c.f.~Fig.~\ref{fig:constructionimpurity}(a)] 
will correspond to  $\lambda_{e-}$ [Eq.~\ref{eq:lepm}], while the wider one will correspond to $\lambda_{e+}$ (with $\lambda_{e+} \!>\!\lambda_{e-}$). This is represented schematically in Fig.~\ref{fig:constructionimpurity}(b), from which we can see that the real-space potentials lead to a defect adjacent to the upper edge of the synthetic dimension. For further confirmation of this, we have calculated the 1D Floquet Hamiltonian [c.f.~Section~\ref{sec:tunableedge}], for values of $y$ both inside and outside the  impurity region. Taking $V_d\!=\!0.13 \text{MHz}$ and other parameters as in Section \ref{sec:scheme}, we indeed have found that the onsite potential in synthetic space increases over a few sites to become comparable to the nearest-neighbor hopping amplitudes around $\lambda_{e+}$ and $\lambda_{e-}$, when outside and inside the impurity region respectively.   

Before continuing, it is also worthwhile highlighting that there is considerable freedom to choose the position and size of this impurity. For example, we can select different values of $y^{-}$ and $y^{+}$ in order to change the width of the defect or move it along the $y$-direction. We can similarly tune $x^-$ in order to increase the width of the defect along the synthetic dimension. However, we cannot easily vary our parameters to move the entire defect away from the upper edge to another position along $\lambda$, such as the lower edge at $\lambda=0$. This is because the type of real-space potential considered here affects all states above a given $\lambda$.     

\begin{figure}
\includegraphics[width = \columnwidth]{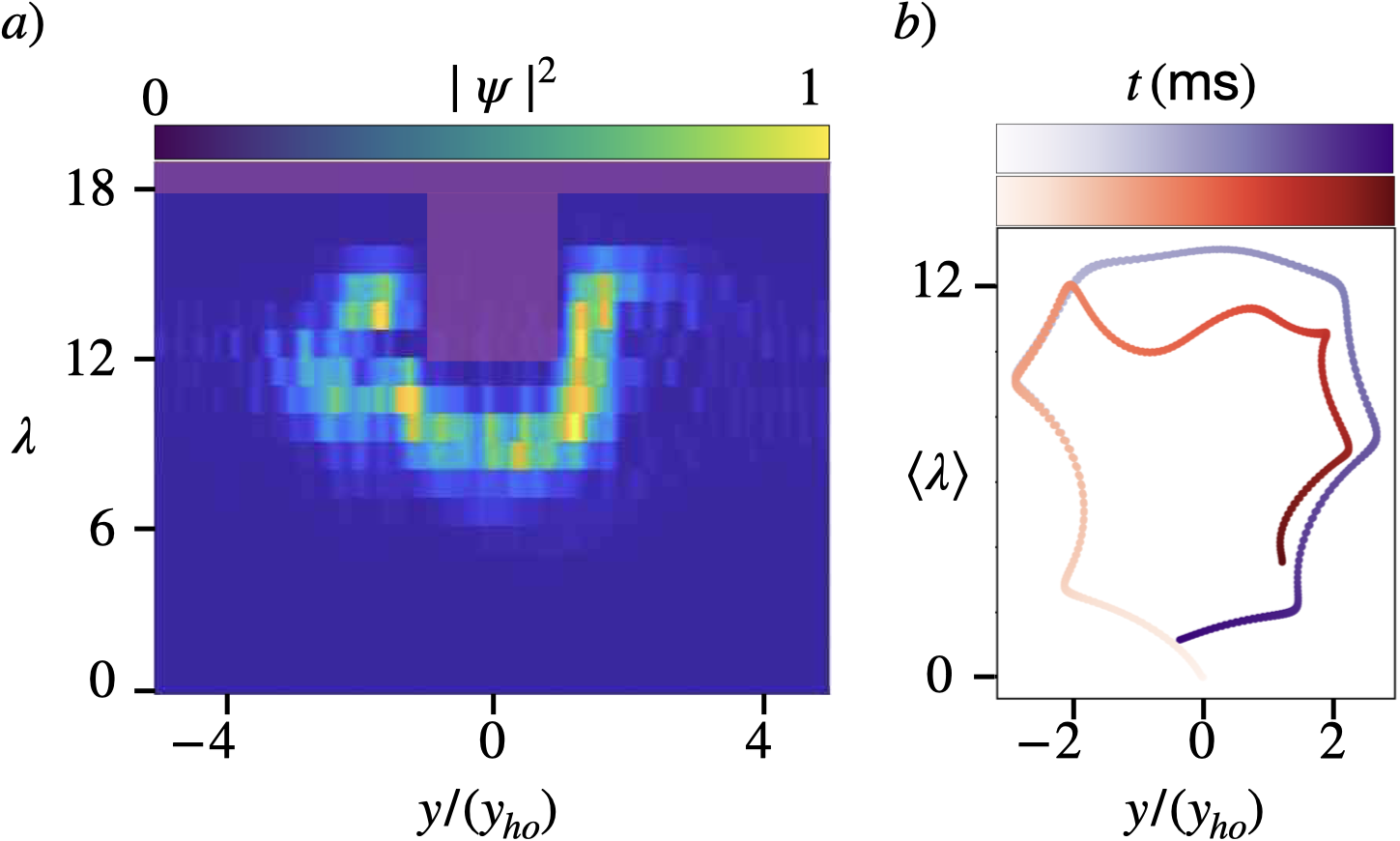}
\caption{a) A snapshot at $t=310 \text{ms}$ of the atomic density for the full time-evolution of an initial Gaussian wave-packet in the presence of the potential from Fig.~\ref{fig:constructionimpurity}. Parameters are as stated in the main text and in Section~\ref{sec:scheme}. The approximate position of the edge and impurity are indicated by purple shading [c.f.~Fig.~\ref{fig:constructionimpurity}(b)]. Here we see that the wave-packet travels robustly around the defect. (b) The corresponding center-of-mass trajectories in the $y\! -\! \lambda$ plane with the impurity ({\it red}) and without the impurity ({\it blue}). As can be seen, both trajectories are identical before reaching the impurity region, at which point  the former trajectory deviates to go around the impurity, before returning to the usual chiral motion , demonstrating the robustness of this behavior. }
\label{fig:7}
\end{figure}

To test the robustness of the chiral edge state behavior observed in our proposal, we now numerically simulate the full time-evolution of an initial Gaussian wave-packet including the real-space potentials described above. A snapshot of the resulting atomic density at $t=310$ms is shown in Fig.~\ref{fig:7}(a), where we can see that the wave-packet robustly travels around the defect. Note, although we outline the impurity and upper edge using $\lambda_{e^{+}}$ and $\lambda_{e^{-}}$ [{\it purple shading}], the atoms feel the some of the potential from the impurity earlier as expected from the Floquet matrix elements. It is also worth re-iterating that the peak of the atomic density is always seen a few sites lower than the edge due to the wave-packet's width [c.f.~Section~\ref{section:1a)}]. The parameters used here are $V_d\!=\! 0.13$MHz, with all others being the same as those in Fig.~\ref{fig:constructionimpurity} and in Section~\ref{sec:scheme}. The corresponding center-of-mass trajectory in the $\lambda \!-\!y$ plane is also then shown in Fig.~\ref{fig:7}(b) for the case of without ({\it blue}) and with ({\it red}) the defect. As can be seen, the defect causes a deviation in the latter trajectory, but the overall chiral behavior is still clearly visible, demonstrating the robustness of this motion. 

\begin{figure}
\includegraphics[width = \columnwidth]{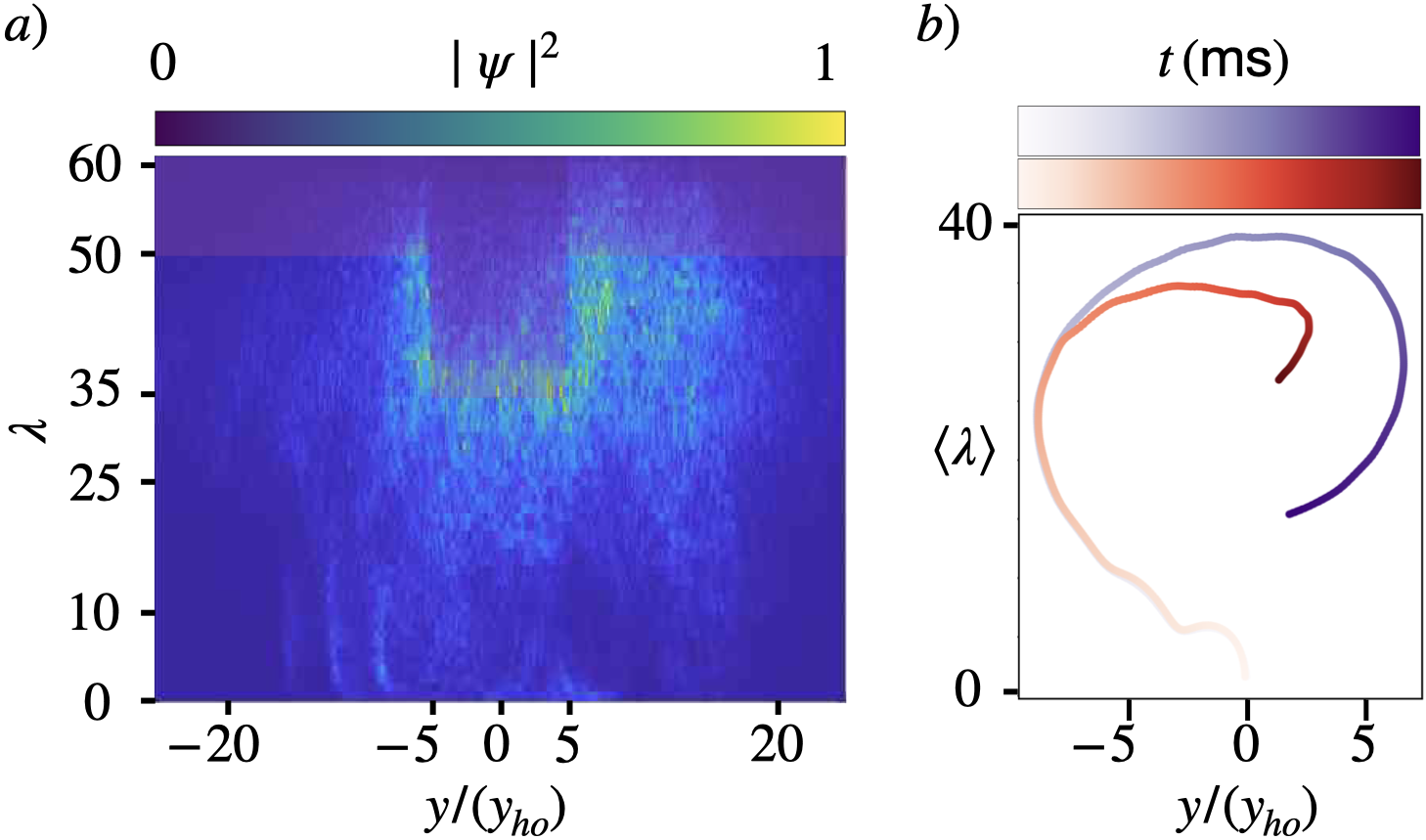}
\caption{ a) A snap shot at $t=300 \text{ms}$ for the atomic density of an initial thermal cloud.  Parameters are as stated in the main text and in Section~\ref{sec:scheme}. The approximate position, calculated via Eq. \ref{eq:le}, of the edge and impurity are indicated by purple shading. We again observe that most of the atoms travel robustly around the defect; however, in this case there is more noise due to a lower fraction of the atoms coupling into the edge state. b) The corresponding center-of-mass trajectories with (\textit{red}) and without ({blue}) the impurity for a total run-time of $t_f=451 \text{ms}$. Here we can see that the two trajectories diverge after reaching the defect with both continuing in a clockwise orbit, but, due to the spread of the thermal cloud, the rough outline of the impurity can not be directly observed.}
\label{fig:thermimp}
\end{figure}

Having understood the idealised case, we will perform more experimentally-realistic simulations for a thermal cloud [see Section~\ref{sec:scheme}]. 
A snapshot of the atomic density is shown in Fig.~\ref{fig:thermimp}(a) for $t=300$ms, with $x^+=\sqrt{100} x_{ho}$, $x^-=\sqrt{70}x_{ho}$, $y^-=-5y_{ho}\!$, $y^+\!=5y_{ho}\!$ and $V_d\!=\! 0.13$MHz, and other parameters as in Section~\ref{sec:scheme}. As can be seen, the impurity again acts as expected with most of the atoms traveling around the defect. However, as compared to the Gaussian wave-packet [c.f.~Fig.~\ref{fig:7}], there is more atomic density spread through the bulk and remaining around the lower edge of the system. This is because a smaller fraction of the atoms are initially well-coupled into the edge state for the thermal simulations, as also discussed in Section~\ref{section:1a)}. More sophisticated state preparation would therefore be needed if coupling a larger fraction of the wave-packet is required experimentally. Finally, we also show in Fig.~\ref{fig:thermimp}(b), the corresponding center-of-mass trajectory with ({\it red}) and without ({\it blue}) the defect. As can be seen, the two trajectories again diverge when reaching the impurity; however, the increased spread of atoms in the thermal simulations obscure the outline of the defect. Nevertheless, the majority of the atoms continue moving in a clockwise orbit, indicating the robustness of the chiral motion in the presence of an impurity. 

\section{Bulk Physics\label{section:4}}

An advantage of using atomic trap states is that the resulting synthetic dimension can be very long and consist of many tens of sites, allowing us to employ a single set-up to explore the different phenomena that can occur on the edge or in the bulk of a quantum Hall system. In this section, we focus on how to observe the characteristic semi-classical bulk physics associated with the 2D coupled wire model 
with our scheme, firstly in idealised synthetic systems, and then for more realistic experimental settings. 

As has been previously studied theoretically~\cite{Ozawa:2017PRL,oliver2023artificial} and experimentally~\cite{Chalopin2020, Roell2023}, a semi-classical wave-packet in the bulk of a 2D coupled wire model will undergo cyclotron orbits, similar to those found in Landau levels. A key signature of these orbits is that they are clockwise/anti-clockwise depending on the sign of the magnetic field strength, $\phi_0$. Considering the equations of motion for
a classical particle subject
to a magnetic field, also leads to an estimate for the cyclotron frequency as~\cite{Ozawa:2017PRL}
\begin{equation}
\omega_c = \frac{\phi_0}{\mid m_{\text{eff} }\mid}
\label{eq:omegac}
\end{equation}
where the effective mass $m_{\text{eff}}\!=\!\sqrt{m_d m_c}$ is determined by the geometric mean of the effective masses $m_d$ and $m_c$ associated with the discrete and continuous directions, respectively~\cite{oliver2023artificial}. The latter effective mass, $m_c$, simply corresponds to the mass of the associated atom, while the former, $m_d$, can be estimated from the 1D tight-binding dispersion as
\begin{equation}
m_{d} = -\frac{1}{2 J \cos \langle k_d \rangle },\label{Eq:meff}
\end{equation}
where $\langle k_d \rangle $ is the center-of-mass momentum along the discrete direction and $J$ is the corresponding (uniform) hopping amplitude. 

However, as compared to the ideal model above, our scheme has several complications, as discussed further below, including non-uniform hopping amplitudes along our discrete direction $\lambda$; a weak harmonic trap along our continuous direction $y$; and initial states which are naturally localised at the edge (i.e. near $\lambda\!=\! 0$) rather than in the bulk of the system, due to the energetic nature of the $\lambda$-state synthetic dimension [c.f.~Section.~\ref{sec:scheme}]. To overcome the latter issue, we propose to employ an initial ``preparation phase" in which we turn off the artificial magnetic flux and use a detuning of the driving frequency, $\Delta \!\equiv \! \omega - \omega_D$, to impose a force [c.f.~Eq.~\ref{eq:main}] and hence induce a Bloch oscillation to move atoms up the synthetic dimension~\cite{oliver2023bloch}. Once the atoms are sufficiently far away from the edge of the synthetic space, we then set $\Delta\!=\!0$ and turn on the artificial magnetic flux, $\phi_0$, in order to look for bulk cyclotron-orbit behavior.

  \begin{figure}
    \centering
      \includegraphics[width=0.98\columnwidth]{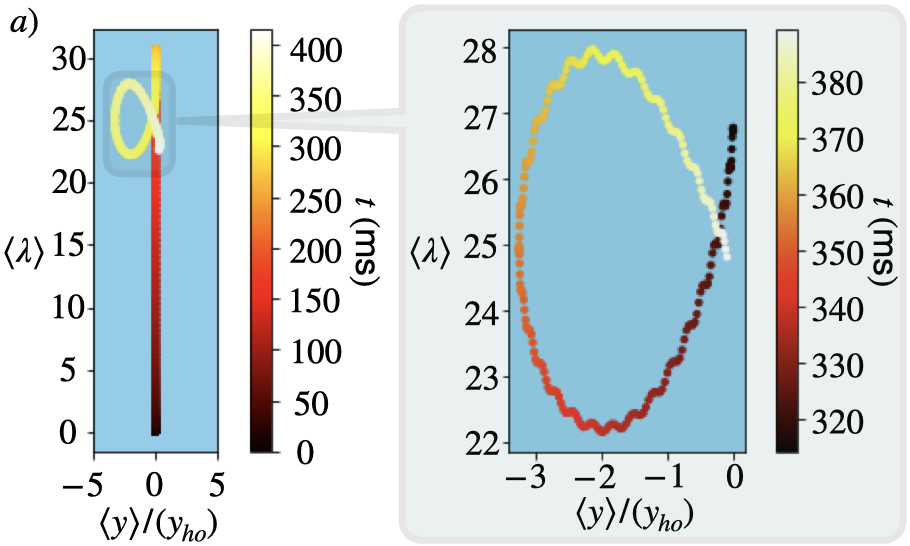}%
      
      \includegraphics[width=0.98\columnwidth]{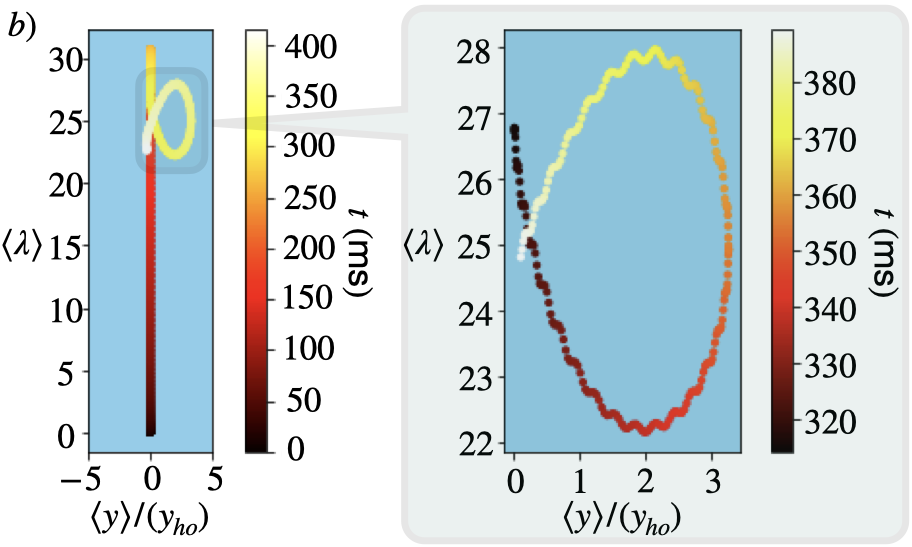}%

    \caption{  \label{fig:blochcom} The numerical evolution of the center-of-mass in the 
    $\lambda \!-\! y$ plane for an initial Gaussian wave-packet under a two-step protocol consisting of an initial preparation phase, where a non-zero detuning induces a Bloch oscillation and pushes the wave-packet into the bulk, followed by a second phase in which a non-zero artificial magnetic flux generates cyclotron-like orbits, as also highlighted in the right-hand panels. Parameters are as given in the main text, with $\phi_0 = \pm 0.2 (\mu \text{m})^{-1}$ in the second phase of the protocol, for the upper and lower panels respectively. As can be seen, reversing the sign of the flux, reverses the sense of the orbit, as expected for cyclotron motion.   }
  \end{figure}

To first demonstrate the qualitative signatures of cyclotron orbits in our scheme, we simulate the full time-dynamics under Eq.~\ref{eq:full} for an initial Gaussian wave-packet, with parameters as in Section~\ref{sec:scheme}. During the initial preparation phase, we set $\Delta \!=\! 10$Hz from $t\!=\!0$ until $t_p\!=\! 314$ms; thereafter we set  $\Delta \!=\! 0$ and turn on the artificial magnetic flux. The resulting center-of-mass trajectory in the synthetic $\lambda\!-\!y$ plane is shown in Fig.~\ref{fig:blochcom} with (a) $\phi_0 = 0.2 \mu m^{-1}$ and (b) $\phi_0 = -0.2 \mu m^{-1}$. As can be seen, during the preparation phase, the center-of-mass first rises steadily up the synthetic dimension, while maintaining constant $\langle y\rangle$. As a Bloch oscillation corresponds to periodic motion along $\lambda$, the wave-packet reaches a maximum value of $\langle \lambda \rangle$ (set by the size of the detuning as well as other parameters~\cite{oliver2023bloch}) before starting to descend the synthetic dimension again. We then quench the Hamiltonian by switching off the detuning and turning on the magnetic flux. By changing $t_p$ and $\Delta$, we can control both the final center-of-mass position $\lambda$ at the end of the preparation phase as well as the corresponding semiclassical center-of-mass momentum $\langle k_{\lambda} \rangle$ along the synthetic dimension. Following the quench, the wave-packet then moves in an ellipse-like trajectory in an opposite sense for opposite magnetic flux [see insets of Fig.~\ref{fig:blochcom}], as expected for a cyclotron-like orbit. Note that we also observe small and fast oscillations in these trajectories, corresponding to micromotion in the full time dynamics going beyond the effective Hamiltonian~\cite{Price2017}.

\begin{figure}
\centering
      \includegraphics[width=1\columnwidth]{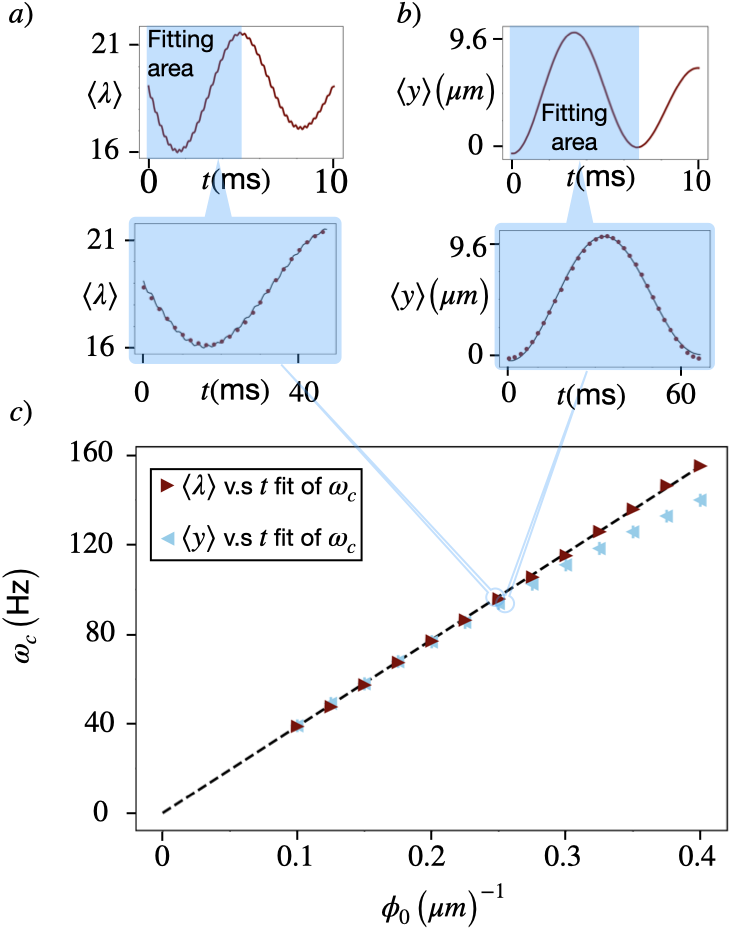}
       \caption{\label{fig:cyclotronsimple} Extracting the cyclotron frequency for an idealised scheme, which neglects the harmonic trap along $y$ and assumes the wave-packet can be prepared directly in the bulk (all parameters given in the main text). (a) \& (b) The numerical center-of-mass along $\lambda$ and $y$ respectively as a function of time, showing characteristic sinusoidal oscillations. For convenience, the latter is plotted in units of $y_\text{ho}$ [see main text], although $\omega_y=0$ for the simulations. The data within the window highlighted in blue is fitted according to Eq.~\ref{Eq:fitfucntion}. The result of this fit is plotted in the lower panels ({\it maroon dots}), showing good agreement with the full dynamics ({\it solid black lines}). Note that the data for $\langle \lambda \rangle$ is smoothed by a moving average before fitting to remove the effects of micromotion. (c) The corresponding values for the cyclotron frequency obtained by fitting $\langle \lambda \rangle$  (\textit{maroon triangles}) and $\langle y \rangle $  (\textit{sky blue triangles}) for different values of $\phi_0$. Errors from the fits are plotted but do not exceed the size of the markers. Also plotted is a simple analytical prediction ({\it black dashed line}) for the cyclotron frequency from Eq.~\ref{eq:omegac} and Eq.~\ref{Eq:meff}, as further discussed in the main text.}
  \end{figure}
  
\begin{figure}
\centering
      \includegraphics[width=1\columnwidth]{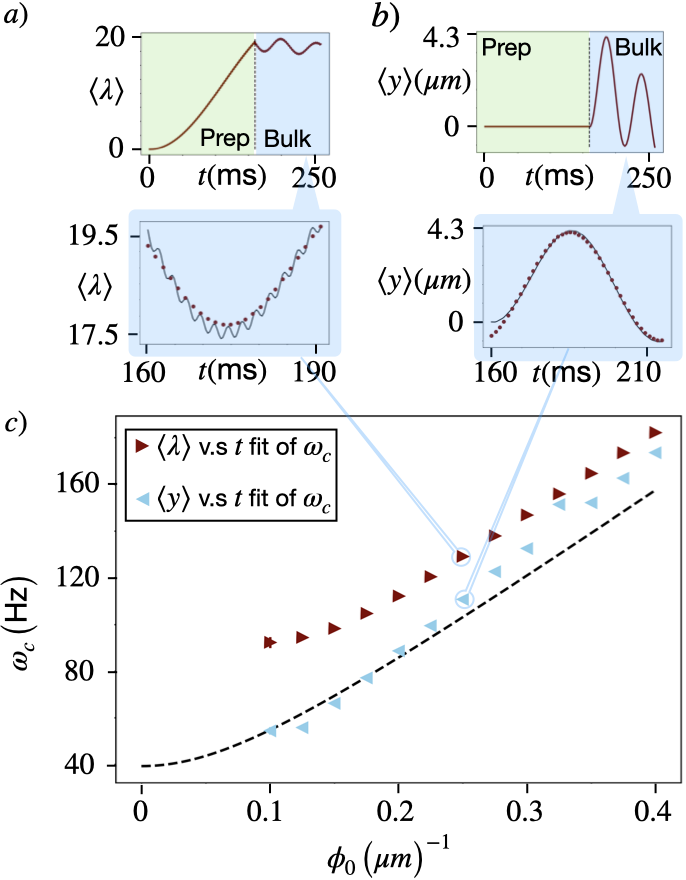}
\caption{\label{fig:cycltron_widthprep}  Extracting the cyclotron frequency when including the weak harmonic trap and the initial preparation stage (all parameters in the main text). (a)\&(b) The numerical center-of-mass along $\lambda$ and $y$ respectively, with the latter in units of $y_\text{ho}$. The initial preparation stage is highlighted in green, and is used to push the wave-packet from the bottom of the synthetic dimension into the bulk. This is followed by the second phase highlighted in blue, in which we turn off the detuning and switch on the magnetic field to induce cyclotron-like motion. The initial oscillations within the second phase are fitted according to Eq.~\ref{Eq:fitfucntion}, with the result plotted in the lower panels ({\it maroon dots}), alongside the full dynamics ({\it solid black lines}). Note that the data for $\langle \lambda \rangle$ is again smoothed  before fitting to remove micromotion effects. (c) The corresponding extracted cyclotron frequencies for $\langle \lambda \rangle$  (\textit{maroon triangles}) and $\langle y \rangle $  (\textit{sky blue triangles}) for different values of $\phi_0$. Errors from the fits are plotted but are small compared to the size of the markers. Also plotted is a simple analytical prediction ({\it black dashed line}) taken from Eq.~\ref{eq:omegac2} and Eq.~\ref{Eq:meff}, as discussed in the main text. Overall, there is good qualitative agreement, although with deviations due to the more complicated trajectories, shown e.g. in panels (a)\&(b).}
  \end{figure}

To now quantitatively study these orbits, we start from a simplified model in which we neglect both the weak harmonic trap along $y$ and the initial preparation stage. This means that we choose an initial Gaussian state [Eq.\ref{Eq:gaussian_initstate}], here with $\lambda_0 = 19.1$ and $y_0=0.1$, such that the wave-packet initially resides entirely within the bulk. We also choose $k_{\lambda}=2$, reflecting that the wave-packet can have a non-zero momentum along $\lambda$. The other wave-packet parameters are chosen as $\sigma_{\lambda} = 4.3$,  $\sigma_{y} = 0.35 y_{ho}$ and $k_y = 0$ to ensure good visibility of the desired cyclotron motion numerically. Using this initial state, we then simulate the full time-dynamics for different $\phi_0 \in \left[0.1, 0.4 \right] \left(\mu \text{m}\right)^{-1}$ as shown in Fig.~\ref{fig:cyclotronsimple}. All parameters are as stated in section \ref{sec:scheme} expect for $\kappa = -100\text{Hz}$.

For each choice of the magnetic flux, $\phi_0$, we 
track the center-of-mass trajectories $\langle\lambda \rangle$ and $\langle y\rangle$ with respect to time; an example of this is shown for $\phi_0 = 0.25 (\mu \text{m})^{-1}$ in the upper panels of Fig.~\ref{fig:cyclotronsimple}(a) and (b). As expected, we observe that both $\langle\lambda \rangle$ and $\langle y\rangle$ evolve sinusoidally in time, corresponding to a cyclotron-like orbit. To estimate the corresponding cyclotron frequency, $\omega_c$, we fit each trajectory according to the functional form
  \begin{equation}
  f = A\cos(\omega_c t + \phi) + h_0, 
  \label{Eq:fitfucntion}
  \end{equation} 
  where $A$, $\omega_c$, $\phi$ and $h_0$ are all fitting parameters. Physically, $\phi$ is an initial phase offset, while $A$ and $h_0$  represent the amplitude and initial value for oscillations in either $\langle y\rangle$ or $\langle \lambda \rangle$ as appropriate. When fitting there are two additional issues we have to deal with; firstly, there is micromotion visible as small and fast oscillations in the $\langle \lambda \rangle$ trajectory. To obtain the best estimate for the cyclotron frequency in this case, we therefore remove the micromotion by applying a moving average to the numerical data before fitting. Secondly, the final fit parameters are sensitive to the region of the trajectory for which the fitting is undertaken. To reduce arbitrariness, we therefore define our fitting region as the time-window extending up to the second extremum of the oscillation (not counting extrema that occur very close to $t\!=\!0$). Examples of this fitting window are highlighted in blue in the upper panels
of (a) and (b). In the corresponding lower panels,  we then zoom in to this window to highlight the resulting good agreement between the numerical data ({\it solid black line}) and the corresponding fit function ({\it maroon dots}). 

In Fig.~\ref{fig:cyclotronsimple}(c), we then plot the cyclotron frequencies obtained by fitting the numerical data separately for $\langle\lambda \rangle$ ({\it maroon triangles}) and $\langle y \rangle$ ({\it light blue triangles}) as a function of the artificial magnetic flux, $\phi_0$. We also plot error bars from the fit for each point but these are small compared to the marker size. We compare our numerical results with the simple analytical prediction from Eq.~\ref{eq:omegac}, taking $m_c=m$ and using Eq.~\ref{Eq:meff} to estimate $m_d$. For the latter, we approximate $J$ as the value of $J_\lambda$ [Eq.~\ref{eq:J}] evaluated at $\lambda=\lambda_0$, and use $\langle k_d \rangle \approx k_\lambda$ from the initial wave-packet. Note that $J_\lambda$ and hence $m_d$ can vary during the cyclotron orbit; however, as $J_\lambda \!\propto\! \sqrt{\lambda}$, we assume this effect can be neglected when only a narrow range of high-$\lambda$ states are explored as in these cases. The resulting analytical prediction is plotted as a dashed black line, and is seen to be in excellent agreement despite these approximations, particularly with the fitted results for $\langle \lambda \rangle$. Note that the fitted results for $\langle y \rangle$ also agree well for small $\phi_0$, but then begin to deviate at higher magnetic fluxes where the fitting performs less well.  
   
Now we will build upon these results for a simplified system and begin to add in more experimental complications, starting with the weak harmonic trap along $y$. This weak trap affects how the cyclotron frequency varies as a function of magnetic field, as can be seen by considering a simple classical Hamiltonian with  minimal coupling ~\begin{equation}
\mathcal{H}_{MC} = \frac{p_x^2}{2m_x} + \frac{\left(p_y - q B x \right)^2}{2m_y} + \frac{1}{2}m_y \omega_y^2 y^2,
\end{equation}
where $q$ is the particle charge, $\mathbf{A} = \left(0, Bx, 0 \right)^T$ is the magnetic vector potential, $B$ is the magnetic field strength and $m_x$ and $m_y$ are the masses along the (continuous) $x$ and $y$ directions respectively. Using Hamilton's equations leads to
\begin{align}
{}&
\begin{pmatrix}
\Ddot{p}_y \\ \Ddot{x} \\ \Ddot{y} \\ \Ddot{p}_x
\end{pmatrix} = \begin{pmatrix}
-\omega_y^2 & qB \omega_y^2 & 0 & 0 \\ \frac{qB}{m_xm_y} & -\frac{(qB)^2}{m_xm_y} & 0 & 0  \\ 0  & 0 & -\omega_y^2 & -\frac{qB}{m_xm_y}\\ 0 & 0 & -qB \omega_y^2 &-\frac{(qB)^2}{m_xm_y}
\end{pmatrix}  \begin{pmatrix}
p_y \\ x \\ y \\ p_x
\end{pmatrix},
\end{align} 
from which we can assume solutions of the form
\begin{equation}
 \begin{pmatrix}
p_y(t) \\ x(t) \\ y(t) \\ p_x(t)
\end{pmatrix}  = \begin{pmatrix}
A_1 \\ A_2 \\ A_3 \\ A_4 
\end{pmatrix} e^{i \omega_c t},
\end{equation}
where $A_1,...A_4$ denote the  time-independent amplitudes.
In addition to two trivial solutions, $\omega_c^2 = 0$, this leads to cyclotron orbits with
\begin{equation}\omega_c =\pm \sqrt{\omega_y^2 + \frac{(qB)^2}{m_{\text{eff}}^2}} , \label{eq:omegac2}
\end{equation}
where $m_{\text{eff}}=\sqrt{m_x m_y}$. Therefore, we may expect that adding a weak harmonic trap along $y$ leads to an offset in the cyclotron frequency if $\omega_y$ is sufficiently small enough. 

To test this numerically, we again investigate the full time-dynamics for an initial Gaussian wave-packet, but now including both the weak trap along $y$ as well as the preparation stage of the protocol described above. Examples of the resulting numerical evolution of the center-of-mass along $\lambda$ and $y$ are shown in Fig.~\ref{fig:cycltron_widthprep}(a)\&(b) respectively. Here, we have chosen an initial state  [Eq.~\ref{Eq:gaussian_initstate}] with $\lambda_0 \!=\! y_0\! =\!k_y \!=\!k_{\lambda} \!=\!0$, $\sigma_y \!= \!0.5 y_{ho}$ and $\sigma_{\lambda} \!= \!0.5$. We have also set the detuning $\Delta \!= \!12.5 \text{Hz}$, as well as  $\kappa \!=\!-100 \text{Hz}$ and $\omega \!= \!2 \pi \times 200 \text{Hz}$. The initial preparation phase [{\it light green region} in (a)\&(b)] with $\phi_0\!=\!0$ lasts until $t_p \!= \!2/|\Delta|$, such that the semiclassical wave-packet has a final center-of-mass momentum $\langle k_{\lambda} \rangle\! \approx \!2$ [c.f.~Fig.~\ref{fig:cyclotronsimple}] as well as a final center-of-mass position in the bulk. Due to the weak trap, the width of the wave-packet along $y$ also expands and contracts during the dynamics; for the above parameters, we have numerically selected $\omega_y \!=\! 2\pi \times 6.25$Hz as leading to a reasonably narrow final width and thus clear oscillations in the second phase of the evolution [{\it light blue region} in (a)\&(b)]. In this second phase, $\Delta\!=\!0$ with a non-zero value of $\phi_0$, here chosen as $\phi_0\!=\! 0.25 \left(\mu \text{m}\right)^{-1}$. Note that the trajectories are now more complicated than for the idealised case in Fig.~\ref{fig:cycltron_widthprep}, for example, due to the breathing of the wave-packet due to the weak trap.     

To estimate a cyclotron frequency, we follow the same procedure as for Fig.~\ref{fig:cycltron_widthprep} by  fitting the numerical data according to a sinusoidal function [Eq.~\ref{Eq:fitfucntion}]. Similar to above, we define the fitting window as extending from $t_p$, the time of the quench, up to the second extremum in the oscillation (not counting extrema that occur very near to $t_p$). Examples of the resulting fits are shown ({\it maroon dots}) in the lower panels of Fig. \ref{fig:cycltron_widthprep}~(a)\&(b), along with the full numerical trajectories ({\it black solid line}). Due to the more complicated evolution noted above, these fits are not as accurate as for the previous idealised case, but still appear to capture the main features of the data. The corresponding estimates for $\omega_c$ are then extracted and plotted in Fig. \ref{fig:cycltron_widthprep}~(c) for different values of $\phi_0$. Errors from the fit are plotted, but are small compared to the marker size. Here, we have also plotted the classical analytical prediction ({\it dashed black line}) of Eq.~\ref{eq:omegac2}, with $q B\!=\!\phi_0$ and $m_{\text{eff}}=\sqrt{m_d m}$, where $m_d$ is estimated from Eq.~\ref{Eq:meff}. For the latter, we have used that  $\langle k_{\lambda} \rangle \approx 2$ and approximated $J$ as $J_\lambda$ [Eq.~\ref{eq:J}] evaluated at the final center-of-mass position along the synthetic dimension at the end of the preparation phase. Note that we observe significant quantitative discrepancies between the different sets of results; this is because the more complicated trajectories make our results more dependent on the choice of fitting window. These discrepancies can be significantly reduced by choosing an optimal fitting window for each trajectory individually; we have chosen not to do this to avoid the arbitrariness introduced by such a procedure. Nevertheless, despite these issues, we observe that the simple analytical formula exhibits similar trends to the data, including a flattening out of the cyclotron frequency at small $\phi_0$, due to the weak trap, as well as the expected linear dependence of $\omega_c$ on the magnetic field at larger values of $\phi_0$. Note, that in this case $\omega_y$ is sufficiently small to simply act as an offset.

\begin{figure}
    \centering

\includegraphics[width=\columnwidth]{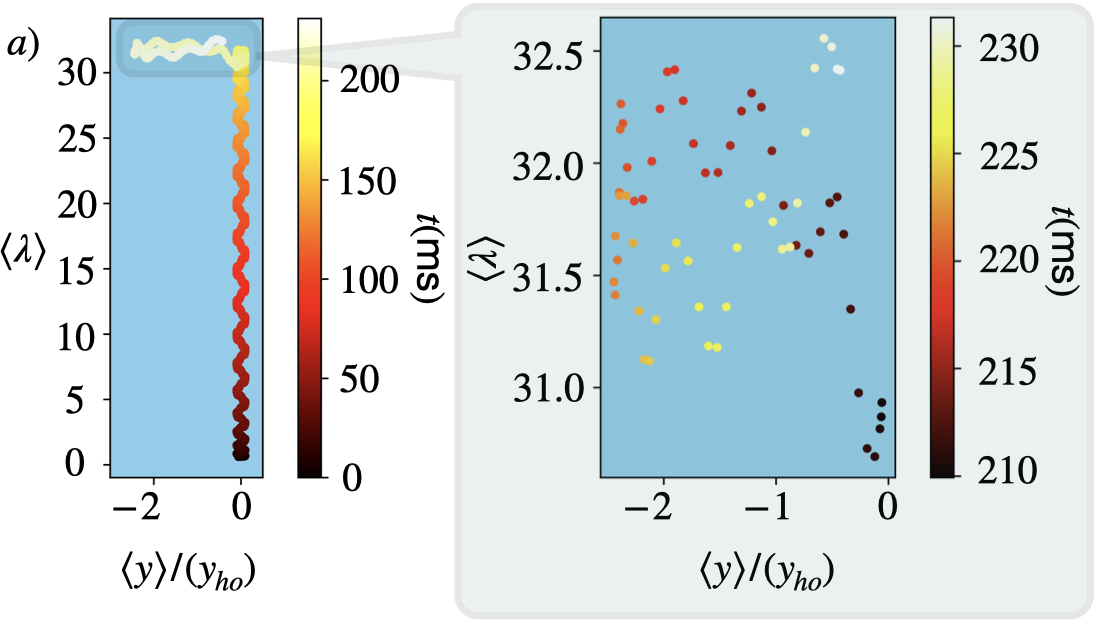}

\includegraphics[width=\columnwidth]{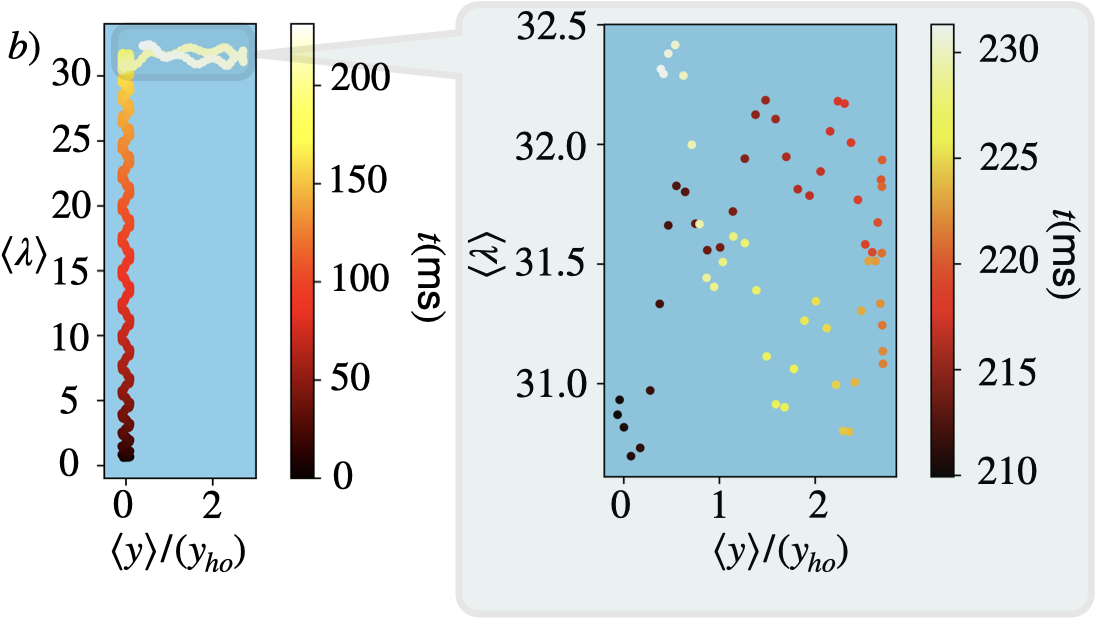}
\caption{\label{Fig:thermalbulk} The numerical evolution of the center-of-mass in the 
    $\lambda \!-\! y$ plane for a thermal cloud [c.f.~Section~\ref{sec:scheme}] under a two-step protocol consisting of an initial preparation phase, where a non-zero detuning induces a Bloch oscillation and pushes the atoms into the bulk, followed by a second phase in which a non-zero artificial magnetic flux generates cyclotron-like orbits, as also highlighted in the right-hand panels. Parameters are as given in the main text, with $\phi_0 = \pm 0.3 (\mu \text{m})^{-1}$ in the second phase of the protocol, for the upper and lower panels respectively. As can be seen, reversing the sign of the flux, reverses the sense of the orbit, as expected for cyclotron motion.  }
\end{figure}

Finally, we find that qualitative signatures of cyclotron-like motion can still be observed when including thermal effects [c.f.~Section~\ref{sec:scheme}], as shown for the center-of-mass evolution in the $\lambda \! -\! y$ plane in Fig. \ref{Fig:thermalbulk}. Here we have imposed a significantly stronger trap along $y$, set by $\omega_y \!=\! 2 \pi \times 100 \text{Hz}$, in order to ensure the atoms are initially well-localised in the $y$ direction at the initial temperature of $T \!=\! 20 \text{nK}$. Other parameters are as in Section~\ref{sec:scheme}, except for $\Delta\!=\! 15 \text{Hz}$, $\kappa\! =\! -260 \text{Hz}$ and $t_p\!=\! 209 \text{ms}$. During the second phase of the protocol, we set $\Delta\!=\!0$ and $\phi_0 = \pm 0.3 \left(\mu m\right)^{-1}$ for panel (a) and (b) respectively. As can be seen, the initial Bloch oscillation again drives the atoms up the synthetic dimension, although now with additional visible oscillations along $y$ due to the stronger trap in this direction. After the preparation phase, we again observe elliptical cyclotron-like trajectories, as highlighted by the right hand panels in Fig. \ref{Fig:thermalbulk}. Due to the larger $\omega_y$ and larger $\kappa$, the corresponding cyclotron frequency is significantly larger [c.f.~Eq. \ref{eq:omegac2}], leading to smaller orbits and more visible micromotion oscillations along $\lambda$. We also clearly observe that these orbits have the opposite sense for opposite signs of $\phi_0$, as expected for cyclotron-like physics. 
 
 \section{Conclusions} \label{sec:conclusions}

In this paper, we have theoretically proposed how to observe robust chiral edge state physics by exploiting atomic trap states as a synthetic dimension. To achieve this, we combine the synthetic dimension with a real spatial dimension and an artificial magnetic field, so as to realize a version of the 2D quantum Hall coupled wire model. To create a tunable upper edge along the synthetic dimension, we imprint additional real-space potentials in order to effectively simulate a synthetic 2D $\lambda\!-\!y$ plane with closed boundary conditions, and hence to explore chiral edge state motion. We have shown how this upper edge can be controlled, not only in terms of its location, but also its ``softness", allowing us to observe different types of chiral orbits. We can also imprint defects along this upper edge and hence visualize the robustness of the edge state motion. Finally, we have shown how we can  probe bulk cyclotron-like orbits within the same scheme by using suitable state-preparation protocols. 

Experimentally, our proposal could be implemented using, e.g. a digital micromirror device (DMD), and so does not require additional equipment as compared to the previous experiment~\cite{oliver2023bloch}. In our study, we have also used experimentally-relevant parameters and carried out simulations both for an idealized Gaussian wave-packet and a more realistic thermal cloud, paving the way for future experiments. Interestingly, the topological edge states described in this work are physically very different from those in other realizations~\cite{braun2024real, yao2023observation, Mancini2015, Stuhl2015, Chalopin2020}, as they correspond to atoms moving along opposite real-space directions depending on whether they occupy low-energy or high-energy states of a (perpendicularly-oriented) strong harmonic trap. This is therefore a new approach to experimentally controlling different trap states, which can be of interest in areas such as trapped and guided atom interferometry~\cite{Hu2018,Frank2014,Guarrera2015} and quantum thermodynamics~\cite{Vin2016,Quan2007,Uzdin2015}.     

This work also opens up many other interesting avenues for future study. Firstly, the artificial magnetic field in this set-up is highly controllable, as it is determined by the spatially-dependent phase of the applied shaking potential. In the future, this could be modified dynamically, e.g.~to investigate quenches of the magnetic field strength, or spatially, e.g.~to study the effects of magnetic barriers on transport. Secondly, it will be of great interest to explore the role of inter-particle interactions in this model, as the usual mean-field contact interactions in real space correspond to having exotic long-range interactions along the synthetic dimension~\cite{Price2017}, and so may lead to new types of interacting ground states and edge state physics. Finally, it will be interesting to extend our scheme to move defects away from the upper edge in the synthetic dimension, and so to study, more generally, the effects of imperfections and disorder on both edge and bulk physics. It is also possible to extend the width of the defect along the synthetic dimension, so as to create different types of geometries such as tunable quantum Hall interferometers~\cite{carrega2021anyons}. These types of systems are of great interest in solid-state physics, as, for example, they provide a way to probe the fractional statistics of quasiparticles in fractional quantum Hall materials~\cite{nakamura2020direct}, but they have yet to be realised within cold-atom experiments. For an even longer synthetic dimension of hundreds of sites, to become more competitive with new schemes~\cite{braun2024real, yao2023observation,Wang2024}, further work has to be undertaken to overcome current limitations including the anharmonicity of the trapping potentials. \\

 {\it Acknowledgements:} We thank Amit Vashisht and Vincent Boyer for helpful discussions. 
This work is supported by the Royal Society via grants UF160112, RGF\textbackslash{}EA\textbackslash{}180121 and RGF\textbackslash{}R1\textbackslash{}180071, by the Engineering and Physical Sciences Research Council [grant number EP/W016141/1] and the CDT of Topological Design [grant number EP/S02297X/1]. Work in Brussels is supported by the ERC Grant LATIS, the EOS project CHEQS, the Fonds de la Recherche Scientifique (F.R.S.-FNRS) and the Fondation ULB. G.S. has received funding from the Academy of Finland under Project No. 13354165. Some of the computations described in this paper were performed using the University of Birmingham's Bluebear HPC service, which provides a High Performance Computing service to the University's research community. See http://www.birmingham.ac.uk/bear for more details.

\appendix

\section{Choice of Driving Potential} \label{sec:appdriving}

As mentioned in the main text, many different forms of the driving potential can be chosen to generate couplings along the synthetic dimension of harmonic trap states. In particular, any suitable potential with an odd spatial dependence in $x$ can be utilised~\cite{Price2017}; the recent experiment in Ref.~\cite{oliver2023bloch}, for example, employed the driving potential 
\begin{equation}
 V_{\text{alt}}(x, t) = -V_0 \Theta(x\sin(\omega_D t + \phi)) .
 \label{Eq:blochdrive}
\end{equation} 
One natural option to study topological chiral edge states would therefore be to extend this type of shaking potential by setting $\phi = \phi_0 y$. However, experimentally, it is also important that a substantial amount of the atomic density is coupled into the edge state to ensure a clear signature can be observed. Previously, in Ref.~\cite{oliver2023bloch}, it was found that only $\approx 50 \%$ of the atomic density was coupled into the synthetic dimension by this type of shaking potential. Similarly, when simulating the full time-evolution of a Gaussian wave-packet under the full Hamiltonian of Eq.~\ref{eq:full} with the shaking potential $V$ replaced by $ \mathcal{V}_{\text{alt}}$ [Eq.~\ref{Eq:blochdrive}], we have observed that a sizable fraction of the wave-packet remains around $\lambda=0$ throughout the dynamics. In contrast, for the driving protocol used in the main text [Eq.~\ref{Eq.couplingpotential}], most of the atomic density does move up the synthetic dimension, and hence chirally around the system, allowing for a large experimental signature. 

\begin{figure}
    \centering
      \includegraphics[width=1\columnwidth]{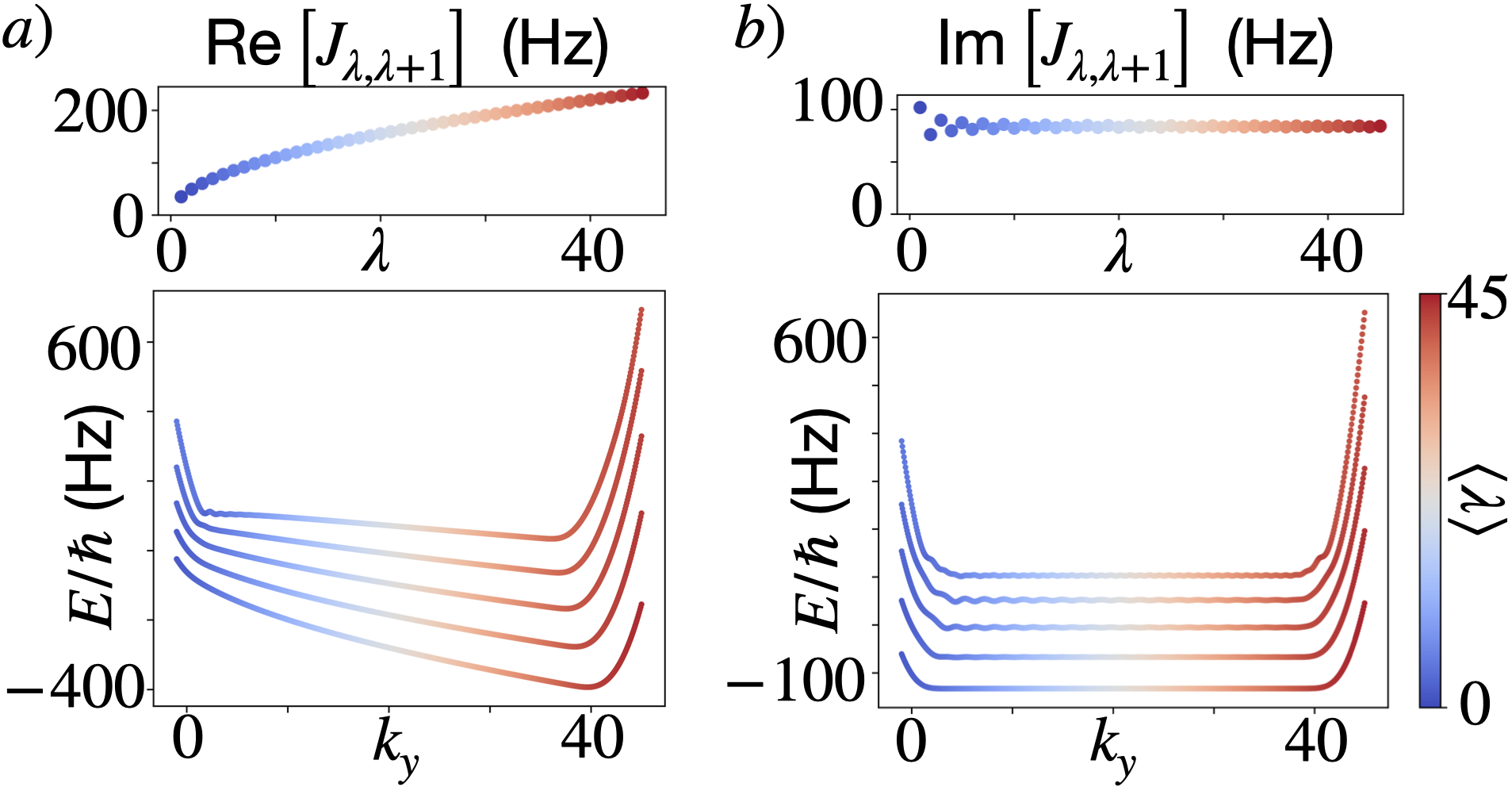}%
    \caption{  ({\it Upper panels}) Numerically-calculated nearest-neighbor matrix elements of the
effective 1D Floquet Hamiltonian using (a) the driving potential from the main text [Eq.~\ref{Eq.couplingpotential}] and (b) from the experiment in Ref.~\cite{oliver2023bloch}. Parameters are as given for the Gaussian wave-packet in Section~\ref{sec:scheme}. Note that for (a) these hopping amplitudes are purely real and vary as $\sqrt{\lambda}$ [c.f.~Eq.~\ref{eq:J}], while for (b) they are purely imaginary and approximately constant, except for at small $\lambda$. ({\it Lower panels}) The 
2D effective band-structures for the corresponding coupled resulting coupled wire models [Eq.~\ref{eq:cw}] for $N_\lambda=45$ sites along the synthetic dimension. Each state is colored according to its expectation value $\langle \lambda \rangle$ with respect to the synthetic dimension. Both cases show localized chiral edge states. For (a) the bulk states are also tilted with respect to $k_y$, similar to Landau levels under an electric field. For (b), the bulk levels are flat but oscillations appear in the band-structure near the edge states around $k_y \!=\!0$ where states are associated with $\langle \lambda \rangle \approx 0$.   }
    \label{fig:suppband}
  \end{figure}

To explore why these driving potentials lead to such different results, we start by numerically calculating the 1D effective Floquet Hamiltonians that describe the stroboscopic motion along the $x$ direction (i.e. with $y=0$) [c.f.~Section~\ref{section:1a)}]. The resulting nearest-neighbor hopping terms are plotted in the upper panels of Fig.~\ref{fig:suppband} when using (a) the driving protocol from the main text [Eq.~\ref{Eq.couplingpotential}], or (b) the driving potential from the recent experiment~\cite{oliver2023bloch} [Eq.~\ref{Eq:blochdrive}].Experimentally-realistic parameters are chosen as stated in Section~\ref{sec:scheme} but in the case of  (b) parameters $V_0= -48 \text{Hz}$ and $\omega = 2 \pi \times 167 \text{Hz}$ are chosen to be closer to those used in the previous experiment. For the main-text shaking potential, the hopping amplitudes can be approximated with $J_\lambda \propto \sqrt{\lambda}$ [Eq.~\ref{eq:J}], as can also be seen in Fig.~\ref{fig:suppband}(a). For the previously-realized experimental potential, on the other hand, the hopping amplitudes can be approximated as $J_\lambda \propto i J$, where $J$ is a constant~\cite{oliver2023bloch}, as can be seen for larger values of $\lambda$ in Fig.~\ref{fig:suppband}(a). However, the latter approximate form breaks down towards $\lambda=0$, where oscillations in the hopping amplitudes are observed~\cite{oliver2023bloch}.

To visualise the effect of these oscillations, we can calculate the energies of the corresponding effective 2D coupled-wire models
\begin{equation}
\mathcal{H}_{\text{cw}} = \frac{ k_y^2}{2m} + \sum_{\lambda} J_{\lambda}e^{i \phi_0 y} | \lambda -1, y\rangle \langle \lambda, y |  + h.c. 
\label{eq:cw}
\end{equation}
where $k_y$ is the momentum along the $y$ direction and $J_{\lambda}$ is given by the above numerically-calculated hoppings. The resulting band-structures are shown in the lower panels of Fig.~\ref{fig:suppband} (a) and (b) respectively, for a finite number of sites ($N_\lambda =45$) along the synthetic dimension. Here, the colorbar indicates the expectation value $\langle \lambda \rangle$ of the position along the synthetic dimension for each state. As can be seen, chiral edge physics is clearly visible in both cases as the states at $k_y\! \!< \!0$ are localised on the lower edge of the system (around $\!\lambda\!=\!0$) with a negative group velocity along the $y$ direction, while those with $k_y \!>\!40$ are localised on the upper edge of the system (around $\lambda\!=\!45$) with a positive group velocity~\cite{Kane2002,Budich_2017}. The $\sqrt{\lambda}$-dependence in the hopping amplitudes of panel (a) results in a tilting of the effective bulk bands, similar to the effect of applying an electric field along the synthetic dimension. In panel (b), conversely, the bulk bands are flat, similar to usual Landau levels. However, on closer inspection, there are small local minima in the bandstructure around $k_y\! =\!0$, arising from the oscillations in the hopping amplitudes at low $\lambda$.  

    \begin{figure}
    \centering
      \includegraphics[width=0.5\columnwidth]{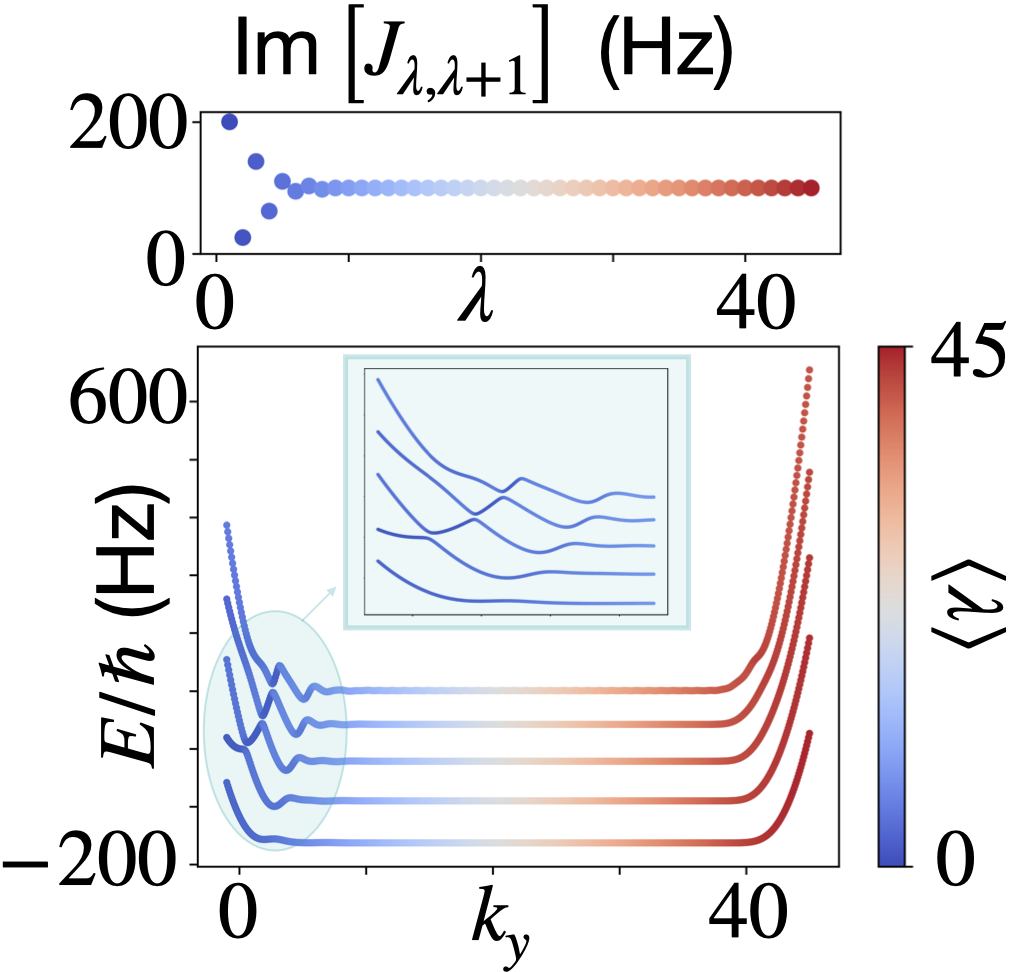}%
    \caption{  ({\it Upper panel}) Nearest-neighbor matrix elements chosen for a toy Hamiltonian. The hopping amplitudes are chosen as uniform for $\lambda \!>\! 7$, but with large oscillations for small $\lambda$, motivated by Fig.~\ref{fig:suppband}(b). Note, we calculate these hopping amplitudes for more than $\lambda = 45$ sites to remove small finite size effects. 
({\it Lower panel}) The corresponding coupled-wire model band-structure [Eq.~\ref{eq:cw}], with parameters and colorbar as in Fig.~\ref{fig:suppband}. For this toy Hamiltonian, we again
see flat Landau-like bands in the bulk but now with larger oscillations around $k_y\!=\!0$, corresponding to states localised at the lower edge in $\lambda$. As shown in the inset, zooming into this region, there are local minima and states which are no longer chiral. We can therefore expect that preparing an initial wave-packet around $\lambda\!=\!0$ will populate both chiral and non-chiral states, decreasing the desired experimental signal. } 
    \label{fig:supptoy}
  \end{figure}

To see this even more clearly, we construct by hand a toy Hamiltonian with hopping amplitudes that have even larger oscillations near $\lambda=0$ [see upper panel in Fig.~\ref{fig:supptoy}]. We have checked in our numerical simulations of the dynamics that this leads to even fewer atoms moving chirally around the system. We have also recalculated the corresponding 2D band-structure, as shown in the lower panel of Fig.~\ref{fig:supptoy}. As expected, the oscillations in the hopping amplitudes lead to even larger distortions in the band-structure around $k_y \!= \!0$ [see also inset in Fig.~\ref{fig:supptoy}] as compared to Fig.~\ref{fig:suppband}(b). In particular, we observe the appearance of local minima and states with positive group velocities along the $y$ direction. These minima occur near where the wave-packet is initially prepared and so may explain why a large percentage of the atomic density does not exhibit the desired chiral behavior. Populating states around $\lambda\!=\!0$ will, in general, lead to the occupation of both chiral and non-chiral states, decreasing the desired experimental signal. This suggests that potentials such as the experimental driving potential utilized in Ref.~\cite{oliver2023bloch} are not as good candidates for investigating chiral edge state motion as the potential used in the main text, for which there are no oscillations in the effective hopping amplitudes at small $\lambda$.   

\section{Flexibility of driving potential}

Throughout this paper, we have focused on an ideal yet experimentally implementable potential, which we believe can be realized by utilizing a digital micro-mirror device. However, we are not limited to just this potential as long as the nearest neighbor hopping amplitudes of the resulting Floquet Hamiltonian avoid the issues discussed in section \ref{sec:appdriving}. Here we introduce a family of potentials that could be trialed to produce similar results discussed in the main paper. Our aim is to provide a less rigid route for the implementation of these results. Note that other potentials not captured by this description could also be useful.

\begin{figure}
\includegraphics[width=\columnwidth]{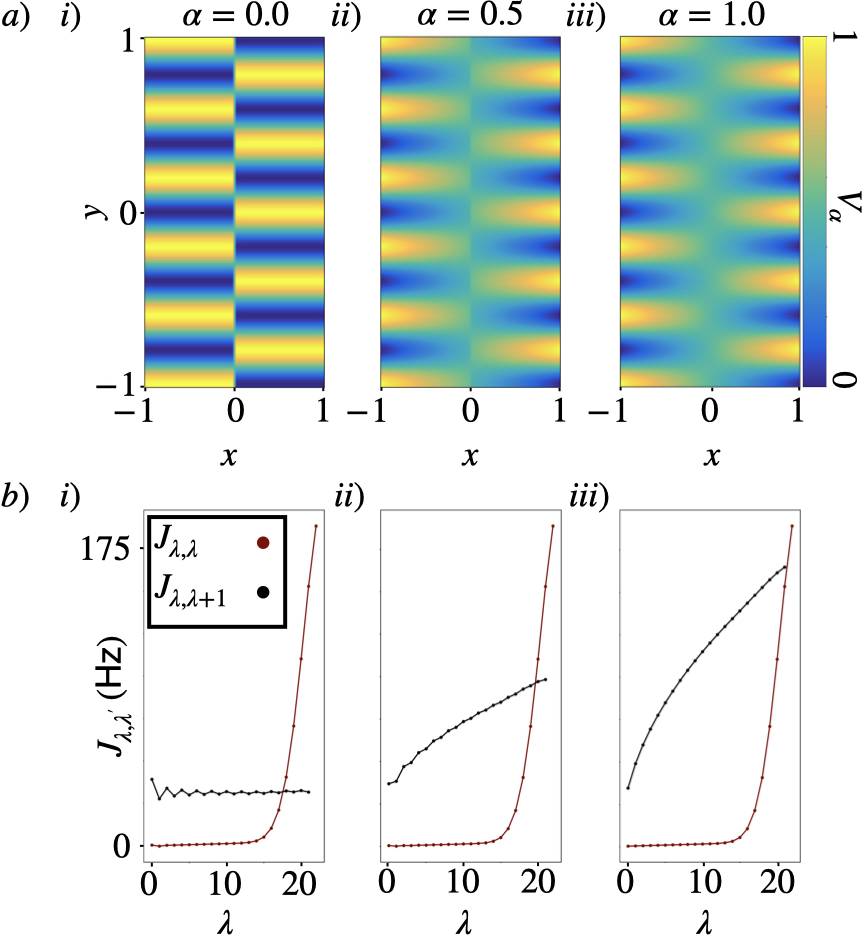}
\caption{ \label{Fig:DMDprojections} Interpolation between an alternative and ideal potential used in the main paper utilizing parameter $\alpha$ [Eq.~\ref{Eq:Valpha}]. a) Contour plots of the driving potentials for different values $\alpha$ illustrating how the potential transitions from discrete \textit{``fingers"} ($\alpha=0$) to linear\textit{``fingers"} ($\alpha=1$). i), ii) and iii) are the cases for $\alpha=0.0, 0.5$ and $1.0$ respectively. b) i) ii) and iii) Are the resulting Floquet Hamiltonian matrix hopping elements resulting from the choices of $\alpha$ in a)i), ii) and iii) respectively with a square well edge implemented in all cases [Eq.~\ref{Eq:floqalpha}]. Parameters are as given for the Gaussian wave-packet in Section~\ref{sec:scheme} with $V_0 = -98\text{Hz}$, $V_d = 590\text{Hz}$ and $N_{\lambda}=23$. We see that in i) the real nearest neighbor hopping amplitudes, \textit{``black dots"}, are reminiscent of the imaginary amplitudes from experimental potential ref. \cite{oliver2023bloch}. iii) Are the hopping amplitudes of the idealized potential - as previously discussed in Appendix \ref{sec:appdriving}. ii) We observe, for $\alpha=0.5$, that the nearest neighbor hopping amplitudes qualitatively appear halfway between the constant amplitudes with oscillations for low $\lambda$ as in i) and the square root behavior in iii). This highlights that the interpolation works as desired. In all cases, the onsite potential \textit{``maroon dots"} induced by the square well edge remain unaltered meaning a tunable edge can be implemented for all choices of $\alpha$.}
\end{figure}

Specifically, we interpolate, via a parameter $\alpha \in [0, 1]$, between our ideal potential and another driving potential with some of the drawbacks of the experimental potential discussed in the previous section,
\begin{eqnarray}
&&{}V_{\alpha}(x, y, t) = \left[V_0\left( 1 - \alpha \right)(2\Theta(x) - 1) +\alpha \kappa x\right] \times\nonumber\\& &\cos(\omega_d t + \phi_0 y) 
\label{Eq:Valpha}
\end{eqnarray}
where all parameters are as stated previously. For $\alpha = 0$ we obtain a driving potential with discrete interlocking \textit{``fingers"} [Fig.~\ref{Fig:DMDprojections}ai)]. This potential is likely to be easier to implement as it is a less linear driving potential, meaning less time averaging/half toning is required from the DMD - this relative simplicity is why we opted to include it here. For $\alpha = 1$ we obtain the familiar idealized driving potential with linear \textit{``fingers"}  [Fig.~\ref{Fig:DMDprojections}aii)]. As $\alpha$ increases the \textit{``fingers''} progressively shift from discrete to linear, with $\alpha=0.5$ being a mixture of both cases [Fig.~\ref{Fig:DMDprojections}aii)] . 

We now calculate the 1D Floquet Hamiltonian for the three cases above to verify that the interpolation has the desired effect on the effective hopping amplitudes. We use the same method as in section \ref{sec:tunableedge} with the new potential,
\begin{equation}
\mathcal{H}_{\text{Floq}}^{\alpha} = \frac{\hat{p}_x^2}{2m} + \frac{m}{2}\omega^2x^2 + V_{\alpha}(x, y=0, t) + V_e(x)
\label{Eq:floqalpha}
\end{equation}
where the parameters are unaltered from those used in section \ref{sec:scheme} and $V_0=-98\text{Hz}$. We set up the square well potential with a depth of $V_d= 590\text{Hz}$ and $x_{\text{width}} = 12 x_{ho}$. The resulting matrix hopping amplitudes are calculated for $\alpha=0.0, 0.5, 1.0$ [Fig \ref{Fig:DMDprojections}b)i), ii) and iii) respectively] such that they correspond to the driving potentials in figure \ref{Fig:DMDprojections}a). In all cases the onsite hopping amplitudes (\textit{``maroon dots"}) induced by the square well trap remain unaltered with only the nearest neighbor hopping amplitudes (\textit{``black dots"}) being affected by the different driving protocols. For $\alpha =0.0$ we observe nearest neighbor hopping amplitudes that oscillate for low $\lambda$ but stabilize to a constant value for sufficiently large values of $\lambda$. Note that these amplitudes are essentially a real version to the imaginary amplitudes seen in the experimental potential used in Ref. \cite{oliver2023bloch} and discussed in section \ref{sec:appdriving} - thus the aforementioned drawbacks effect this potential too. In case of $\alpha=1.0$ is the idealized potential and as such nearest neighbor hopping amplitude are characterized by square root behavior as in section \ref{sec:appdriving}. Importantly, the hopping amplitudes produced from $\alpha=0.5$ seem to be `half-way' between the case i) and iii) with some oscillations for low $\lambda$ and an a slower square root like increase with respect to $\lambda$. This confirms that the resulting interpolation works in that the hopping amplitudes linearly shift between i) and iii) as expected.

\begin{figure}
\includegraphics[width=\columnwidth]{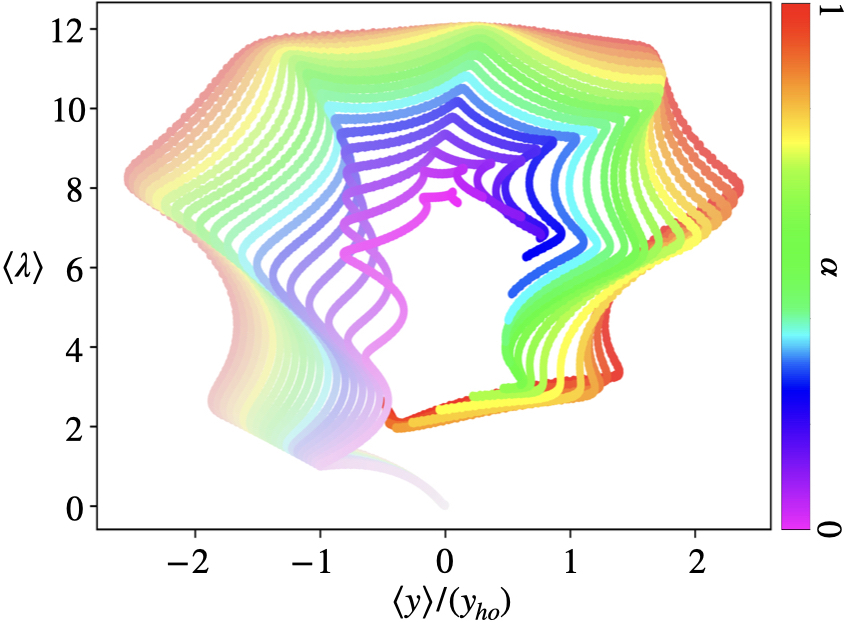}
\caption{\label{Fig:alphatrajectories} Twenty Gaussian wave-packet trajectories resulting from a sweep of equally spaced values of $\alpha \in \left[ 0, 1\right] $ which are identified by an individual color from a rainbow spectrum. The parameters are described in section \ref{sec:scheme} with the square well edge implemented at $\lambda_e = 18$. The color of each trajectory increases in saturation with time from $t=0$ to $t=600$ms. We observe that the trajectories incrementally perform worst when compared to the expected edge-state dynamics, as $\alpha$ is reduced from 1 due to the lack of coupling into the edge state discussed in \ref{sec:appdriving}. Despite this for a range of $\alpha$ the edge state trajectory can still be observed although with reduced measurable observables.}
\end{figure}

To investigate the range of values of $\alpha$ where a reasonable edge-state trajectory can be observed we perform twenty dynamical simulations utilizing a Gaussian wave-packet with parameters described before in section \ref{sec:scheme} for equispaced values of $\alpha$ between $0$ and $1$. For all trajectories, a square well potential of width $x_{\text{width}}= 12 x_{ho}$ and height $V_d = 1.3$MHz is utilized and we drive the system with $V_0 = -98 \text{Hz}$ ($\kappa$ has the same value as before). We designate each of the trajectories a color from a rainbow spectrum with $\alpha =1$ being red and $\alpha=0$ being magenta [Fig: \ref{Fig:alphatrajectories}]. Each trajectory runs for a total time of $t=600$ms with the time being indicated by an increasing saturation of color along a given trajectory. We observe that as $\alpha$ is reduced from $1$ (red) the observables, including the shift along $y$, maximum $\lambda$ occupied and the proportion of the trajectory completed in the given time, are reduced. This reduction in observables comes from the wave-packet not coupling as well to the edge-state due to the oscillations in the nearest neighbor hopping amplitudes as discussed in section \ref{sec:appdriving}. However, despite these oscillations, we still observe clear edge-state trajectories for a range of $\alpha \in[0.4, 1.0]$, which then breaks down roughly at the purple colors $\alpha<0.4$. This demonstrates that much of the physics discussed in the main paper is accessible for a range of potentials described by Eq. \ref{Eq:Valpha} indicating some flexibility in the choice of drive if the potential is less linearized.

Finally, we emphasize again that the ideal drive discussed in the main paper does provide the cleanest results and is the implementation we favor. However, in addition to its practical use to the specific experimental protocol, the framework provided in this section could be utilized to determine if another driving potential an experimental group has implemented could be transferred to study these physical phenomena. 

\begin{figure*}
    \centering
      \includegraphics[width=1\textwidth]{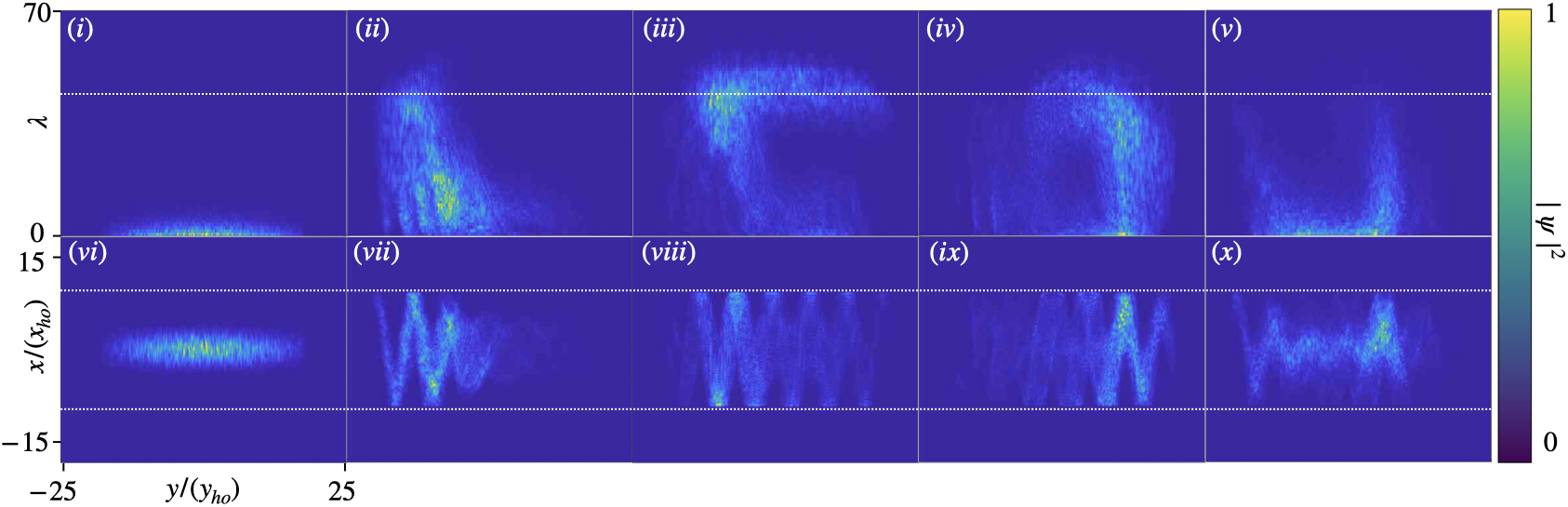}%
    \caption{ \label{Fig:realvsexpobserable} 
    (i)-(v) 
Snapshots for the numerical evolution of the atomic density, $|\psi|^2$, for a thermal cloud in the synthetic $\lambda\!-\!y$ plane with parameters as discussed in the  text and as stated in section~\ref{sec:scheme}. In this case the snapshots are taken such that the micromotion is included. The snapshots correspond to times $t\!=\! 0, 110, 200, 321, 440 \text{ms}$ respectively, showing the chiral motion of the atomic density around the edge of the synthetic plane. (vi)-(x) Corresponding snapshots of the atomic density in real $x \!-\! y$ space for the same times as  the upper panels. As can be seen, the thermal cloud moves to the left along $y$ while expanding along $x$, as it climbs the synthetic dimension. It then moves back along $y$ while maintaining the width along $x$, before shrinking again as it descends the synthetic dimension.  }
  \end{figure*}

\section{Evolution of the Density for Thermal Simulations}
\label{sec:appreal}

In Fig.~\ref{Fig:gauss_synth_real} in the main text, we presented snapshots for the evolution of the atomic density, $|\psi|^2$, for an initial Gaussian wave-packet. Here we present analogous results in Fig.~\ref{Fig:realvsexpobserable} for the thermal simulations. For simplicity we select the same parameters as for the thermal simulations in Section~\ref{section:1a)}, with $\lambda_e=45$ [c.f.~Eq.~\ref{eq:le}]. The snapshots show successive time, firstly, in (i)-(v) synthetic $\lambda\!-\!y$ space and, secondly, in (vi)-(x) real $x\!-\!y$ space. 

In synthetic space, we can clearly observe the clockwise trajectory of the atomic cloud, as expected for the chiral edge state behavior in a 2D quantum Hall system. Note that in contrast to the simulations for the Gaussian wave-packet, the initial cloud is very wide along $y$, with the consequence that the atomic density becomes smeared out along the edge state orbit, with some atoms remaining near $\lambda\!=\!0$ throughout the dynamics. As also briefly discussed in the main text, the initial width of the thermal cloud is controlled by the (experimentally-motivated) choices of temperature and weak trap frequency, and so cannot be independently optimized to decrease these unwanted effects. Nevertheless, the signatures of the desired chiral motion are still readily apparent in the full dynamics. 

The above chiral features are also reflected in the real-space motion in the $x\!-\!y$ plane as shown in the lower panels of Fig.~\ref{Fig:realvsexpobserable}. Here, we can see that the thermal cloud moves first to the left along $y$, while expanding along $x$ in a characteristic zig-zag pattern. The width of this pattern reaches a maximum when the atoms feel the additional square-well potential, corresponding to approaching the upper-edge along the synthetic dimension. The cloud then moves back along the $y$ direction, before shrinking again as the atoms return to near $\lambda\!=\!0$, and then move back towards the center of the plane.  

\section{Extracting the Upper Edge for the Thermal Simulations}
\label{sec:appupper}

    \begin{figure}
    \centering
      \includegraphics[width=1\columnwidth]{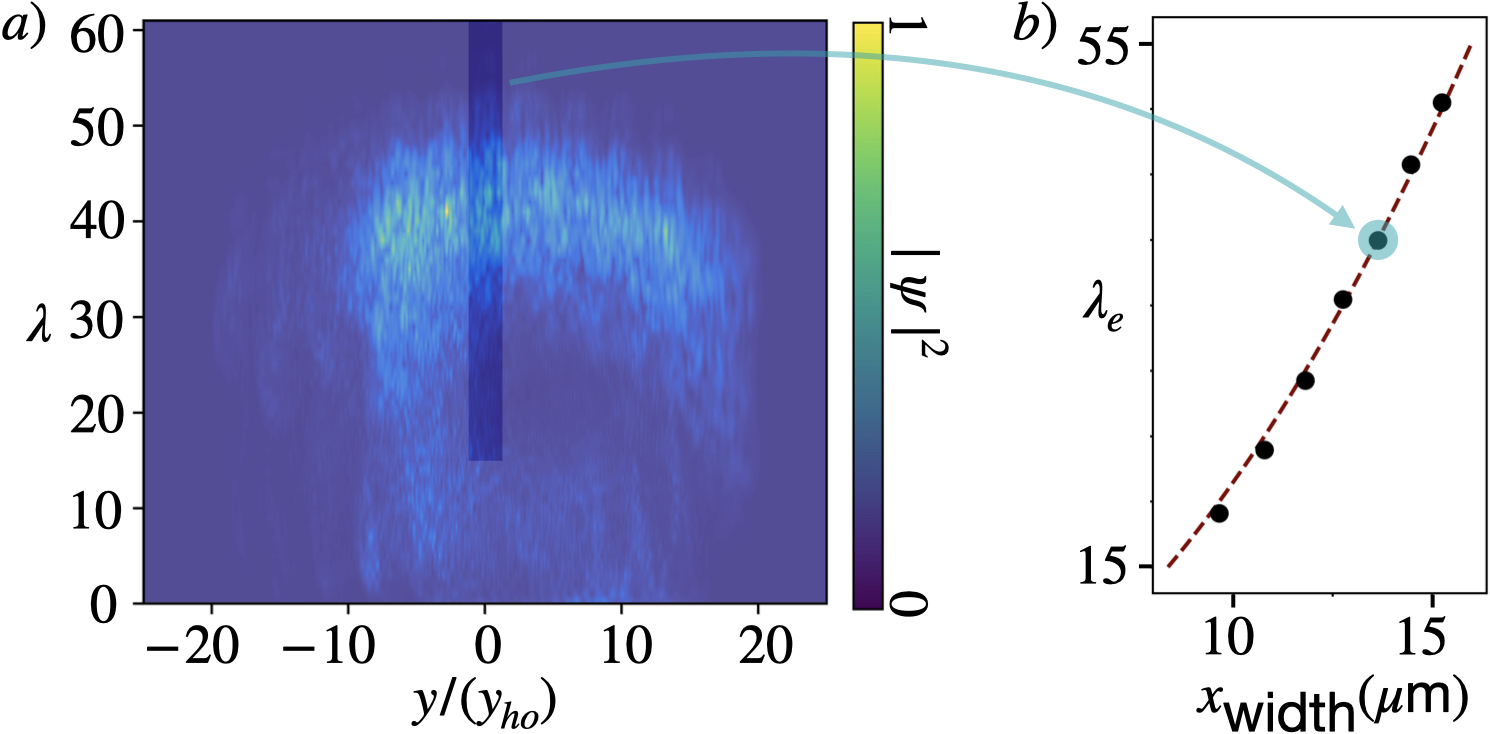}\caption{\label{fig:datanalysis} (a) Snapshot of the numerical evolution of the atomic density for a thermal cloud at $t=388 \text{ms}$ for the example of $x_{\text{width}}=\sqrt{320}x_{\text{ho}}$ and all other parameters as given in the main text. As can be seen, the atomic density is spread around the boundary of the synthetic plane. To estimate the location of the upper edge, we therefore select a sub-section of the plane, as highlighted by the non-faded region, and calculate the corresponding local $\langle \lambda \rangle$. This is then an estimate for $\lambda_e$ which is re-plotted in panel (b), taken from Fig.~\ref{fig:thermal} in the main text. }
  \end{figure}

In Fig.~\ref{fig:thermal} in the main text, we have estimated the location of the upper edge along the synthetic dimension from our numerical simulations of a thermal cloud. However, as shown in Appendix~\ref{sec:appreal}, the atoms are very spread out along the chiral orbit in synthetic space due to the large initial width of the cloud along the $y$ direction. To deal with this, we have modified our data analysis as we now discuss in more detail.

We shall illustrate our analysis procedure for the specific example of $x_{\text{width}}= \sqrt{320} x_{\text{ho}}$, which is one of those studied in the main text. We begin by calculating the center of the mass $\langle \lambda \rangle$ of the cloud throughout the dynamical simulation and then choose a snapshot of the density in synthetic space where the maximum value of $\langle \lambda \rangle$ has been reached. Such a snapshot is shown, for example, in Fig.~\ref{fig:datanalysis}(a) corresponding to $t= 388 \text{ms}$. As can be seen, the wave-packet is almost entirely delocalised around the the edge of the synthetic system; as a consequence the value of $\langle \lambda \rangle$ that has been calculated from the full cloud will underestimate the position of the upper edge significantly. Note that this is unlike in the Gaussian wave-packet simulations where the wave-packet remains well-localised during the trajectory [c.f.~Fig.~\ref{Fig:gauss_synth_real}]. To circumvent this issue, we therefore re-calculate the center of mass $\langle \lambda \rangle$ for a small sub-section of the synthetic plane, which is highlighted by the non-faded region in  Fig.~\ref{fig:datanalysis}(a). For all the simulations analyzed, this sub-section is fixed as $\lambda \in \left[ 15, 60\right], y \in \left[ -1.23, 1.28 \right]y_{\text{ho}}$. In this way, we only consider atoms which are traveling along the upper edge of the synthetic dimension. This new calculated value of $\langle \lambda \rangle$ is then our estimate of $\lambda_e$, and is the value plotted in Fig.~\ref{fig:datanalysis}(b) and Fig.~\ref{fig:thermal}. As can be seen, this analysis provides good agreement with the analytical prediction as also discussed in the main paper. 

     \section{Additional Real-Space Potentials}
    \label{sec:appedge}
    
    In this Appendix, we detail the forms of the real-space potentials which have been used in the main text to tune the softness of the upper edge in Section~\ref{section:1b} and to imprint an impurity in Section~\ref{section:3}. 
    
    Firstly, to tune the softness of the edge, we superpose two top hat functions
    \begin{align}
    \mathcal{S}_1  {}&= \Theta\left(x - x^{-}\right) - \Theta\left(x- x^{+}\right) \notag \\&+\Theta\left(x + x^{+}\right) -\Theta\left(x- x^{-}\right),
    \label{eq:s1twosquarewells}
    \end{align} 
    where $x^{+}>x^{-}$. We also have the square well potential [Eq.~\ref{Eq.potentialedge}] with $x_{e} = x^{+}$ 
    \begin{equation}
  V_e = V_d[ 1 +   \Theta\left(x- x^{+}\right)  -\Theta\left(x+ x^{+}\right)].
    \end{equation} 
    The total additional real-space potential is then  
    \begin{equation}
    V(x)= V_{d}\mathcal{S}_1\frac{\left(\mid x\mid^{\gamma} -\left(x^{-}\right)^{\gamma} \right)}{\left( x^{+} \right)^{
    \gamma}- \left(x^{-}\right)^{\gamma}} + V_e.
    \label{Eq.softedge}
    \end{equation}
    This potential is designed so that the first term vanishes at $|x^{-}|$ and then rises with a power law, of power $\gamma$, to be equal to $V_d$ at $|x^{+}|$ as plotted in Fig.~\ref{fig:soften}(a). 
    
    Secondly, to add an impurity, we design an additional top hat function as
    \begin{align}
    \mathcal{S}_{y}\left(y\right) ={}&  \Theta \left(y^{+} -y\right) - \Theta \left(y^{-}-y\right).
    \end{align} 
    We then combine this with Eq.~\ref{eq:s1twosquarewells} to give
    \begin{equation}
    V_{\text{impurity}}(x, y) = V_d \mathcal{S}_y \otimes \mathcal{S}_1
    \label{eq:impuirtyreal}
    \end{equation} 
    where $V_d$ determines the impurity height. This process essentially, creates two boxes in real space where $(x^{-},y^{-})$, $( x^{+},y^{-})$, $( x^{-},y^{+})$, $(x^{+},y^{+})$ enumerates the coordinates of the corners of one of these boxes, with the second box being a reflection of the first over the y axis. Finally, to create the upper edge in the synthetic dimension, we again add the square-well potential with width $2x^{+}$ and depth $V_d$. The resulting total additional real-space potential can be seen in Fig.~\ref{fig:constructionimpurity}(a).

\bibliography{bibliography}

\begin{thebibliography}{70}%
\makeatletter
\providecommand \@ifxundefined [1]{%
 \@ifx{#1\undefined}
}%
\providecommand \@ifnum [1]{%
 \ifnum #1\expandafter \@firstoftwo
 \else \expandafter \@secondoftwo
 \fi
}%
\providecommand \@ifx [1]{%
 \ifx #1\expandafter \@firstoftwo
 \else \expandafter \@secondoftwo
 \fi
}%
\providecommand \natexlab [1]{#1}%
\providecommand \enquote  [1]{``#1''}%
\providecommand \bibnamefont  [1]{#1}%
\providecommand \bibfnamefont [1]{#1}%
\providecommand \citenamefont [1]{#1}%
\providecommand \href@noop [0]{\@secondoftwo}%
\providecommand \href [0]{\begingroup \@sanitize@url \@href}%
\providecommand \@href[1]{\@@startlink{#1}\@@href}%
\providecommand \@@href[1]{\endgroup#1\@@endlink}%
\providecommand \@sanitize@url [0]{\catcode `\\12\catcode `\$12\catcode `\&12\catcode `\#12\catcode `\^12\catcode `\_12\catcode `\%12\relax}%
\providecommand \@@startlink[1]{}%
\providecommand \@@endlink[0]{}%
\providecommand \url  [0]{\begingroup\@sanitize@url \@url }%
\providecommand \@url [1]{\endgroup\@href {#1}{\urlprefix }}%
\providecommand \urlprefix  [0]{URL }%
\providecommand \Eprint [0]{\href }%
\providecommand \doibase [0]{http://dx.doi.org/}%
\providecommand \selectlanguage [0]{\@gobble}%
\providecommand \bibinfo  [0]{\@secondoftwo}%
\providecommand \bibfield  [0]{\@secondoftwo}%
\providecommand \translation [1]{[#1]}%
\providecommand \BibitemOpen [0]{}%
\providecommand \bibitemStop [0]{}%
\providecommand \bibitemNoStop [0]{.\EOS\space}%
\providecommand \EOS [0]{\spacefactor3000\relax}%
\providecommand \BibitemShut  [1]{\csname bibitem#1\endcsname}%
\let\auto@bib@innerbib\@empty
\bibitem [{\citenamefont {Hasan}\ and\ \citenamefont {Kane}(2010)}]{RMP_TI}%
  \BibitemOpen
  \bibfield  {author} {\bibinfo {author} {\bibfnamefont {M.~Z.}\ \bibnamefont {Hasan}}\ and\ \bibinfo {author} {\bibfnamefont {C.~L.}\ \bibnamefont {Kane}},\ }\href@noop {} {\bibfield  {journal} {\bibinfo  {journal} {Rev. Mod. Phys.}\ }\textbf {\bibinfo {volume} {82}},\ \bibinfo {pages} {3045} (\bibinfo {year} {2010})}\BibitemShut {NoStop}%
\bibitem [{\citenamefont {Qi}\ and\ \citenamefont {Zhang}(2011)}]{RMP_TI2}%
  \BibitemOpen
  \bibfield  {author} {\bibinfo {author} {\bibfnamefont {X.-L.}\ \bibnamefont {Qi}}\ and\ \bibinfo {author} {\bibfnamefont {S.-C.}\ \bibnamefont {Zhang}},\ }\href@noop {} {\bibfield  {journal} {\bibinfo  {journal} {Rev. Mod. Phys.}\ }\textbf {\bibinfo {volume} {83}},\ \bibinfo {pages} {1057} (\bibinfo {year} {2011})}\BibitemShut {NoStop}%
\bibitem [{\citenamefont {Klitzing}\ \emph {et~al.}(1980)\citenamefont {Klitzing}, \citenamefont {Dorda},\ and\ \citenamefont {Pepper}}]{klitzing1980new}%
  \BibitemOpen
  \bibfield  {author} {\bibinfo {author} {\bibfnamefont {K.~v.}\ \bibnamefont {Klitzing}}, \bibinfo {author} {\bibfnamefont {G.}~\bibnamefont {Dorda}}, \ and\ \bibinfo {author} {\bibfnamefont {M.}~\bibnamefont {Pepper}},\ }\href@noop {} {\bibfield  {journal} {\bibinfo  {journal} {Physical review letters}\ }\textbf {\bibinfo {volume} {45}},\ \bibinfo {pages} {494} (\bibinfo {year} {1980})}\BibitemShut {NoStop}%
\bibitem [{\citenamefont {von Klitzing}(1986)}]{VonKlitzing1986}%
  \BibitemOpen
  \bibfield  {author} {\bibinfo {author} {\bibfnamefont {K.}~\bibnamefont {von Klitzing}},\ }\href@noop {} {\bibfield  {journal} {\bibinfo  {journal} {Rev. Mod. Phys.}\ }\textbf {\bibinfo {volume} {58}},\ \bibinfo {pages} {519} (\bibinfo {year} {1986})}\BibitemShut {NoStop}%
\bibitem [{\citenamefont {Chiu}\ \emph {et~al.}(2016)\citenamefont {Chiu}, \citenamefont {Teo}, \citenamefont {Schnyder},\ and\ \citenamefont {Ryu}}]{Chiu:2016RMP}%
  \BibitemOpen
  \bibfield  {author} {\bibinfo {author} {\bibfnamefont {C.-K.}\ \bibnamefont {Chiu}}, \bibinfo {author} {\bibfnamefont {J.~C.~Y.}\ \bibnamefont {Teo}}, \bibinfo {author} {\bibfnamefont {A.~P.}\ \bibnamefont {Schnyder}}, \ and\ \bibinfo {author} {\bibfnamefont {S.}~\bibnamefont {Ryu}},\ }\href@noop {} {\bibfield  {journal} {\bibinfo  {journal} {Rev. Mod. Phys.}\ }\textbf {\bibinfo {volume} {88}},\ \bibinfo {pages} {035005} (\bibinfo {year} {2016})}\BibitemShut {NoStop}%
\bibitem [{\citenamefont {Goldman}\ \emph {et~al.}(2014)\citenamefont {Goldman}, \citenamefont {Juzeli{\=u}nas}, \citenamefont {{\"O}hberg},\ and\ \citenamefont {Spielman}}]{Goldman:2014ROPP}%
  \BibitemOpen
  \bibfield  {author} {\bibinfo {author} {\bibfnamefont {N.}~\bibnamefont {Goldman}}, \bibinfo {author} {\bibfnamefont {G.}~\bibnamefont {Juzeli{\=u}nas}}, \bibinfo {author} {\bibfnamefont {P.}~\bibnamefont {{\"O}hberg}}, \ and\ \bibinfo {author} {\bibfnamefont {I.~B.}\ \bibnamefont {Spielman}},\ }\href@noop {} {\bibfield  {journal} {\bibinfo  {journal} {Reports on Progress in Physics}\ }\textbf {\bibinfo {volume} {77}},\ \bibinfo {pages} {126401} (\bibinfo {year} {2014})}\BibitemShut {NoStop}%
\bibitem [{\citenamefont {Goldman}\ \emph {et~al.}(2016{\natexlab{a}})\citenamefont {Goldman}, \citenamefont {Budich},\ and\ \citenamefont {Zoller}}]{Goldman:2016NatPhys}%
  \BibitemOpen
  \bibfield  {author} {\bibinfo {author} {\bibfnamefont {N.}~\bibnamefont {Goldman}}, \bibinfo {author} {\bibfnamefont {J.}~\bibnamefont {Budich}}, \ and\ \bibinfo {author} {\bibfnamefont {P.}~\bibnamefont {Zoller}},\ }\href@noop {} {\bibfield  {journal} {\bibinfo  {journal} {Nat. {Phys.}}\ }\textbf {\bibinfo {volume} {12}},\ \bibinfo {pages} {639} (\bibinfo {year} {2016}{\natexlab{a}})}\BibitemShut {NoStop}%
\bibitem [{\citenamefont {Cooper}\ \emph {et~al.}(2019)\citenamefont {Cooper}, \citenamefont {Dalibard},\ and\ \citenamefont {Spielman}}]{cooper2018}%
  \BibitemOpen
  \bibfield  {author} {\bibinfo {author} {\bibfnamefont {N.~R.}\ \bibnamefont {Cooper}}, \bibinfo {author} {\bibfnamefont {J.}~\bibnamefont {Dalibard}}, \ and\ \bibinfo {author} {\bibfnamefont {I.~B.}\ \bibnamefont {Spielman}},\ }\href {\doibase 10.1103/RevModPhys.91.015005} {\bibfield  {journal} {\bibinfo  {journal} {Rev. Mod. Phys.}\ }\textbf {\bibinfo {volume} {91}},\ \bibinfo {pages} {015005} (\bibinfo {year} {2019})}\BibitemShut {NoStop}%
\bibitem [{\citenamefont {Lu}\ \emph {et~al.}(2016)\citenamefont {Lu}, \citenamefont {Joannopoulos},\ and\ \citenamefont {Solja{\v{c}}i{\'c}}}]{lu2016topological}%
  \BibitemOpen
  \bibfield  {author} {\bibinfo {author} {\bibfnamefont {L.}~\bibnamefont {Lu}}, \bibinfo {author} {\bibfnamefont {J.~D.}\ \bibnamefont {Joannopoulos}}, \ and\ \bibinfo {author} {\bibfnamefont {M.}~\bibnamefont {Solja{\v{c}}i{\'c}}},\ }\href@noop {} {\bibfield  {journal} {\bibinfo  {journal} {Nat. {Phys.}}\ }\textbf {\bibinfo {volume} {12}},\ \bibinfo {pages} {626} (\bibinfo {year} {2016})}\BibitemShut {NoStop}%
\bibitem [{\citenamefont {Khanikaev}\ and\ \citenamefont {Shvets}(2017)}]{khanikaev2017two}%
  \BibitemOpen
  \bibfield  {author} {\bibinfo {author} {\bibfnamefont {A.~B.}\ \bibnamefont {Khanikaev}}\ and\ \bibinfo {author} {\bibfnamefont {G.}~\bibnamefont {Shvets}},\ }\href@noop {} {\bibfield  {journal} {\bibinfo  {journal} {Nature Photonics}\ }\textbf {\bibinfo {volume} {11}},\ \bibinfo {pages} {763} (\bibinfo {year} {2017})}\BibitemShut {NoStop}%
\bibitem [{\citenamefont {Ozawa}\ and\ \citenamefont {Price}(2019)}]{Ozawa_2019}%
  \BibitemOpen
  \bibfield  {author} {\bibinfo {author} {\bibfnamefont {T.}~\bibnamefont {Ozawa}}\ and\ \bibinfo {author} {\bibfnamefont {H.~M.}\ \bibnamefont {Price}},\ }\href {\doibase 10.1038/s42254-019-0045-3} {\bibfield  {journal} {\bibinfo  {journal} {Nature Reviews Physics}\ }\textbf {\bibinfo {volume} {1}},\ \bibinfo {pages} {349–357} (\bibinfo {year} {2019})}\BibitemShut {NoStop}%
\bibitem [{\citenamefont {Price}\ \emph {et~al.}(2022)\citenamefont {Price}, \citenamefont {Chong}, \citenamefont {Khanikaev}, \citenamefont {Schomerus}, \citenamefont {Maczewsky}, \citenamefont {Kremer}, \citenamefont {Heinrich}, \citenamefont {Szameit}, \citenamefont {Zilberberg}, \citenamefont {Yang} \emph {et~al.}}]{price2022roadmap}%
  \BibitemOpen
  \bibfield  {author} {\bibinfo {author} {\bibfnamefont {H.}~\bibnamefont {Price}}, \bibinfo {author} {\bibfnamefont {Y.}~\bibnamefont {Chong}}, \bibinfo {author} {\bibfnamefont {A.}~\bibnamefont {Khanikaev}}, \bibinfo {author} {\bibfnamefont {H.}~\bibnamefont {Schomerus}}, \bibinfo {author} {\bibfnamefont {L.~J.}\ \bibnamefont {Maczewsky}}, \bibinfo {author} {\bibfnamefont {M.}~\bibnamefont {Kremer}}, \bibinfo {author} {\bibfnamefont {M.}~\bibnamefont {Heinrich}}, \bibinfo {author} {\bibfnamefont {A.}~\bibnamefont {Szameit}}, \bibinfo {author} {\bibfnamefont {O.}~\bibnamefont {Zilberberg}}, \bibinfo {author} {\bibfnamefont {Y.}~\bibnamefont {Yang}},  \emph {et~al.},\ }\href@noop {} {\bibfield  {journal} {\bibinfo  {journal} {Journal of Physics: Photonics}\ }\textbf {\bibinfo {volume} {4}},\ \bibinfo {pages} {032501} (\bibinfo {year} {2022})}\BibitemShut {NoStop}%
\bibitem [{\citenamefont {Boada}\ \emph {et~al.}(2012)\citenamefont {Boada}, \citenamefont {Celi}, \citenamefont {Latorre},\ and\ \citenamefont {Lewenstein}}]{Boada2012}%
  \BibitemOpen
  \bibfield  {author} {\bibinfo {author} {\bibfnamefont {O.}~\bibnamefont {Boada}}, \bibinfo {author} {\bibfnamefont {A.}~\bibnamefont {Celi}}, \bibinfo {author} {\bibfnamefont {J.~I.}\ \bibnamefont {Latorre}}, \ and\ \bibinfo {author} {\bibfnamefont {M.}~\bibnamefont {Lewenstein}},\ }\href@noop {} {\bibfield  {journal} {\bibinfo  {journal} {Phys. Rev. Lett.}\ }\textbf {\bibinfo {volume} {108}},\ \bibinfo {pages} {133001} (\bibinfo {year} {2012})}\BibitemShut {NoStop}%
\bibitem [{\citenamefont {Celi}\ \emph {et~al.}(2014)\citenamefont {Celi}, \citenamefont {Massignan}, \citenamefont {Ruseckas}, \citenamefont {Goldman}, \citenamefont {Spielman}, \citenamefont {Juzeli\ifmmode~\bar{u}\else \={u}\fi{}nas},\ and\ \citenamefont {Lewenstein}}]{Celi2014}%
  \BibitemOpen
  \bibfield  {author} {\bibinfo {author} {\bibfnamefont {A.}~\bibnamefont {Celi}}, \bibinfo {author} {\bibfnamefont {P.}~\bibnamefont {Massignan}}, \bibinfo {author} {\bibfnamefont {J.}~\bibnamefont {Ruseckas}}, \bibinfo {author} {\bibfnamefont {N.}~\bibnamefont {Goldman}}, \bibinfo {author} {\bibfnamefont {I.~B.}\ \bibnamefont {Spielman}}, \bibinfo {author} {\bibfnamefont {G.}~\bibnamefont {Juzeli\ifmmode~\bar{u}\else \={u}\fi{}nas}}, \ and\ \bibinfo {author} {\bibfnamefont {M.}~\bibnamefont {Lewenstein}},\ }\href {https://link.aps.org/doi/10.1103/PhysRevLett.112.043001} {\bibfield  {journal} {\bibinfo  {journal} {Phys. Rev. Lett.}\ }\textbf {\bibinfo {volume} {112}},\ \bibinfo {pages} {043001} (\bibinfo {year} {2014})}\BibitemShut {NoStop}%
\bibitem [{\citenamefont {Mancini}\ \emph {et~al.}(2015)\citenamefont {Mancini}, \citenamefont {Pagano}, \citenamefont {Cappellini}, \citenamefont {Livi}, \citenamefont {Rider}, \citenamefont {Catani}, \citenamefont {Sias}, \citenamefont {Zoller}, \citenamefont {Inguscio}, \citenamefont {Dalmonte},\ and\ \citenamefont {Fallani}}]{Mancini2015}%
  \BibitemOpen
  \bibfield  {author} {\bibinfo {author} {\bibfnamefont {M.}~\bibnamefont {Mancini}}, \bibinfo {author} {\bibfnamefont {G.}~\bibnamefont {Pagano}}, \bibinfo {author} {\bibfnamefont {G.}~\bibnamefont {Cappellini}}, \bibinfo {author} {\bibfnamefont {L.}~\bibnamefont {Livi}}, \bibinfo {author} {\bibfnamefont {M.}~\bibnamefont {Rider}}, \bibinfo {author} {\bibfnamefont {J.}~\bibnamefont {Catani}}, \bibinfo {author} {\bibfnamefont {C.}~\bibnamefont {Sias}}, \bibinfo {author} {\bibfnamefont {P.}~\bibnamefont {Zoller}}, \bibinfo {author} {\bibfnamefont {M.}~\bibnamefont {Inguscio}}, \bibinfo {author} {\bibfnamefont {M.}~\bibnamefont {Dalmonte}}, \ and\ \bibinfo {author} {\bibfnamefont {L.}~\bibnamefont {Fallani}},\ }\href {https://www.science.org/doi/abs/10.1126/science.aaa8736} {\bibfield  {journal} {\bibinfo  {journal} {Science}\ }\textbf {\bibinfo {volume} {349}},\ \bibinfo {pages} {1510} (\bibinfo {year} {2015})}\BibitemShut {NoStop}%
\bibitem [{\citenamefont {Stuhl}\ \emph {et~al.}(2015)\citenamefont {Stuhl}, \citenamefont {Lu}, \citenamefont {Aycock}, \citenamefont {Genkina},\ and\ \citenamefont {Spielman}}]{Stuhl2015}%
  \BibitemOpen
  \bibfield  {author} {\bibinfo {author} {\bibfnamefont {B.~K.}\ \bibnamefont {Stuhl}}, \bibinfo {author} {\bibfnamefont {H.-I.}\ \bibnamefont {Lu}}, \bibinfo {author} {\bibfnamefont {L.~M.}\ \bibnamefont {Aycock}}, \bibinfo {author} {\bibfnamefont {D.}~\bibnamefont {Genkina}}, \ and\ \bibinfo {author} {\bibfnamefont {I.~B.}\ \bibnamefont {Spielman}},\ }\href {https://www.science.org/doi/abs/10.1126/science.aaa8515} {\bibfield  {journal} {\bibinfo  {journal} {Science}\ }\textbf {\bibinfo {volume} {349}},\ \bibinfo {pages} {1514} (\bibinfo {year} {2015})}\BibitemShut {NoStop}%
\bibitem [{\citenamefont {Livi}\ \emph {et~al.}(2016)\citenamefont {Livi}, \citenamefont {Cappellini}, \citenamefont {Diem}, \citenamefont {Franchi}, \citenamefont {Clivati}, \citenamefont {Frittelli}, \citenamefont {Levi}, \citenamefont {Calonico}, \citenamefont {Catani}, \citenamefont {Inguscio},\ and\ \citenamefont {Fallani}}]{Livi2016}%
  \BibitemOpen
  \bibfield  {author} {\bibinfo {author} {\bibfnamefont {L.~F.}\ \bibnamefont {Livi}}, \bibinfo {author} {\bibfnamefont {G.}~\bibnamefont {Cappellini}}, \bibinfo {author} {\bibfnamefont {M.}~\bibnamefont {Diem}}, \bibinfo {author} {\bibfnamefont {L.}~\bibnamefont {Franchi}}, \bibinfo {author} {\bibfnamefont {C.}~\bibnamefont {Clivati}}, \bibinfo {author} {\bibfnamefont {M.}~\bibnamefont {Frittelli}}, \bibinfo {author} {\bibfnamefont {F.}~\bibnamefont {Levi}}, \bibinfo {author} {\bibfnamefont {D.}~\bibnamefont {Calonico}}, \bibinfo {author} {\bibfnamefont {J.}~\bibnamefont {Catani}}, \bibinfo {author} {\bibfnamefont {M.}~\bibnamefont {Inguscio}}, \ and\ \bibinfo {author} {\bibfnamefont {L.}~\bibnamefont {Fallani}},\ }\href {https://link.aps.org/doi/10.1103/PhysRevLett.117.220401} {\bibfield  {journal} {\bibinfo  {journal} {Phys. Rev. Lett.}\ }\textbf {\bibinfo {volume} {117}},\ \bibinfo {pages} {220401} (\bibinfo {year} {2016})}\BibitemShut {NoStop}%
\bibitem [{\citenamefont {Kolkowitz}\ \emph {et~al.}(2017)\citenamefont {Kolkowitz}, \citenamefont {Bromley}, \citenamefont {Bothwell}, \citenamefont {Wall}, \citenamefont {Marti}, \citenamefont {Koller}, \citenamefont {Zhang}, \citenamefont {Rey},\ and\ \citenamefont {Ye}}]{Kolkowitz2017}%
  \BibitemOpen
  \bibfield  {author} {\bibinfo {author} {\bibfnamefont {S.}~\bibnamefont {Kolkowitz}}, \bibinfo {author} {\bibfnamefont {S.}~\bibnamefont {Bromley}}, \bibinfo {author} {\bibfnamefont {T.}~\bibnamefont {Bothwell}}, \bibinfo {author} {\bibfnamefont {M.~L.}\ \bibnamefont {Wall}}, \bibinfo {author} {\bibfnamefont {G.~E.}\ \bibnamefont {Marti}}, \bibinfo {author} {\bibfnamefont {A.~P.}\ \bibnamefont {Koller}}, \bibinfo {author} {\bibfnamefont {X.}~\bibnamefont {Zhang}}, \bibinfo {author} {\bibfnamefont {A.~M.}\ \bibnamefont {Rey}}, \ and\ \bibinfo {author} {\bibfnamefont {J.}~\bibnamefont {Ye}},\ }\href {https://www.nature.com/articles/nature20811} {\bibfield  {journal} {\bibinfo  {journal} {Nature}\ }\textbf {\bibinfo {volume} {542}},\ \bibinfo {pages} {66 } (\bibinfo {year} {2017})}\BibitemShut {NoStop}%
\bibitem [{\citenamefont {Roell}\ \emph {et~al.}(2023)\citenamefont {Roell}, \citenamefont {Laskar}, \citenamefont {Huybrechts},\ and\ \citenamefont {Weitz}}]{Roell2023}%
  \BibitemOpen
  \bibfield  {author} {\bibinfo {author} {\bibfnamefont {R.~V.}\ \bibnamefont {Roell}}, \bibinfo {author} {\bibfnamefont {A.~W.}\ \bibnamefont {Laskar}}, \bibinfo {author} {\bibfnamefont {F.~R.}\ \bibnamefont {Huybrechts}}, \ and\ \bibinfo {author} {\bibfnamefont {M.}~\bibnamefont {Weitz}},\ }\href@noop {} {\bibfield  {journal} {\bibinfo  {journal} {Phys. Rev. A}\ }\textbf {\bibinfo {volume} {107}},\ \bibinfo {pages} {043302} (\bibinfo {year} {2023})}\BibitemShut {NoStop}%
\bibitem [{\citenamefont {Chalopin}\ \emph {et~al.}(2020)\citenamefont {Chalopin}, \citenamefont {Satoor}, \citenamefont {Evrard}, \citenamefont {Mahkalov}, \citenamefont {Dalibard}, \citenamefont {Lopes},\ and\ \citenamefont {Nascimbene}}]{Chalopin2020}%
  \BibitemOpen
  \bibfield  {author} {\bibinfo {author} {\bibfnamefont {T.}~\bibnamefont {Chalopin}}, \bibinfo {author} {\bibfnamefont {T.}~\bibnamefont {Satoor}}, \bibinfo {author} {\bibfnamefont {A.}~\bibnamefont {Evrard}}, \bibinfo {author} {\bibfnamefont {V.}~\bibnamefont {Mahkalov}}, \bibinfo {author} {\bibfnamefont {J.}~\bibnamefont {Dalibard}}, \bibinfo {author} {\bibfnamefont {R.}~\bibnamefont {Lopes}}, \ and\ \bibinfo {author} {\bibfnamefont {S.}~\bibnamefont {Nascimbene}},\ }\href {https://www.nature.com/articles/s41567-020-0942-5} {\bibfield  {journal} {\bibinfo  {journal} {Nat. Phys.}\ }\textbf {\bibinfo {volume} {16}},\ \bibinfo {pages} {1017–1021} (\bibinfo {year} {2020})}\BibitemShut {NoStop}%
\bibitem [{\citenamefont {Chen}\ \emph {et~al.}(2024{\natexlab{a}})\citenamefont {Chen}, \citenamefont {Huang}, \citenamefont {Velkovsky}, \citenamefont {Ozawa}, \citenamefont {Price}, \citenamefont {Covey},\ and\ \citenamefont {Gadway}}]{chen2024interaction}%
  \BibitemOpen
  \bibfield  {author} {\bibinfo {author} {\bibfnamefont {T.}~\bibnamefont {Chen}}, \bibinfo {author} {\bibfnamefont {C.}~\bibnamefont {Huang}}, \bibinfo {author} {\bibfnamefont {I.}~\bibnamefont {Velkovsky}}, \bibinfo {author} {\bibfnamefont {T.}~\bibnamefont {Ozawa}}, \bibinfo {author} {\bibfnamefont {H.}~\bibnamefont {Price}}, \bibinfo {author} {\bibfnamefont {J.~P.}\ \bibnamefont {Covey}}, \ and\ \bibinfo {author} {\bibfnamefont {B.}~\bibnamefont {Gadway}},\ }\href {\doibase 10.1038/s41567-024-02714-7} {\bibfield  {journal} {\bibinfo  {journal} {Nat. Phys.}\ }\textbf {\bibinfo {volume} {21}},\ \bibinfo {pages} {221} (\bibinfo {year} {2024}{\natexlab{a}})}\BibitemShut {NoStop}%
\bibitem [{\citenamefont {Chen}\ \emph {et~al.}(2024{\natexlab{b}})\citenamefont {Chen}, \citenamefont {Huang}, \citenamefont {Velkovsky}, \citenamefont {Hazzard}, \citenamefont {Covey},\ and\ \citenamefont {Gadway}}]{chen2024strongly}%
  \BibitemOpen
  \bibfield  {author} {\bibinfo {author} {\bibfnamefont {T.}~\bibnamefont {Chen}}, \bibinfo {author} {\bibfnamefont {C.}~\bibnamefont {Huang}}, \bibinfo {author} {\bibfnamefont {I.}~\bibnamefont {Velkovsky}}, \bibinfo {author} {\bibfnamefont {K.~R.}\ \bibnamefont {Hazzard}}, \bibinfo {author} {\bibfnamefont {J.~P.}\ \bibnamefont {Covey}}, \ and\ \bibinfo {author} {\bibfnamefont {B.}~\bibnamefont {Gadway}},\ }\href@noop {} {\bibfield  {journal} {\bibinfo  {journal} {Nat. {Commun.}}\ }\textbf {\bibinfo {volume} {15}},\ \bibinfo {pages} {2675} (\bibinfo {year} {2024}{\natexlab{b}})}\BibitemShut {NoStop}%
\bibitem [{\citenamefont {Lu}\ \emph {et~al.}(2024)\citenamefont {Lu}, \citenamefont {Wang}, \citenamefont {Kanungo}, \citenamefont {Dunning},\ and\ \citenamefont {Killian}}]{lu2024probing}%
  \BibitemOpen
  \bibfield  {author} {\bibinfo {author} {\bibfnamefont {Y.}~\bibnamefont {Lu}}, \bibinfo {author} {\bibfnamefont {C.}~\bibnamefont {Wang}}, \bibinfo {author} {\bibfnamefont {S.~K.}\ \bibnamefont {Kanungo}}, \bibinfo {author} {\bibfnamefont {F.~B.}\ \bibnamefont {Dunning}}, \ and\ \bibinfo {author} {\bibfnamefont {T.~C.}\ \bibnamefont {Killian}},\ }\href {\doibase 10.1103/PhysRevA.110.023318} {\bibfield  {journal} {\bibinfo  {journal} {Phys. Rev. A}\ }\textbf {\bibinfo {volume} {110}},\ \bibinfo {pages} {023318} (\bibinfo {year} {2024})}\BibitemShut {NoStop}%
\bibitem [{\citenamefont {Lienhard}\ \emph {et~al.}(2020)\citenamefont {Lienhard}, \citenamefont {Scholl}, \citenamefont {Weber}, \citenamefont {Barredo}, \citenamefont {de~L{\'e}s{\'e}leuc}, \citenamefont {Bai}, \citenamefont {Lang}, \citenamefont {Fleischhauer}, \citenamefont {B{\"u}chler}, \citenamefont {Lahaye},\ and\ \citenamefont {Browaeys}}]{lienhard2020realization}%
  \BibitemOpen
  \bibfield  {author} {\bibinfo {author} {\bibfnamefont {V.}~\bibnamefont {Lienhard}}, \bibinfo {author} {\bibfnamefont {P.}~\bibnamefont {Scholl}}, \bibinfo {author} {\bibfnamefont {S.}~\bibnamefont {Weber}}, \bibinfo {author} {\bibfnamefont {D.}~\bibnamefont {Barredo}}, \bibinfo {author} {\bibfnamefont {S.}~\bibnamefont {de~L{\'e}s{\'e}leuc}}, \bibinfo {author} {\bibfnamefont {R.}~\bibnamefont {Bai}}, \bibinfo {author} {\bibfnamefont {N.}~\bibnamefont {Lang}}, \bibinfo {author} {\bibfnamefont {M.}~\bibnamefont {Fleischhauer}}, \bibinfo {author} {\bibfnamefont {H.}~\bibnamefont {B{\"u}chler}}, \bibinfo {author} {\bibfnamefont {T.}~\bibnamefont {Lahaye}}, \ and\ \bibinfo {author} {\bibfnamefont {A.}~\bibnamefont {Browaeys}},\ }\href {https://journals.aps.org/prx/abstract/10.1103/PhysRevX.10.021031} {\bibfield  {journal} {\bibinfo  {journal} {Phys. Rev. X}\ }\textbf {\bibinfo {volume} {10}},\ \bibinfo {pages} {021031} (\bibinfo {year} {2020})}\BibitemShut {NoStop}%
\bibitem [{\citenamefont {Kanungo}\ \emph {et~al.}(2022)\citenamefont {Kanungo}, \citenamefont {Whalen}, \citenamefont {Lu}, \citenamefont {Yuan}, \citenamefont {Dasgupta}, \citenamefont {Dunning}, \citenamefont {Hazzard},\ and\ \citenamefont {Killian}}]{Kanungo2021}%
  \BibitemOpen
  \bibfield  {author} {\bibinfo {author} {\bibfnamefont {S.}~\bibnamefont {Kanungo}}, \bibinfo {author} {\bibfnamefont {J.}~\bibnamefont {Whalen}}, \bibinfo {author} {\bibfnamefont {Y.}~\bibnamefont {Lu}}, \bibinfo {author} {\bibfnamefont {M.}~\bibnamefont {Yuan}}, \bibinfo {author} {\bibfnamefont {S.}~\bibnamefont {Dasgupta}}, \bibinfo {author} {\bibfnamefont {F.}~\bibnamefont {Dunning}}, \bibinfo {author} {\bibfnamefont {K.}~\bibnamefont {Hazzard}}, \ and\ \bibinfo {author} {\bibfnamefont {T.}~\bibnamefont {Killian}},\ }\href@noop {} {\bibfield  {journal} {\bibinfo  {journal} {Nat. {Commun.}}\ }\textbf {\bibinfo {volume} {13}},\ \bibinfo {pages} {972} (\bibinfo {year} {2022})}\BibitemShut {NoStop}%
\bibitem [{\citenamefont {Ozawa}\ and\ \citenamefont {Carusotto}(2017)}]{Ozawa:2017PRL}%
  \BibitemOpen
  \bibfield  {author} {\bibinfo {author} {\bibfnamefont {T.}~\bibnamefont {Ozawa}}\ and\ \bibinfo {author} {\bibfnamefont {I.}~\bibnamefont {Carusotto}},\ }\href {https://link.aps.org/doi/10.1103/PhysRevLett.118.013601} {\bibfield  {journal} {\bibinfo  {journal} {Phys. Rev. Lett.}\ }\textbf {\bibinfo {volume} {118}},\ \bibinfo {pages} {013601} (\bibinfo {year} {2017})}\BibitemShut {NoStop}%
\bibitem [{\citenamefont {Price}\ \emph {et~al.}(2017)\citenamefont {Price}, \citenamefont {Ozawa},\ and\ \citenamefont {Goldman}}]{Price2017}%
  \BibitemOpen
  \bibfield  {author} {\bibinfo {author} {\bibfnamefont {H.}~\bibnamefont {Price}}, \bibinfo {author} {\bibfnamefont {T.}~\bibnamefont {Ozawa}}, \ and\ \bibinfo {author} {\bibfnamefont {N.}~\bibnamefont {Goldman}},\ }\href {https://link.aps.org/doi/10.1103/PhysRevA.95.023607} {\bibfield  {journal} {\bibinfo  {journal} {Phys. Rev. A}\ }\textbf {\bibinfo {volume} {95}},\ \bibinfo {pages} {023607} (\bibinfo {year} {2017})}\BibitemShut {NoStop}%
\bibitem [{\citenamefont {Meier}\ \emph {et~al.}(2016)\citenamefont {Meier}, \citenamefont {An},\ and\ \citenamefont {Gadway}}]{Gadway2016}%
  \BibitemOpen
  \bibfield  {author} {\bibinfo {author} {\bibfnamefont {E.}~\bibnamefont {Meier}}, \bibinfo {author} {\bibfnamefont {F.}~\bibnamefont {An}}, \ and\ \bibinfo {author} {\bibfnamefont {B.}~\bibnamefont {Gadway}},\ }\href {https://www.nature.com/articles/ncomms13986} {\bibfield  {journal} {\bibinfo  {journal} {Nat. Commun.}\ }\textbf {\bibinfo {volume} {7}},\ \bibinfo {pages} {13986} (\bibinfo {year} {2016})}\BibitemShut {NoStop}%
\bibitem [{\citenamefont {Viebahn}\ \emph {et~al.}(2019)\citenamefont {Viebahn}, \citenamefont {Sbroscia}, \citenamefont {Carter}, \citenamefont {Yu},\ and\ \citenamefont {Schneider}}]{Viebahn2019}%
  \BibitemOpen
  \bibfield  {author} {\bibinfo {author} {\bibfnamefont {K.}~\bibnamefont {Viebahn}}, \bibinfo {author} {\bibfnamefont {M.}~\bibnamefont {Sbroscia}}, \bibinfo {author} {\bibfnamefont {E.}~\bibnamefont {Carter}}, \bibinfo {author} {\bibfnamefont {J.}~\bibnamefont {Yu}}, \ and\ \bibinfo {author} {\bibfnamefont {U.}~\bibnamefont {Schneider}},\ }\href {https://link.aps.org/doi/10.1103/PhysRevLett.122.110404} {\bibfield  {journal} {\bibinfo  {journal} {Phys. Rev. Lett.}\ }\textbf {\bibinfo {volume} {122}},\ \bibinfo {pages} {110404} (\bibinfo {year} {2019})}\BibitemShut {NoStop}%
\bibitem [{\citenamefont {An}\ \emph {et~al.}(2021)\citenamefont {An}, \citenamefont {Sundar}, \citenamefont {Hou}, \citenamefont {Luo}, \citenamefont {Meier}, \citenamefont {Zhang}, \citenamefont {Hazzard},\ and\ \citenamefont {Gadway}}]{An2021}%
  \BibitemOpen
  \bibfield  {author} {\bibinfo {author} {\bibfnamefont {F.~A.}\ \bibnamefont {An}}, \bibinfo {author} {\bibfnamefont {B.}~\bibnamefont {Sundar}}, \bibinfo {author} {\bibfnamefont {J.}~\bibnamefont {Hou}}, \bibinfo {author} {\bibfnamefont {X.-W.}\ \bibnamefont {Luo}}, \bibinfo {author} {\bibfnamefont {E.~J.}\ \bibnamefont {Meier}}, \bibinfo {author} {\bibfnamefont {C.}~\bibnamefont {Zhang}}, \bibinfo {author} {\bibfnamefont {K.~R.~A.}\ \bibnamefont {Hazzard}}, \ and\ \bibinfo {author} {\bibfnamefont {B.}~\bibnamefont {Gadway}},\ }\href {\doibase 10.1103/PhysRevLett.127.130401} {\bibfield  {journal} {\bibinfo  {journal} {Phys. Rev. Lett.}\ }\textbf {\bibinfo {volume} {127}},\ \bibinfo {pages} {130401} (\bibinfo {year} {2021})}\BibitemShut {NoStop}%
\bibitem [{\citenamefont {Agrawal}\ \emph {et~al.}(2024)\citenamefont {Agrawal}, \citenamefont {Paladugu},\ and\ \citenamefont {Gadway}}]{PRXQuantum.5.010310}%
  \BibitemOpen
  \bibfield  {author} {\bibinfo {author} {\bibfnamefont {S.}~\bibnamefont {Agrawal}}, \bibinfo {author} {\bibfnamefont {S.~N.~M.}\ \bibnamefont {Paladugu}}, \ and\ \bibinfo {author} {\bibfnamefont {B.}~\bibnamefont {Gadway}},\ }\href {\doibase 10.1103/PRXQuantum.5.010310} {\bibfield  {journal} {\bibinfo  {journal} {PRX Quantum}\ }\textbf {\bibinfo {volume} {5}},\ \bibinfo {pages} {010310} (\bibinfo {year} {2024})}\BibitemShut {NoStop}%
\bibitem [{\citenamefont {Bouhiron}\ \emph {et~al.}(2024)\citenamefont {Bouhiron}, \citenamefont {Fabre}, \citenamefont {Liu}, \citenamefont {Redon}, \citenamefont {Mittal}, \citenamefont {Satoor}, \citenamefont {Lopes},\ and\ \citenamefont {Nascimbene}}]{bouhiron2024realization}%
  \BibitemOpen
  \bibfield  {author} {\bibinfo {author} {\bibfnamefont {J.-B.}\ \bibnamefont {Bouhiron}}, \bibinfo {author} {\bibfnamefont {A.}~\bibnamefont {Fabre}}, \bibinfo {author} {\bibfnamefont {Q.}~\bibnamefont {Liu}}, \bibinfo {author} {\bibfnamefont {Q.}~\bibnamefont {Redon}}, \bibinfo {author} {\bibfnamefont {N.}~\bibnamefont {Mittal}}, \bibinfo {author} {\bibfnamefont {T.}~\bibnamefont {Satoor}}, \bibinfo {author} {\bibfnamefont {R.}~\bibnamefont {Lopes}}, \ and\ \bibinfo {author} {\bibfnamefont {S.}~\bibnamefont {Nascimbene}},\ }\href@noop {} {\bibfield  {journal} {\bibinfo  {journal} {Science}\ }\textbf {\bibinfo {volume} {384}},\ \bibinfo {pages} {223} (\bibinfo {year} {2024})}\BibitemShut {NoStop}%
\bibitem [{\citenamefont {Cai}\ \emph {et~al.}(2019)\citenamefont {Cai}, \citenamefont {Liu}, \citenamefont {Wu}, \citenamefont {He}, \citenamefont {Zhu}, \citenamefont {Zhang},\ and\ \citenamefont {Wang}}]{cai2019experimental}%
  \BibitemOpen
  \bibfield  {author} {\bibinfo {author} {\bibfnamefont {H.}~\bibnamefont {Cai}}, \bibinfo {author} {\bibfnamefont {J.}~\bibnamefont {Liu}}, \bibinfo {author} {\bibfnamefont {J.}~\bibnamefont {Wu}}, \bibinfo {author} {\bibfnamefont {Y.}~\bibnamefont {He}}, \bibinfo {author} {\bibfnamefont {S.}~\bibnamefont {Zhu}}, \bibinfo {author} {\bibfnamefont {J.}~\bibnamefont {Zhang}}, \ and\ \bibinfo {author} {\bibfnamefont {D.}~\bibnamefont {Wang}},\ }\href {https://journals.aps.org/prl/abstract/10.1103/PhysRevLett.122.023601} {\bibfield  {journal} {\bibinfo  {journal} {Phys. Rev. Lett.}\ }\textbf {\bibinfo {volume} {122}},\ \bibinfo {pages} {023601} (\bibinfo {year} {2019})}\BibitemShut {NoStop}%
\bibitem [{\citenamefont {Sundar}\ \emph {et~al.}(2018)\citenamefont {Sundar}, \citenamefont {Gadway},\ and\ \citenamefont {Hazzard}}]{Sundar2018}%
  \BibitemOpen
  \bibfield  {author} {\bibinfo {author} {\bibfnamefont {B.}~\bibnamefont {Sundar}}, \bibinfo {author} {\bibfnamefont {B.}~\bibnamefont {Gadway}}, \ and\ \bibinfo {author} {\bibfnamefont {K.}~\bibnamefont {Hazzard}},\ }\href {https://www.nature.com/articles/s41598-018-21699-x} {\bibfield  {journal} {\bibinfo  {journal} {Sci. Rep.}\ }\textbf {\bibinfo {volume} {8}},\ \bibinfo {pages} {3422} (\bibinfo {year} {2018})}\BibitemShut {NoStop}%
\bibitem [{\citenamefont {Kang}\ \emph {et~al.}(2020)\citenamefont {Kang}, \citenamefont {Han},\ and\ \citenamefont {Shin}}]{kang2020creutz}%
  \BibitemOpen
  \bibfield  {author} {\bibinfo {author} {\bibfnamefont {J.~H.}\ \bibnamefont {Kang}}, \bibinfo {author} {\bibfnamefont {J.~H.}\ \bibnamefont {Han}}, \ and\ \bibinfo {author} {\bibfnamefont {Y.}~\bibnamefont {Shin}},\ }\href {https://iopscience.iop.org/article/10.1088/1367-2630/ab61d7} {\bibfield  {journal} {\bibinfo  {journal} {New Journal of Physics}\ }\textbf {\bibinfo {volume} {22}},\ \bibinfo {pages} {013023} (\bibinfo {year} {2020})}\BibitemShut {NoStop}%
\bibitem [{\citenamefont {Lustig}\ \emph {et~al.}(2019)\citenamefont {Lustig}, \citenamefont {Weimann}, \citenamefont {Plotnik}, \citenamefont {Lumer}, \citenamefont {Bandres}, \citenamefont {Szameit},\ and\ \citenamefont {Segev}}]{lustig2019photonic}%
  \BibitemOpen
  \bibfield  {author} {\bibinfo {author} {\bibfnamefont {E.}~\bibnamefont {Lustig}}, \bibinfo {author} {\bibfnamefont {S.}~\bibnamefont {Weimann}}, \bibinfo {author} {\bibfnamefont {Y.}~\bibnamefont {Plotnik}}, \bibinfo {author} {\bibfnamefont {Y.}~\bibnamefont {Lumer}}, \bibinfo {author} {\bibfnamefont {M.}~\bibnamefont {Bandres}}, \bibinfo {author} {\bibfnamefont {A.}~\bibnamefont {Szameit}}, \ and\ \bibinfo {author} {\bibfnamefont {M.}~\bibnamefont {Segev}},\ }\href {https://www.nature.com/articles/s41586-019-0943-7/} {\bibfield  {journal} {\bibinfo  {journal} {Nature}\ }\textbf {\bibinfo {volume} {567}},\ \bibinfo {pages} {356} (\bibinfo {year} {2019})}\BibitemShut {NoStop}%
\bibitem [{\citenamefont {Dutt}\ \emph {et~al.}(2020)\citenamefont {Dutt}, \citenamefont {Lin}, \citenamefont {Yuan}, \citenamefont {Minkov}, \citenamefont {Xiao},\ and\ \citenamefont {Fan}}]{dutt2020single}%
  \BibitemOpen
  \bibfield  {author} {\bibinfo {author} {\bibfnamefont {A.}~\bibnamefont {Dutt}}, \bibinfo {author} {\bibfnamefont {Q.}~\bibnamefont {Lin}}, \bibinfo {author} {\bibfnamefont {L.}~\bibnamefont {Yuan}}, \bibinfo {author} {\bibfnamefont {M.}~\bibnamefont {Minkov}}, \bibinfo {author} {\bibfnamefont {M.}~\bibnamefont {Xiao}}, \ and\ \bibinfo {author} {\bibfnamefont {S.}~\bibnamefont {Fan}},\ }\href {https://www.science.org/doi/10.1126/science.aaz3071} {\bibfield  {journal} {\bibinfo  {journal} {Science}\ }\textbf {\bibinfo {volume} {367}},\ \bibinfo {pages} {59} (\bibinfo {year} {2020})}\BibitemShut {NoStop}%
\bibitem [{\citenamefont {Bal{\v{c}}ytis}\ \emph {et~al.}(2022)\citenamefont {Bal{\v{c}}ytis}, \citenamefont {Ozawa}, \citenamefont {Ota}, \citenamefont {Iwamoto}, \citenamefont {Maeda},\ and\ \citenamefont {Baba}}]{balvcytis2021synthetic}%
  \BibitemOpen
  \bibfield  {author} {\bibinfo {author} {\bibfnamefont {A.}~\bibnamefont {Bal{\v{c}}ytis}}, \bibinfo {author} {\bibfnamefont {T.}~\bibnamefont {Ozawa}}, \bibinfo {author} {\bibfnamefont {Y.}~\bibnamefont {Ota}}, \bibinfo {author} {\bibfnamefont {S.}~\bibnamefont {Iwamoto}}, \bibinfo {author} {\bibfnamefont {J.}~\bibnamefont {Maeda}}, \ and\ \bibinfo {author} {\bibfnamefont {T.}~\bibnamefont {Baba}},\ }\href@noop {} {\bibfield  {journal} {\bibinfo  {journal} {Science {Advances}}\ }\textbf {\bibinfo {volume} {8}},\ \bibinfo {pages} {eabk0468} (\bibinfo {year} {2022})}\BibitemShut {NoStop}%
\bibitem [{\citenamefont {Chen}\ \emph {et~al.}(2021)\citenamefont {Chen}, \citenamefont {Yang}, \citenamefont {Qin}, \citenamefont {Li}, \citenamefont {Wang}, \citenamefont {Han}, \citenamefont {Zhang}, \citenamefont {Liu}, \citenamefont {Wang}, \citenamefont {Long}, \citenamefont {Zhang},\ and\ \citenamefont {Peixiang}}]{chen2021real}%
  \BibitemOpen
  \bibfield  {author} {\bibinfo {author} {\bibfnamefont {H.}~\bibnamefont {Chen}}, \bibinfo {author} {\bibfnamefont {N.}~\bibnamefont {Yang}}, \bibinfo {author} {\bibfnamefont {C.}~\bibnamefont {Qin}}, \bibinfo {author} {\bibfnamefont {W.}~\bibnamefont {Li}}, \bibinfo {author} {\bibfnamefont {B.}~\bibnamefont {Wang}}, \bibinfo {author} {\bibfnamefont {T.}~\bibnamefont {Han}}, \bibinfo {author} {\bibfnamefont {C.}~\bibnamefont {Zhang}}, \bibinfo {author} {\bibfnamefont {W.}~\bibnamefont {Liu}}, \bibinfo {author} {\bibfnamefont {K.}~\bibnamefont {Wang}}, \bibinfo {author} {\bibfnamefont {H.}~\bibnamefont {Long}}, \bibinfo {author} {\bibfnamefont {X.}~\bibnamefont {Zhang}}, \ and\ \bibinfo {author} {\bibfnamefont {L.}~\bibnamefont {Peixiang}},\ }\href {https://www.nature.com/articles/s41377-021-00494-w} {\bibfield  {journal} {\bibinfo  {journal} {Light Sci. Appl.}\ }\textbf {\bibinfo {volume} {10}},\ \bibinfo {pages} {1} (\bibinfo {year} {2021})}\BibitemShut {NoStop}%
\bibitem [{\citenamefont {Lustig}\ and\ \citenamefont {Segev}(2021)}]{lustig2021topological}%
  \BibitemOpen
  \bibfield  {author} {\bibinfo {author} {\bibfnamefont {E.}~\bibnamefont {Lustig}}\ and\ \bibinfo {author} {\bibfnamefont {M.}~\bibnamefont {Segev}},\ }\href@noop {} {\bibfield  {journal} {\bibinfo  {journal} {{Advances} in Optics and Photonics}\ }\textbf {\bibinfo {volume} {13}},\ \bibinfo {pages} {426} (\bibinfo {year} {2021})}\BibitemShut {NoStop}%
\bibitem [{\citenamefont {Oliver}\ \emph {et~al.}(2023{\natexlab{a}})\citenamefont {Oliver}, \citenamefont {Smith}, \citenamefont {Easton}, \citenamefont {Salerno}, \citenamefont {Guarrera}, \citenamefont {Goldman}, \citenamefont {Barontini},\ and\ \citenamefont {Price}}]{oliver2023bloch}%
  \BibitemOpen
  \bibfield  {author} {\bibinfo {author} {\bibfnamefont {C.}~\bibnamefont {Oliver}}, \bibinfo {author} {\bibfnamefont {A.}~\bibnamefont {Smith}}, \bibinfo {author} {\bibfnamefont {T.}~\bibnamefont {Easton}}, \bibinfo {author} {\bibfnamefont {G.}~\bibnamefont {Salerno}}, \bibinfo {author} {\bibfnamefont {V.}~\bibnamefont {Guarrera}}, \bibinfo {author} {\bibfnamefont {N.}~\bibnamefont {Goldman}}, \bibinfo {author} {\bibfnamefont {G.}~\bibnamefont {Barontini}}, \ and\ \bibinfo {author} {\bibfnamefont {H.~M.}\ \bibnamefont {Price}},\ }\href@noop {} {\bibfield  {journal} {\bibinfo  {journal} {Physical Review Research}\ }\textbf {\bibinfo {volume} {5}},\ \bibinfo {pages} {033001} (\bibinfo {year} {2023}{\natexlab{a}})}\BibitemShut {NoStop}%
\bibitem [{\citenamefont {Oliver}\ \emph {et~al.}(2023{\natexlab{b}})\citenamefont {Oliver}, \citenamefont {Mukherjee}, \citenamefont {Rechstman}, \citenamefont {Carusotto},\ and\ \citenamefont {Price}}]{oliver2023artificial}%
  \BibitemOpen
  \bibfield  {author} {\bibinfo {author} {\bibfnamefont {C.}~\bibnamefont {Oliver}}, \bibinfo {author} {\bibfnamefont {S.}~\bibnamefont {Mukherjee}}, \bibinfo {author} {\bibfnamefont {M.~C.}\ \bibnamefont {Rechstman}}, \bibinfo {author} {\bibfnamefont {I.}~\bibnamefont {Carusotto}}, \ and\ \bibinfo {author} {\bibfnamefont {H.~M.}\ \bibnamefont {Price}},\ }\href@noop {} {\bibfield  {journal} {\bibinfo  {journal} {Science {Advances}}\ }\textbf {\bibinfo {volume} {9}},\ \bibinfo {pages} {eadj0360} (\bibinfo {year} {2023}{\natexlab{b}})}\BibitemShut {NoStop}%
\bibitem [{\citenamefont {Dinh}\ \emph {et~al.}(2024)\citenamefont {Dinh}, \citenamefont {Bal{\v{c}}ytis}, \citenamefont {Ozawa}, \citenamefont {Ota}, \citenamefont {Ren}, \citenamefont {Baba}, \citenamefont {Iwamoto}, \citenamefont {Mitchell},\ and\ \citenamefont {Nguyen}}]{dinh2024reconfigurable}%
  \BibitemOpen
  \bibfield  {author} {\bibinfo {author} {\bibfnamefont {H.~X.}\ \bibnamefont {Dinh}}, \bibinfo {author} {\bibfnamefont {A.}~\bibnamefont {Bal{\v{c}}ytis}}, \bibinfo {author} {\bibfnamefont {T.}~\bibnamefont {Ozawa}}, \bibinfo {author} {\bibfnamefont {Y.}~\bibnamefont {Ota}}, \bibinfo {author} {\bibfnamefont {G.}~\bibnamefont {Ren}}, \bibinfo {author} {\bibfnamefont {T.}~\bibnamefont {Baba}}, \bibinfo {author} {\bibfnamefont {S.}~\bibnamefont {Iwamoto}}, \bibinfo {author} {\bibfnamefont {A.}~\bibnamefont {Mitchell}}, \ and\ \bibinfo {author} {\bibfnamefont {T.~G.}\ \bibnamefont {Nguyen}},\ }\href@noop {} {\bibfield  {journal} {\bibinfo  {journal} {{Commun.} Physics}\ }\textbf {\bibinfo {volume} {7}},\ \bibinfo {pages} {185} (\bibinfo {year} {2024})}\BibitemShut {NoStop}%
\bibitem [{\citenamefont {Leefmans}\ \emph {et~al.}(2022)\citenamefont {Leefmans}, \citenamefont {Dutt}, \citenamefont {Williams}, \citenamefont {Yuan}, \citenamefont {Parto}, \citenamefont {Nori}, \citenamefont {Fan},\ and\ \citenamefont {Marandi}}]{leefmans2022topological}%
  \BibitemOpen
  \bibfield  {author} {\bibinfo {author} {\bibfnamefont {C.}~\bibnamefont {Leefmans}}, \bibinfo {author} {\bibfnamefont {A.}~\bibnamefont {Dutt}}, \bibinfo {author} {\bibfnamefont {J.}~\bibnamefont {Williams}}, \bibinfo {author} {\bibfnamefont {L.}~\bibnamefont {Yuan}}, \bibinfo {author} {\bibfnamefont {M.}~\bibnamefont {Parto}}, \bibinfo {author} {\bibfnamefont {F.}~\bibnamefont {Nori}}, \bibinfo {author} {\bibfnamefont {S.}~\bibnamefont {Fan}}, \ and\ \bibinfo {author} {\bibfnamefont {A.}~\bibnamefont {Marandi}},\ }\href@noop {} {\bibfield  {journal} {\bibinfo  {journal} {Nat. {Phys.}}\ }\textbf {\bibinfo {volume} {18}},\ \bibinfo {pages} {442} (\bibinfo {year} {2022})}\BibitemShut {NoStop}%
\bibitem [{\citenamefont {Ehrhardt}\ \emph {et~al.}(2023)\citenamefont {Ehrhardt}, \citenamefont {Weidemann}, \citenamefont {Maczewsky}, \citenamefont {Heinrich},\ and\ \citenamefont {Szameit}}]{ehrhardt2023perspective}%
  \BibitemOpen
  \bibfield  {author} {\bibinfo {author} {\bibfnamefont {M.}~\bibnamefont {Ehrhardt}}, \bibinfo {author} {\bibfnamefont {S.}~\bibnamefont {Weidemann}}, \bibinfo {author} {\bibfnamefont {L.~J.}\ \bibnamefont {Maczewsky}}, \bibinfo {author} {\bibfnamefont {M.}~\bibnamefont {Heinrich}}, \ and\ \bibinfo {author} {\bibfnamefont {A.}~\bibnamefont {Szameit}},\ }\href@noop {} {\bibfield  {journal} {\bibinfo  {journal} {Laser \& Photonics Reviews}\ }\textbf {\bibinfo {volume} {17}},\ \bibinfo {pages} {2200518} (\bibinfo {year} {2023})}\BibitemShut {NoStop}%
\bibitem [{\citenamefont {Wang}\ \emph {et~al.}(2024{\natexlab{a}})\citenamefont {Wang}, \citenamefont {Wu}, \citenamefont {Jiang}, \citenamefont {Cai}, \citenamefont {Li}, \citenamefont {Mei}, \citenamefont {Qi}, \citenamefont {Zhou},\ and\ \citenamefont {Duan}}]{PhysRevLett.132.130601}%
  \BibitemOpen
  \bibfield  {author} {\bibinfo {author} {\bibfnamefont {Y.}~\bibnamefont {Wang}}, \bibinfo {author} {\bibfnamefont {Y.-K.}\ \bibnamefont {Wu}}, \bibinfo {author} {\bibfnamefont {Y.}~\bibnamefont {Jiang}}, \bibinfo {author} {\bibfnamefont {M.-L.}\ \bibnamefont {Cai}}, \bibinfo {author} {\bibfnamefont {B.-W.}\ \bibnamefont {Li}}, \bibinfo {author} {\bibfnamefont {Q.-X.}\ \bibnamefont {Mei}}, \bibinfo {author} {\bibfnamefont {B.-X.}\ \bibnamefont {Qi}}, \bibinfo {author} {\bibfnamefont {Z.-C.}\ \bibnamefont {Zhou}}, \ and\ \bibinfo {author} {\bibfnamefont {L.-M.}\ \bibnamefont {Duan}},\ }\href {\doibase 10.1103/PhysRevLett.132.130601} {\bibfield  {journal} {\bibinfo  {journal} {Phys. Rev. Lett.}\ }\textbf {\bibinfo {volume} {132}},\ \bibinfo {pages} {130601} (\bibinfo {year} {2024}{\natexlab{a}})}\BibitemShut {NoStop}%
\bibitem [{\citenamefont {Wang}\ \emph {et~al.}(2021)\citenamefont {Wang}, \citenamefont {Dutt}, \citenamefont {Yang}, \citenamefont {Wojcik}, \citenamefont {Vu\v{c}kovi\'{c}},\ and\ \citenamefont {Fan}}]{Wang2021}%
  \BibitemOpen
  \bibfield  {author} {\bibinfo {author} {\bibfnamefont {K.}~\bibnamefont {Wang}}, \bibinfo {author} {\bibfnamefont {A.}~\bibnamefont {Dutt}}, \bibinfo {author} {\bibfnamefont {Y.}~\bibnamefont {Yang}}, \bibinfo {author} {\bibfnamefont {C.}~\bibnamefont {Wojcik}}, \bibinfo {author} {\bibfnamefont {J.}~\bibnamefont {Vu\v{c}kovi\'{c}}}, \ and\ \bibinfo {author} {\bibfnamefont {S.}~\bibnamefont {Fan}},\ }\href {https://doi.org/10.1126/science.abf6568} {\bibfield  {journal} {\bibinfo  {journal} {Science}\ }\textbf {\bibinfo {volume} {371}},\ \bibinfo {pages} {6535} (\bibinfo {year} {2021})}\BibitemShut {NoStop}%
\bibitem [{\citenamefont {Price}\ \emph {et~al.}(2015)\citenamefont {Price}, \citenamefont {Zilberberg}, \citenamefont {Ozawa}, \citenamefont {Carusotto},\ and\ \citenamefont {Goldman}}]{price2015}%
  \BibitemOpen
  \bibfield  {author} {\bibinfo {author} {\bibfnamefont {H.~M.}\ \bibnamefont {Price}}, \bibinfo {author} {\bibfnamefont {O.}~\bibnamefont {Zilberberg}}, \bibinfo {author} {\bibfnamefont {T.}~\bibnamefont {Ozawa}}, \bibinfo {author} {\bibfnamefont {I.}~\bibnamefont {Carusotto}}, \ and\ \bibinfo {author} {\bibfnamefont {N.}~\bibnamefont {Goldman}},\ }\href {https://link.aps.org/doi/10.1103/PhysRevLett.115.195303} {\bibfield  {journal} {\bibinfo  {journal} {Phys. Rev. Lett.}\ }\textbf {\bibinfo {volume} {115}},\ \bibinfo {pages} {195303} (\bibinfo {year} {2015})}\BibitemShut {NoStop}%
\bibitem [{\citenamefont {Braun}\ \emph {et~al.}(2024)\citenamefont {Braun}, \citenamefont {Saint-Jalm}, \citenamefont {Hesse}, \citenamefont {Arceri}, \citenamefont {Bloch},\ and\ \citenamefont {Aidelsburger}}]{braun2024real}%
  \BibitemOpen
  \bibfield  {author} {\bibinfo {author} {\bibfnamefont {C.}~\bibnamefont {Braun}}, \bibinfo {author} {\bibfnamefont {R.}~\bibnamefont {Saint-Jalm}}, \bibinfo {author} {\bibfnamefont {A.}~\bibnamefont {Hesse}}, \bibinfo {author} {\bibfnamefont {J.}~\bibnamefont {Arceri}}, \bibinfo {author} {\bibfnamefont {I.}~\bibnamefont {Bloch}}, \ and\ \bibinfo {author} {\bibfnamefont {M.}~\bibnamefont {Aidelsburger}},\ }\href@noop {} {\bibfield  {journal} {\bibinfo  {journal} {Nat. {Phys.}}\ ,\ \bibinfo {pages} {1}} (\bibinfo {year} {2024})}\BibitemShut {NoStop}%
\bibitem [{\citenamefont {Yao}\ \emph {et~al.}(2024)\citenamefont {Yao}, \citenamefont {Chi}, \citenamefont {Mukherjee}, \citenamefont {Shaffer}, \citenamefont {Zwierlein},\ and\ \citenamefont {Fletcher}}]{yao2023observation}%
  \BibitemOpen
  \bibfield  {author} {\bibinfo {author} {\bibfnamefont {R.}~\bibnamefont {Yao}}, \bibinfo {author} {\bibfnamefont {S.}~\bibnamefont {Chi}}, \bibinfo {author} {\bibfnamefont {B.}~\bibnamefont {Mukherjee}}, \bibinfo {author} {\bibfnamefont {A.}~\bibnamefont {Shaffer}}, \bibinfo {author} {\bibfnamefont {M.}~\bibnamefont {Zwierlein}}, \ and\ \bibinfo {author} {\bibfnamefont {R.~J.}\ \bibnamefont {Fletcher}},\ }\href {\doibase 10.1038/s41567-024-02617-7} {\bibfield  {journal} {\bibinfo  {journal} {Nature Physics}\ }\textbf {\bibinfo {volume} {20}},\ \bibinfo {pages} {1726–1731} (\bibinfo {year} {2024})}\BibitemShut {NoStop}%
\bibitem [{\citenamefont {Wang}\ \emph {et~al.}(2024{\natexlab{b}})\citenamefont {Wang}, \citenamefont {Aidelsburger}, \citenamefont {Dalibard}, \citenamefont {Eckardt},\ and\ \citenamefont {Goldman}}]{Wang2024}%
  \BibitemOpen
  \bibfield  {author} {\bibinfo {author} {\bibfnamefont {B.}~\bibnamefont {Wang}}, \bibinfo {author} {\bibfnamefont {M.}~\bibnamefont {Aidelsburger}}, \bibinfo {author} {\bibfnamefont {J.}~\bibnamefont {Dalibard}}, \bibinfo {author} {\bibfnamefont {A.}~\bibnamefont {Eckardt}}, \ and\ \bibinfo {author} {\bibfnamefont {N.}~\bibnamefont {Goldman}},\ }\href {\doibase 10.1103/PhysRevLett.132.163402} {\bibfield  {journal} {\bibinfo  {journal} {Phys. Rev. Lett.}\ }\textbf {\bibinfo {volume} {132}},\ \bibinfo {pages} {163402} (\bibinfo {year} {2024}{\natexlab{b}})}\BibitemShut {NoStop}%
\bibitem [{\citenamefont {Kane}\ \emph {et~al.}(2002)\citenamefont {Kane}, \citenamefont {Mukhopadhyay},\ and\ \citenamefont {Lubensky}}]{Kane2002}%
  \BibitemOpen
  \bibfield  {author} {\bibinfo {author} {\bibfnamefont {C.~L.}\ \bibnamefont {Kane}}, \bibinfo {author} {\bibfnamefont {R.}~\bibnamefont {Mukhopadhyay}}, \ and\ \bibinfo {author} {\bibfnamefont {T.~C.}\ \bibnamefont {Lubensky}},\ }\href {\doibase 10.1103/PhysRevLett.88.036401} {\bibfield  {journal} {\bibinfo  {journal} {Phys. Rev. Lett.}\ }\textbf {\bibinfo {volume} {88}},\ \bibinfo {pages} {036401} (\bibinfo {year} {2002})}\BibitemShut {NoStop}%
\bibitem [{\citenamefont {Budich}\ \emph {et~al.}(2017)\citenamefont {Budich}, \citenamefont {Elben}, \citenamefont {{\L}{k{a}}cki}, \citenamefont {Sterdyniak}, \citenamefont {Baranov},\ and\ \citenamefont {Zoller}}]{Budich_2017}%
  \BibitemOpen
  \bibfield  {author} {\bibinfo {author} {\bibfnamefont {J.~C.}\ \bibnamefont {Budich}}, \bibinfo {author} {\bibfnamefont {A.}~\bibnamefont {Elben}}, \bibinfo {author} {\bibfnamefont {M.}~\bibnamefont {{\L}{k{a}}cki}}, \bibinfo {author} {\bibfnamefont {A.}~\bibnamefont {Sterdyniak}}, \bibinfo {author} {\bibfnamefont {M.~A.}\ \bibnamefont {Baranov}}, \ and\ \bibinfo {author} {\bibfnamefont {P.}~\bibnamefont {Zoller}},\ }\href {https://doi.org/10.1103%2Fphysreva.95.043632} {\bibfield  {journal} {\bibinfo  {journal} {Physical Review A}\ }\textbf {\bibinfo {volume} {95}} (\bibinfo {year} {2017})}\BibitemShut {NoStop}%
\bibitem [{\citenamefont {Salerno}\ \emph {et~al.}(2019)\citenamefont {Salerno}, \citenamefont {Price}, \citenamefont {Lebrat}, \citenamefont {H\"ausler}, \citenamefont {Esslinger}, \citenamefont {Corman}, \citenamefont {Brantut},\ and\ \citenamefont {Goldman}}]{Salerno2019}%
  \BibitemOpen
  \bibfield  {author} {\bibinfo {author} {\bibfnamefont {G.}~\bibnamefont {Salerno}}, \bibinfo {author} {\bibfnamefont {H.~M.}\ \bibnamefont {Price}}, \bibinfo {author} {\bibfnamefont {M.}~\bibnamefont {Lebrat}}, \bibinfo {author} {\bibfnamefont {S.}~\bibnamefont {H\"ausler}}, \bibinfo {author} {\bibfnamefont {T.}~\bibnamefont {Esslinger}}, \bibinfo {author} {\bibfnamefont {L.}~\bibnamefont {Corman}}, \bibinfo {author} {\bibfnamefont {J.-P.}\ \bibnamefont {Brantut}}, \ and\ \bibinfo {author} {\bibfnamefont {N.}~\bibnamefont {Goldman}},\ }\href {https://link.aps.org/doi/10.1103/PhysRevX.9.041001} {\bibfield  {journal} {\bibinfo  {journal} {Phys. Rev. X}\ }\textbf {\bibinfo {volume} {9}},\ \bibinfo {pages} {041001} (\bibinfo {year} {2019})}\BibitemShut {NoStop}%
\bibitem [{\citenamefont {Hu}\ \emph {et~al.}(2018)\citenamefont {Hu}, \citenamefont {Niu}, \citenamefont {Jin}, \citenamefont {Chen}, \citenamefont {Dong}, \citenamefont {Schmiedmayer},\ and\ \citenamefont {Zhou}}]{Hu2018}%
  \BibitemOpen
  \bibfield  {author} {\bibinfo {author} {\bibfnamefont {D.}~\bibnamefont {Hu}}, \bibinfo {author} {\bibfnamefont {L.}~\bibnamefont {Niu}}, \bibinfo {author} {\bibfnamefont {S.}~\bibnamefont {Jin}}, \bibinfo {author} {\bibfnamefont {X.}~\bibnamefont {Chen}}, \bibinfo {author} {\bibfnamefont {G.}~\bibnamefont {Dong}}, \bibinfo {author} {\bibfnamefont {J.}~\bibnamefont {Schmiedmayer}}, \ and\ \bibinfo {author} {\bibfnamefont {X.}~\bibnamefont {Zhou}},\ }\href {\doibase 10.1038/s42005-018-0030-7} {\bibfield  {journal} {\bibinfo  {journal} {Commun. Phys.}\ }\textbf {\bibinfo {volume} {1}},\ \bibinfo {pages} {29} (\bibinfo {year} {2018})}\BibitemShut {NoStop}%
\bibitem [{\citenamefont {van Frank}\ \emph {et~al.}(2014)\citenamefont {van Frank}, \citenamefont {Negretti}, \citenamefont {Berrada}, \citenamefont {Bücker}, \citenamefont {Montangero}, \citenamefont {Schaff}, \citenamefont {Schumm}, \citenamefont {Calarco},\ and\ \citenamefont {Schmiedmayer}}]{Frank2014}%
  \BibitemOpen
  \bibfield  {author} {\bibinfo {author} {\bibfnamefont {S.}~\bibnamefont {van Frank}}, \bibinfo {author} {\bibfnamefont {A.}~\bibnamefont {Negretti}}, \bibinfo {author} {\bibfnamefont {T.}~\bibnamefont {Berrada}}, \bibinfo {author} {\bibfnamefont {R.}~\bibnamefont {Bücker}}, \bibinfo {author} {\bibfnamefont {S.}~\bibnamefont {Montangero}}, \bibinfo {author} {\bibfnamefont {J.-F.}\ \bibnamefont {Schaff}}, \bibinfo {author} {\bibfnamefont {T.}~\bibnamefont {Schumm}}, \bibinfo {author} {\bibfnamefont {T.}~\bibnamefont {Calarco}}, \ and\ \bibinfo {author} {\bibfnamefont {J.}~\bibnamefont {Schmiedmayer}},\ }\href {https://www.nature.com/articles/ncomms5009} {\bibfield  {journal} {\bibinfo  {journal} {Nat. Commun.}\ }\textbf {\bibinfo {volume} {5}},\ \bibinfo {pages} {4009} (\bibinfo {year} {2014})}\BibitemShut {NoStop}%
\bibitem [{\citenamefont {Guarrera}\ \emph {et~al.}(2015)\citenamefont {Guarrera}, \citenamefont {Szmuk}, \citenamefont {Reichel},\ and\ \citenamefont {Rosenbusch}}]{Guarrera2015}%
  \BibitemOpen
  \bibfield  {author} {\bibinfo {author} {\bibfnamefont {V.}~\bibnamefont {Guarrera}}, \bibinfo {author} {\bibfnamefont {R.}~\bibnamefont {Szmuk}}, \bibinfo {author} {\bibfnamefont {J.}~\bibnamefont {Reichel}}, \ and\ \bibinfo {author} {\bibfnamefont {P.}~\bibnamefont {Rosenbusch}},\ }\href {https://doi.org/10.1088/1367-2630/17/8/083022} {\bibfield  {journal} {\bibinfo  {journal} {New J. Phys.}\ }\textbf {\bibinfo {volume} {17}},\ \bibinfo {pages} {083022} (\bibinfo {year} {2015})}\BibitemShut {NoStop}%
\bibitem [{\citenamefont {Vinjanampathy}\ and\ \citenamefont {Anders}(2016)}]{Vin2016}%
  \BibitemOpen
  \bibfield  {author} {\bibinfo {author} {\bibfnamefont {S.}~\bibnamefont {Vinjanampathy}}\ and\ \bibinfo {author} {\bibfnamefont {J.}~\bibnamefont {Anders}},\ }\href {https://doi.org/10.1080/00107514.2016.1201896} {\bibfield  {journal} {\bibinfo  {journal} {Contemp. Phys.}\ }\textbf {\bibinfo {volume} {57}},\ \bibinfo {pages} {545} (\bibinfo {year} {2016})}\BibitemShut {NoStop}%
\bibitem [{\citenamefont {Quan}\ \emph {et~al.}(2007)\citenamefont {Quan}, \citenamefont {Liu}, \citenamefont {Sun},\ and\ \citenamefont {Nori}}]{Quan2007}%
  \BibitemOpen
  \bibfield  {author} {\bibinfo {author} {\bibfnamefont {H.~T.}\ \bibnamefont {Quan}}, \bibinfo {author} {\bibfnamefont {Y.-x.}\ \bibnamefont {Liu}}, \bibinfo {author} {\bibfnamefont {C.~P.}\ \bibnamefont {Sun}}, \ and\ \bibinfo {author} {\bibfnamefont {F.}~\bibnamefont {Nori}},\ }\href {https://link.aps.org/doi/10.1103/PhysRevE.76.031105} {\bibfield  {journal} {\bibinfo  {journal} {Phys. Rev. E}\ }\textbf {\bibinfo {volume} {76}},\ \bibinfo {pages} {031105} (\bibinfo {year} {2007})}\BibitemShut {NoStop}%
\bibitem [{\citenamefont {Uzdin}\ \emph {et~al.}(2015)\citenamefont {Uzdin}, \citenamefont {Levy},\ and\ \citenamefont {Kosloff}}]{Uzdin2015}%
  \BibitemOpen
  \bibfield  {author} {\bibinfo {author} {\bibfnamefont {R.}~\bibnamefont {Uzdin}}, \bibinfo {author} {\bibfnamefont {A.}~\bibnamefont {Levy}}, \ and\ \bibinfo {author} {\bibfnamefont {R.}~\bibnamefont {Kosloff}},\ }\href {https://link.aps.org/doi/10.1103/PhysRevX.5.031044} {\bibfield  {journal} {\bibinfo  {journal} {Phys. Rev. X}\ }\textbf {\bibinfo {volume} {5}},\ \bibinfo {pages} {031044} (\bibinfo {year} {2015})}\BibitemShut {NoStop}%
\bibitem [{\citenamefont {Gauthier}\ \emph {et~al.}(2016)\citenamefont {Gauthier}, \citenamefont {Lenton}, \citenamefont {McKay~Parry}, \citenamefont {Baker}, \citenamefont {Davis}, \citenamefont {Rubinsztein-Dunlop},\ and\ \citenamefont {Neely}}]{Gauthier_2016}%
  \BibitemOpen
  \bibfield  {author} {\bibinfo {author} {\bibfnamefont {G.}~\bibnamefont {Gauthier}}, \bibinfo {author} {\bibfnamefont {I.}~\bibnamefont {Lenton}}, \bibinfo {author} {\bibfnamefont {N.}~\bibnamefont {McKay~Parry}}, \bibinfo {author} {\bibfnamefont {M.}~\bibnamefont {Baker}}, \bibinfo {author} {\bibfnamefont {M.~J.}\ \bibnamefont {Davis}}, \bibinfo {author} {\bibfnamefont {H.}~\bibnamefont {Rubinsztein-Dunlop}}, \ and\ \bibinfo {author} {\bibfnamefont {T.~W.}\ \bibnamefont {Neely}},\ }\href {\doibase 10.1364/optica.3.001136} {\bibfield  {journal} {\bibinfo  {journal} {Optica}\ }\textbf {\bibinfo {volume} {3}},\ \bibinfo {pages} {1136} (\bibinfo {year} {2016})}\BibitemShut {NoStop}%
\bibitem [{\citenamefont {Amico}\ \emph {et~al.}(2021)\citenamefont {Amico}, \citenamefont {Boshier}, \citenamefont {Birkl}, \citenamefont {Minguzzi}, \citenamefont {Miniatura}, \citenamefont {Kwek}, \citenamefont {Aghamalyan}, \citenamefont {Ahufinger}, \citenamefont {Anderson}, \citenamefont {Andrei}, \citenamefont {Arnold}, \citenamefont {Baker}, \citenamefont {Bell}, \citenamefont {Bland}, \citenamefont {Brantut}, \citenamefont {Cassettari}, \citenamefont {Chetcuti}, \citenamefont {Chevy}, \citenamefont {Citro}, \citenamefont {De~Palo}, \citenamefont {Dumke}, \citenamefont {Edwards}, \citenamefont {Folman}, \citenamefont {Fortagh}, \citenamefont {Gardiner}, \citenamefont {Garraway}, \citenamefont {Gauthier}, \citenamefont {Günther}, \citenamefont {Haug}, \citenamefont {Hufnagel}, \citenamefont {Keil}, \citenamefont {Ireland}, \citenamefont {Lebrat}, \citenamefont {Li}, \citenamefont {Longchambon}, \citenamefont {Mompart}, \citenamefont {Morsch}, \citenamefont {Naldesi}, \citenamefont {Neely},
  \citenamefont {Olshanii}, \citenamefont {Orignac}, \citenamefont {Pandey}, \citenamefont {Pérez-Obiol}, \citenamefont {Perrin}, \citenamefont {Piroli}, \citenamefont {Polo}, \citenamefont {Pritchard}, \citenamefont {Proukakis}, \citenamefont {Rylands}, \citenamefont {Rubinsztein-Dunlop}, \citenamefont {Scazza}, \citenamefont {Stringari}, \citenamefont {Tosto}, \citenamefont {Trombettoni}, \citenamefont {Victorin}, \citenamefont {Klitzing}, \citenamefont {Wilkowski}, \citenamefont {Xhani},\ and\ \citenamefont {Yakimenko}}]{Amico_2021}%
  \BibitemOpen
  \bibfield  {author} {\bibinfo {author} {\bibfnamefont {L.}~\bibnamefont {Amico}}, \bibinfo {author} {\bibfnamefont {M.}~\bibnamefont {Boshier}}, \bibinfo {author} {\bibfnamefont {G.}~\bibnamefont {Birkl}}, \bibinfo {author} {\bibfnamefont {A.}~\bibnamefont {Minguzzi}}, \bibinfo {author} {\bibfnamefont {C.}~\bibnamefont {Miniatura}}, \bibinfo {author} {\bibfnamefont {L.-C.}\ \bibnamefont {Kwek}}, \bibinfo {author} {\bibfnamefont {D.}~\bibnamefont {Aghamalyan}}, \bibinfo {author} {\bibfnamefont {V.}~\bibnamefont {Ahufinger}}, \bibinfo {author} {\bibfnamefont {D.}~\bibnamefont {Anderson}}, \bibinfo {author} {\bibfnamefont {N.}~\bibnamefont {Andrei}}, \bibinfo {author} {\bibfnamefont {A.~S.}\ \bibnamefont {Arnold}}, \bibinfo {author} {\bibfnamefont {M.}~\bibnamefont {Baker}}, \bibinfo {author} {\bibfnamefont {T.~A.}\ \bibnamefont {Bell}}, \bibinfo {author} {\bibfnamefont {T.}~\bibnamefont {Bland}}, \bibinfo {author} {\bibfnamefont {J.~P.}\ \bibnamefont {Brantut}}, \bibinfo {author} {\bibfnamefont
  {D.}~\bibnamefont {Cassettari}}, \bibinfo {author} {\bibfnamefont {W.~J.}\ \bibnamefont {Chetcuti}}, \bibinfo {author} {\bibfnamefont {F.}~\bibnamefont {Chevy}}, \bibinfo {author} {\bibfnamefont {R.}~\bibnamefont {Citro}}, \bibinfo {author} {\bibfnamefont {S.}~\bibnamefont {De~Palo}}, \bibinfo {author} {\bibfnamefont {R.}~\bibnamefont {Dumke}}, \bibinfo {author} {\bibfnamefont {M.}~\bibnamefont {Edwards}}, \bibinfo {author} {\bibfnamefont {R.}~\bibnamefont {Folman}}, \bibinfo {author} {\bibfnamefont {J.}~\bibnamefont {Fortagh}}, \bibinfo {author} {\bibfnamefont {S.~A.}\ \bibnamefont {Gardiner}}, \bibinfo {author} {\bibfnamefont {B.~M.}\ \bibnamefont {Garraway}}, \bibinfo {author} {\bibfnamefont {G.}~\bibnamefont {Gauthier}}, \bibinfo {author} {\bibfnamefont {A.}~\bibnamefont {Günther}}, \bibinfo {author} {\bibfnamefont {T.}~\bibnamefont {Haug}}, \bibinfo {author} {\bibfnamefont {C.}~\bibnamefont {Hufnagel}}, \bibinfo {author} {\bibfnamefont {M.}~\bibnamefont {Keil}}, \bibinfo {author} {\bibfnamefont
  {P.}~\bibnamefont {Ireland}}, \bibinfo {author} {\bibfnamefont {M.}~\bibnamefont {Lebrat}}, \bibinfo {author} {\bibfnamefont {W.}~\bibnamefont {Li}}, \bibinfo {author} {\bibfnamefont {L.}~\bibnamefont {Longchambon}}, \bibinfo {author} {\bibfnamefont {J.}~\bibnamefont {Mompart}}, \bibinfo {author} {\bibfnamefont {O.}~\bibnamefont {Morsch}}, \bibinfo {author} {\bibfnamefont {P.}~\bibnamefont {Naldesi}}, \bibinfo {author} {\bibfnamefont {T.~W.}\ \bibnamefont {Neely}}, \bibinfo {author} {\bibfnamefont {M.}~\bibnamefont {Olshanii}}, \bibinfo {author} {\bibfnamefont {E.}~\bibnamefont {Orignac}}, \bibinfo {author} {\bibfnamefont {S.}~\bibnamefont {Pandey}}, \bibinfo {author} {\bibfnamefont {A.}~\bibnamefont {Pérez-Obiol}}, \bibinfo {author} {\bibfnamefont {H.}~\bibnamefont {Perrin}}, \bibinfo {author} {\bibfnamefont {L.}~\bibnamefont {Piroli}}, \bibinfo {author} {\bibfnamefont {J.}~\bibnamefont {Polo}}, \bibinfo {author} {\bibfnamefont {A.~L.}\ \bibnamefont {Pritchard}}, \bibinfo {author} {\bibfnamefont {N.~P.}\
  \bibnamefont {Proukakis}}, \bibinfo {author} {\bibfnamefont {C.}~\bibnamefont {Rylands}}, \bibinfo {author} {\bibfnamefont {H.}~\bibnamefont {Rubinsztein-Dunlop}}, \bibinfo {author} {\bibfnamefont {F.}~\bibnamefont {Scazza}}, \bibinfo {author} {\bibfnamefont {S.}~\bibnamefont {Stringari}}, \bibinfo {author} {\bibfnamefont {F.}~\bibnamefont {Tosto}}, \bibinfo {author} {\bibfnamefont {A.}~\bibnamefont {Trombettoni}}, \bibinfo {author} {\bibfnamefont {N.}~\bibnamefont {Victorin}}, \bibinfo {author} {\bibfnamefont {W.~v.}\ \bibnamefont {Klitzing}}, \bibinfo {author} {\bibfnamefont {D.}~\bibnamefont {Wilkowski}}, \bibinfo {author} {\bibfnamefont {K.}~\bibnamefont {Xhani}}, \ and\ \bibinfo {author} {\bibfnamefont {A.}~\bibnamefont {Yakimenko}},\ }\href {\doibase 10.1116/5.0026178} {\bibfield  {journal} {\bibinfo  {journal} {AVS Quantum Science}\ }\textbf {\bibinfo {volume} {3}} (\bibinfo {year} {2021}),\ 10.1116/5.0026178}\BibitemShut {NoStop}%
\bibitem [{\citenamefont {Goldman}\ \emph {et~al.}(2013)\citenamefont {Goldman}, \citenamefont {Dalibard}, \citenamefont {Dauphin}, \citenamefont {Gerbier}, \citenamefont {Lewenstein}, \citenamefont {Zoller},\ and\ \citenamefont {Spielman}}]{Goldman2013}%
  \BibitemOpen
  \bibfield  {author} {\bibinfo {author} {\bibfnamefont {N.}~\bibnamefont {Goldman}}, \bibinfo {author} {\bibfnamefont {J.}~\bibnamefont {Dalibard}}, \bibinfo {author} {\bibfnamefont {A.}~\bibnamefont {Dauphin}}, \bibinfo {author} {\bibfnamefont {F.}~\bibnamefont {Gerbier}}, \bibinfo {author} {\bibfnamefont {M.}~\bibnamefont {Lewenstein}}, \bibinfo {author} {\bibfnamefont {P.}~\bibnamefont {Zoller}}, \ and\ \bibinfo {author} {\bibfnamefont {I.~B.}\ \bibnamefont {Spielman}},\ }\href {\doibase 10.1073/pnas.1300170110} {\bibfield  {journal} {\bibinfo  {journal} {Proceedings of the National Academy of Sciences of the United States of America}\ }\textbf {\bibinfo {volume} {110}},\ \bibinfo {pages} {6736} (\bibinfo {year} {2013})}\BibitemShut {NoStop}%
\bibitem [{\citenamefont {Goldman}\ \emph {et~al.}(2012)\citenamefont {Goldman}, \citenamefont {Beugnon},\ and\ \citenamefont {Gerbier}}]{PRL108255303}%
  \BibitemOpen
  \bibfield  {author} {\bibinfo {author} {\bibfnamefont {N.}~\bibnamefont {Goldman}}, \bibinfo {author} {\bibfnamefont {J.}~\bibnamefont {Beugnon}}, \ and\ \bibinfo {author} {\bibfnamefont {F.}~\bibnamefont {Gerbier}},\ }\href@noop {} {\bibfield  {journal} {\bibinfo  {journal} {Phys. Rev. Lett}\ }\textbf {\bibinfo {volume} {108}},\ \bibinfo {pages} {255303} (\bibinfo {year} {2012})}\BibitemShut {NoStop}%
\bibitem [{\citenamefont {Goldman}\ \emph {et~al.}(2016{\natexlab{b}})\citenamefont {Goldman}, \citenamefont {Jotzu}, \citenamefont {Messer}, \citenamefont {G\"org}, \citenamefont {Desbuquois},\ and\ \citenamefont {Esslinger}}]{PhysRevA.94.043611}%
  \BibitemOpen
  \bibfield  {author} {\bibinfo {author} {\bibfnamefont {N.}~\bibnamefont {Goldman}}, \bibinfo {author} {\bibfnamefont {G.}~\bibnamefont {Jotzu}}, \bibinfo {author} {\bibfnamefont {M.}~\bibnamefont {Messer}}, \bibinfo {author} {\bibfnamefont {F.}~\bibnamefont {G\"org}}, \bibinfo {author} {\bibfnamefont {R.}~\bibnamefont {Desbuquois}}, \ and\ \bibinfo {author} {\bibfnamefont {T.}~\bibnamefont {Esslinger}},\ }\href {\doibase 10.1103/PhysRevA.94.043611} {\bibfield  {journal} {\bibinfo  {journal} {Phys. Rev. A}\ }\textbf {\bibinfo {volume} {94}},\ \bibinfo {pages} {043611} (\bibinfo {year} {2016}{\natexlab{b}})}\BibitemShut {NoStop}%
\bibitem [{\citenamefont {Rudner}\ and\ \citenamefont {Lindner}(2020)}]{rudner2020floquet}%
  \BibitemOpen
  \bibfield  {author} {\bibinfo {author} {\bibfnamefont {M.~S.}\ \bibnamefont {Rudner}}\ and\ \bibinfo {author} {\bibfnamefont {N.~H.}\ \bibnamefont {Lindner}},\ }\href@noop {} {\enquote {\bibinfo {title} {The {Floquet} engineer's handbook},}\ } (\bibinfo {year} {2020}),\ \Eprint {http://arxiv.org/abs/2003.08252} {arXiv:2003.08252} \BibitemShut {NoStop}%
\bibitem [{\citenamefont {Goldman}\ and\ \citenamefont {Dalibard}(2014)}]{PhysRevX.4.031027}%
  \BibitemOpen
  \bibfield  {author} {\bibinfo {author} {\bibfnamefont {N.}~\bibnamefont {Goldman}}\ and\ \bibinfo {author} {\bibfnamefont {J.}~\bibnamefont {Dalibard}},\ }\href {\doibase 10.1103/PhysRevX.4.031027} {\bibfield  {journal} {\bibinfo  {journal} {Phys. Rev. X}\ }\textbf {\bibinfo {volume} {4}},\ \bibinfo {pages} {031027} (\bibinfo {year} {2014})}\BibitemShut {NoStop}%
\bibitem [{\citenamefont {Binanti}\ \emph {et~al.}(2024)\citenamefont {Binanti}, \citenamefont {Goldman},\ and\ \citenamefont {Repellin}}]{PhysRevResearch.6.L012054}%
  \BibitemOpen
  \bibfield  {author} {\bibinfo {author} {\bibfnamefont {F.}~\bibnamefont {Binanti}}, \bibinfo {author} {\bibfnamefont {N.}~\bibnamefont {Goldman}}, \ and\ \bibinfo {author} {\bibfnamefont {C.}~\bibnamefont {Repellin}},\ }\href {\doibase 10.1103/PhysRevResearch.6.L012054} {\bibfield  {journal} {\bibinfo  {journal} {Phys. Rev. Res.}\ }\textbf {\bibinfo {volume} {6}},\ \bibinfo {pages} {L012054} (\bibinfo {year} {2024})}\BibitemShut {NoStop}%
\bibitem [{\citenamefont {Carrega}\ \emph {et~al.}(2021)\citenamefont {Carrega}, \citenamefont {Chirolli}, \citenamefont {Heun},\ and\ \citenamefont {Sorba}}]{carrega2021anyons}%
  \BibitemOpen
  \bibfield  {author} {\bibinfo {author} {\bibfnamefont {M.}~\bibnamefont {Carrega}}, \bibinfo {author} {\bibfnamefont {L.}~\bibnamefont {Chirolli}}, \bibinfo {author} {\bibfnamefont {S.}~\bibnamefont {Heun}}, \ and\ \bibinfo {author} {\bibfnamefont {L.}~\bibnamefont {Sorba}},\ }\href@noop {} {\bibfield  {journal} {\bibinfo  {journal} {Nature Reviews Physics}\ }\textbf {\bibinfo {volume} {3}},\ \bibinfo {pages} {698} (\bibinfo {year} {2021})}\BibitemShut {NoStop}%
\bibitem [{\citenamefont {Nakamura}\ \emph {et~al.}(2020)\citenamefont {Nakamura}, \citenamefont {Liang}, \citenamefont {Gardner},\ and\ \citenamefont {Manfra}}]{nakamura2020direct}%
  \BibitemOpen
  \bibfield  {author} {\bibinfo {author} {\bibfnamefont {J.}~\bibnamefont {Nakamura}}, \bibinfo {author} {\bibfnamefont {S.}~\bibnamefont {Liang}}, \bibinfo {author} {\bibfnamefont {G.~C.}\ \bibnamefont {Gardner}}, \ and\ \bibinfo {author} {\bibfnamefont {M.~J.}\ \bibnamefont {Manfra}},\ }\href@noop {} {\bibfield  {journal} {\bibinfo  {journal} {Nat. {Phys.}}\ }\textbf {\bibinfo {volume} {16}},\ \bibinfo {pages} {931} (\bibinfo {year} {2020})}\BibitemShut {NoStop}%
\end{thebibliography}%

\end{document}